\documentclass[11pt]{article}
\pdfoutput=1
\usepackage{jheppub}
\usepackage{amsmath}
\usepackage{bm}
\usepackage{slashed}
\usepackage{graphicx}
\usepackage{enumerate}

\usepackage{longtable}
\usepackage{pifont}
\usepackage{ytableau}

\newlength{\dummysp}
\newcommand{\beq}{\begin{eqnarray}}
\newcommand{\eeq}{\end{eqnarray}}

\newcommand{\gappeq}{\mathrel{\rlap {\raise.5ex\hbox{$>$}}
{\lower.5ex\hbox{$\sim$}}}}
\newcommand{\lappeq}{\mathrel{\rlap{\raise.5ex\hbox{$<$}}
{\lower.5ex\hbox{$\sim$}}}}

\newcommand{\ben}{\begin{enumerate}}
\newcommand{\een}{\end{enumerate}}

\newcommand{\bit}{\begin{itemize}}
\newcommand{\eit}{\end{itemize}}

\def\[{\left [}
\def\]{\right ]}
\def\({\left (}
\def\){\right )}

\title{Classification of compactified $su(N_c)$ gauge theories with fermions in all representations\\
 }
   
\author[a]{Mohamed M. Anber,}\author[b]{Lo\"ic Vincent-Genod} 

\affiliation[a]{Department of Physics, Lewis \& Clark College, 
Portland, OR 97219, USA}  
\affiliation[b]{Institute de Th\' eorie des Ph\' enomen\` es Physiques, \' Ecole Polytechnique F\' ed\' erale de Lausanne, CH-1015 Lausanne, Switzerland}

\emailAdd{manber@lclark.edu}\emailAdd{loic.vincent-genod@epfl.ch} 

\abstract{We classify $su(N_c)$ gauge theories on $\mathbb R^3\times \mathbb S^1$ with massless fermions in higher representations obeying periodic boundary conditions along $\mathbb S^1$. In particular,  we single out the class of theories that is asymptotically free and weakly coupled in the infrared, and therefore, is amenable to semi-classical treatment. Our study is conducted by carefully identifying the vacua inside the affine Weyl chamber using Verma bases and Frobenius formula techniques.  Theories with fermions in pure representations are generally strongly coupled. The only exceptions are the four-index symmetric representation of $su(2)$ and adjoint representation of $su(N_c)$. However, we find a plethora of admissible theories with fermions in mixed representations. A sub-class of these theories have degenerate perturbative vacua separated by domain walls. In particular, $su(N_c)$ theories with fermions in the mixed representations  adjoint$\oplus$fundamental and adjoint$\oplus$two-index symmetric admit degenerate vacua that spontaneously break the parity ${\cal P}$, charge conjugation ${\cal C}$, and time reversal ${\cal T}$ symmetries. These are the first examples of strictly weakly coupled  gauge theories on $\mathbb R^3 \times \mathbb S^1$ with spontaneously  broken ${\cal C}$, ${\cal P}$, and ${\cal T}$ symmetries.  
  We also compute the fermion zero modes in the background of monopole-instantons. The monopoles and their composites (topological molecules) proliferate in the vacuum leading to the confinement of electric charges. Interestingly enough, some theories have also accidental degenerate vacua, which are not related by any symmetry.  These vacua admit different numbers of fermionic zero modes, and hence, different kinds of topological molecules. The lack of symmetry, however, indicates that such degeneracy might be lifted by higher order corrections. Finally, we study the general phase structure of adjoint$\oplus$fundamental theories in the small circle and decompactification limits. 
 }

\begin{document}

\maketitle

\flushbottom

\section{Introduction}

Confining gauge theories that are analytically calculable in four dimensions are scarce. In fact, Seiberg-Witten theory on $\mathbb R^4$ \cite{Seiberg:1994rs,Seiberg:1994aj} and certain QCD-like theories on $\mathbb R^3 \times \mathbb S^1$ (where $\mathbb S^1$ is a spatial rather than thermal circle) \cite{Unsal:2007jx} are the only two known examples. Indeed, it has been known for a long time \cite{Seiberg:1996nz,Davies:1999uw,Davies:2000nw} that compactifying a gauge theory on a circle provides  a mechanism for the gauge group to spontaneously break to its maximum abelian subgroup, and hence, for the theory to admit monopole-instantons. The monopole-instantons or their composites proliferate in the vacuum  causing electric probe charges to confine \cite{Unsal:2007jx}.  Surprisingly, the class of confining gauge theories on a circle has not yet been fully explored. In the present work we pursue a systematic study of these theories.

It was understood since the pioneering work of Polyakov \cite{Polyakov:1975rs,Polyakov:1976fu} that adjoint scalars are needed in order for the gauge group to break to its maximum abelian subgroup $su(N_c)\rightarrow u(1)^{N_c-1}$, and hence, to have an analytical control over the theory. The second homotopy group $\Pi_2\left(SU(N_c)/U(1)^{N_c-1}\right)=\mathbb Z$ (the set of integers) is not trivial, and therefore, one expects to have stable nonperturbative solutions of the field equations. These are the monopoles (or dyons) in $3+1$-D and monopole-instantons in $3$-D. The monopoles carry magnetic charge, and hence, their proliferation in the vacuum causes electric charges to confine. This is an example of the celebrated  dual superconductivity \cite{'tHooft:1995fi,Mandelstam:1974vf,Hosotani:1977cp}. The original mechanism of the monopole condensation in nonabelian theories was introduced by Polyakov in the context of the $3$-D Georgi-Glashow model, while lifting the mechanism to four dimensions was realized in the seminal work of Seiberg and Witten on ${\cal N}=2$ super Yang-Mills. A crucial ingredient in this theory is the scalars in the supermultiplet that cause the gauge group to abelianize, and hence, the monopoles to form.  Then, one can obtain confined electric charges in the IR after softly breaking ${\cal N}=2$ down to ${\cal N}=1$.

${\cal N}=2$ super Yang-Mills is not the only $3+1$-D model that offers an analytical understanding of the dual superconductivity picture in four dimensions. A theory with less amount of supersymmetry, yet  under analytical control, is ${\cal N}=1$ super Yang-Mills formulated on a circle $\mathbb S^1$ \cite{Seiberg:1996nz}, i.e., the theory lives on $\mathbb R^3\times \mathbb S^1$. This is a Yang-Mills theory endowed with a single adjoint Weyl fermion obeying periodic boundary conditions along the circle\footnote{This theory is different from thermal Yang-Mills where the fermions obey anti-periodic boundary conditions.}. If we take the circle circumference $L$ to be much smaller than the strong scale of the theory, i.e., $N_cL\Lambda_{\scriptsize{QCD}}\ll1$, then the theory enters its weakly coupled regime. This can be understood as follows. The gauge component along the compact direction is an adjoint scalar that causes the theory to abelianize, i.e., the gauge group breaks spontaneously to its maximum abelian subgroup. In turn, the Higgs mechanism gives mass of order $1/N_cL$ to all the particles that run inside the vacuum polarization graph, and hence, the four-dimensional coupling of the theory ceases to run at scale $\sim 1/N_cL$. Again, as in the Polyakov model,  one finds that the path integral of the weakly coupled theory is dominated by monopole-instantons that generate a superpotential and in turn a mass gap \cite{Davies:1999uw}. It has also been argued that this theory is continuously connected to super Yang-Mills on $\mathbb R^4$ as we take the radius of the circle to infinity \footnote{This continuity also holds in non-supersymmetric theories with fundamental fermions after turning on a nontrivial flavor space  holonomy \cite{Cherman:2016hcd}.}\cite{Poppitz:2012sw,Cossu:2009sq}.

It was believed for a long time that supersymmetry is inevitable to lift the Polyakov model from $3$-D to $3+1$-D. This belief was due to  lack of understanding of the microscopic mechanism that is responsible for confinement in ${\cal N}=1$ supersymmetry\footnote{However, see \cite{Yung:1987zp} for an early work on the effective scalar potential in supersymmetric theories as  a dilute gas of instanton-anti-instanton molecules.} on $\mathbb R^3 \times \mathbb S^1$. In 2007, it was realized that the mass gap generation in ${\cal N}=1$ super Yang-Mills on $\mathbb R^3 \times \mathbb S^1$ is due to the proliferation of a new kind of topological ``molecules" in the vacuum, dubbed {\em magnetic bions} \cite{Unsal:2007jx}. These are stable correlated events made of two monopole-instantons and carry zero topological charge and two units of magnetic charges. Once these molecules were identified, it was immediately realized that the confinement mechanism transcends the supersymmetric theory to QCD(adj) \cite{Unsal:2007jx}, which is a  Yang-Mills theory endowed with  Weyl fermions in the adjoint representation. For small values of the circle radius and relatively small number of massless Weyl fermions, QCD(adj) on $\mathbb R^3 \times \mathbb S^1$ is a weakly coupled theory in the abelian regime and is under analytical control. Thus, QCD(adj) on $\mathbb R^3 \times \mathbb S^1$ represents a novel setup that lifts the Polyakov confinement mechanism (or the dual superconductivity mechanism) from three to four dimensions without the need to invoke supersymmetry.

Unlike thermal Yang-Mills,  $su(N_c)$ QCD(adj) on $\mathbb R^3 \times \mathbb S^1$ has a preserved center symmetry, thanks to the adjoint fermions. As we dimensionally reduce the theory from four to three dimensions an effective potential of the adjoint scalar \footnote{The adjoint scalar is a Wilson line wrapping the $\mathbb S^1$ circle.} is generated. This potential results from integrating out a tower of Kaluza-Klein excitations of both gauge and fermion fluctuations. While the former tend to break the center, the latter tend to preserve it. Finally, the preservation of the center symmetry results from the winning of the fermions over the gauge fields\footnote{This is true for $n_G>1$ adjoint Weyl fermions. In the supersymmetric case, $n_G=1$, the effective potential vanishes to all orders in perturbation theory and the stability of the center is attributed to nonperturbative contributions; see \cite{Poppitz:2012sw} for details.}. The  unbroken center symmetry guarantees that the resulting three dimensional effective potential will cause $su(N_c)$ to break spontaneously to the maximum abelian subgroup $u(1)^{N_c-1}$. However, it was also realized that a preserved center symmetry is not a necessary condition for the gauge group to abelianize \cite{Poppitz:2009tw}.  A less restrictive condition is that the global minimum of the effective potential should lie inside the fundamental domain of the field space (the affine Weyl chamber). A minimum that lies on the boundary of the fundamental domain means that the groups $su(N_c)$ breaks partially to $u(1)^{b}$ subgroup, where $0\leq b<N_c-1$, leaving behind a nonabelian part that is strongly coupled.  

During the last decade, there has been a tremendous amount of effort to study compactified gauge theories. This includes confinement/deconfinement phase transitions \cite{Anber:2011gn,Anber:2012ig,Anber:2013doa}, quantum/thermal continuity between compactified super Yang-Mills and hot pure Yang-Mills \cite{Poppitz:2012sw,Anber:2013sga,Poppitz:2013zqa,Anber:2014lba}, the global structure \cite{Anber:2015wha},  QCD under external magnetic field \cite{Anber:2013tra} and at finite density \cite{Kanazawa:2017mgw}. Recently, there has also been a progress in the  lattice simulations of compactified super Yang-Mills \cite{Bergner:2014dua}. All these studies have shown a striking qualitative agreement between theories on $\mathbb R^3\times \mathbb S^1$ and their cousins on $\mathbb R^4$. We strongly believe that this agreement justifies considering this class of theories seriously.

In the present work we systematically study  $su(N_c)$ QCD-like theories on $\mathbb R^3\times \mathbb S^1$ with massless fermions in higher dimensional representations. In particular, we study theories in pure representations $\cal R$ and mixed representations $G\oplus {\cal R}$, where $G$ is the adjoint representation. Our aim is to single out the admissible theories that are amenable to semi-classical analysis. This is the beginning of a series of works that closely examine the confining class of theories defined over a circle. The ultimate goal of these studies is to shed more light on the role of fermions in confining gauge theories and understand the microscopic structure of the topological molecules responsible for confinement. 

By admissible theories we mean theories that satisfy the following four criteria: (a) they are asymptotically free, (b) they are free of gauge and Witten anomalies, (c) they have global minima inside (and not on the boundary of) the affine Weyl chamber, and (d) they do not have light or massless charged fermions under $u(1)^{N_c-1}$ in the infrared. As a byproduct, we also determine the number of the fermion zero modes attached to the  monopole-instantons in all admissible theories. The zero modes play an important role in forming the topological molecules, a subject that we pursue in a future work. 

Determining the global minimum of the effective potential as well as the fermion zero modes requires the computation of traces in general representations. These computations are carried out using two different methods. In the first method we make use of the Frobenius formula, which gives the trace of a representation ${\cal R}$ in terms of the trace of the fundamental (defining) representation. In the other method we calculate the trace by explicitly constructing the weights of ${\cal R}$ using Verma bases. We used both methods as a cross check on our calculations.

For the reader who is interested only in the final product, our results are discussed in Section \ref{The admissible class of theories} and can be summarized as follows:
\begin{enumerate}

\item Pure representations: the admissible theories are those with fermions in the adjoint representations. The four-index symmetric representation of $su(2)$ is the only addition. 

\item Mixed representations: here we find a plethora of admissible theories that are discussed in Section  \ref{The admissible class of theories} and displayed in Table \ref{summary of admissible theories}.   

\end{enumerate}
  We also display the number of the fermion zero modes attached to the monopole-instantons that correspond to the simple and affine roots of the algebra $\{\bm \alpha_0,\bm\alpha_1,...,\bm \alpha_{N_c-1}\}$.

Interestingly enough, we find a subclass of the admissible theories that have  degenerate vacua. For example, theories with an odd number of colors and fermions in the fundamental representation (with an appropriate number of adjoint fermions to ensure the existence of the vacua inside the affine Weyl chamber) have two degenerate vacua that break the parity ${\cal P}$, charge conjugation ${\cal C}$, and time reversal ${\cal T}$ symmetries. Also, theories with an even number of colors and fermions in the two-index symmetric representation (again with adjoint fermions) have two degenerate vacua that break the center ${\mathbb Z_2}$, ${\cal P}$, ${\cal C}$, and ${\cal T}$ symmetries.  In addition, some theories have also accidental degenerate vacua, which are not related by any symmetry.  These vacua admit different number of fermionic zero modes, and hence, different kinds of topological molecules. The lack of symmetry, however, may indicate that such degeneracy can be lifted by higher order corrections. Detailed discussion of these theories is given in Section \ref{The admissible class of theories}. 

Our work is organized as follows. In Section \ref{Theory and formulation} we formulate our problem and fix the notation.  In Section \ref{Asymptotically and anomaly free theories} we sort out the asymptotically free and anomaly free gauge field theories. Next, we discuss the generation of the effective potential in Section \ref{Integrating out the Kaluza Klein tower: the effective potential}.  In Section \ref{Computation of the traces} we implement the Frobenius formula and Verma bases to compute the traces in a general representation. Then, we  search for the global minima of the effective potential and display our results in Section \ref{Numerical investigation}. The IR flow of the effective $3$-D coupling constant of our theories is discussed in Section \ref{The 3D coupling constant}.   The admissible class of theories, which are mathematically well-defined and well-behaved in the IR, is discussed in Section \ref{The admissible class of theories}. The fermion zero modes in the background of monopole-instantons are calculated in Section \ref{Monopole-instantons and fermions zero modes}. In Section \ref{fermions in adj plus fund} we give a detailed description of the phase structure of theories with fermions in the  adjoint$\oplus$fundamental representation, discuss various interesting physical phenomena, and study the decompactification limit. This serves as a prototype example of the rich phase structure of theories with mixed representations. We finally conclude in Section \ref{Discussion and Final Remarks} by giving a brief summary of our work and plans for future directions. In Appendices \ref{Lie Algebra and Conventions} to \ref{Computing the index using Frobenius formula} we review Lie Algebra, explain the convention, and list a few useful results in group theory that we use throughout this work.

\section{Theory and formulation}
\label{Theory and formulation}

We consider $su(N_c)$ Yang-Mills theory endowed with $n_{G}$  massless adjoint Weyl fermions, $\psi$, along with  $n_{\cal R}$  massless Weyl fermions,  $\chi_{\cal R}$, in a general representation ${\cal R}$ of $su(N_c)$. We consider the theory on $\mathbb R^3 \times \mathbb S^1$, where $\mathbb S^1$ is a circle of radius $R$. The fermions are given periodic boundary conditions along the circle. The action of the system is given  by
\begin{eqnarray}
S=\frac{1}{g^2}\int_{\mathbb R^3 \times \mathbb S^1}\mbox{tr}\left[-\frac{1}{2}F_{MN}F^{MN}+i\bar\psi^I\bar \sigma^M D_{M}^{G}\psi^I+i\bar\chi^I_{{\cal R}}\bar \sigma^M D_{M}^{{\cal R}}\chi^I_{{\cal R}}\right]\,,
\label{the main Lagrangian}
\end{eqnarray}
where the Latin letters $M,N=0,1,2,3$ denote the spacetime dimensions and the circle is taken along the spatial $x^3$ direction: $x^3\equiv x^3+2\pi R$. The index I denotes the flavor, $\bar \sigma^\mu=\left(1,-\sigma^i\right)$, and $\{\sigma^i\}$ are the Pauli matrices. The covariant derivative is  $D_{M}^{{\cal R}}=\partial_M +i A_M^a t^a_{\cal R}$, where  $\{t_{{\cal R}}^a\}$, $a=1,2,...,N_c^2-1$, are the generators of $su(N_c)$ in the representation ${\cal R}$. The generators satisfy the Lie algebra $\left[t^a,t^b\right]=if^{abc}t^c$, where $f^{abc}$ are the group structure constants. The field strength is given by $F_{MN}=\partial_M A_N-\partial_N A_M-i\left[A_N,A_M\right]$, where $A_N$ are the gauge fields. We assume that the theory is asymptotically free, and hence, the theory is strongly coupled in the infrared at scale $\Lambda_{\scriptsize\mbox{QCD}}$. At this stage the reader might refer to Appendix \ref{Lie Algebra and Conventions} for a review of Lie Algebra and the conventions used in this work. 

  If the radius of the circle is taken to be much smaller than $1/N_c\Lambda_{\scriptsize\mbox{QCD}}$, then the theory is weakly coupled and we can apply perturbation theory. Let us denote by $X$ any of the gauge or fermion fields. Then, we can always decompose these fields in the Weyl-Cartan bases $\{\bm H, E_{\bm \beta}, E_{-\bm\beta }\}$, as follows:
\begin{eqnarray}
X=X^at^a=\bm X\cdot \bm H+\sum_{\bm \beta_{+}}X_{\bm \beta}E_{\bm \beta}+\sum_{\bm \beta_{+}}X_{\bm \beta}E_{-\bm \beta}\,,
\label{expansion in Cartan basis}
\end{eqnarray}	
where $\{\bm \beta_+\}$ is the set of all positive roots, $\bm H$ are the Cartan generators, $E_{\pm\bm \beta}$ are the raising (lowering) operators, and the bold face vector  $\bm X=\left(X_1,X_2,...,X_{N_c-1}\right)$ denotes the field components along the Cartan generators. Confinement is an infrared phenomenon, and hence, we are interested in length scales much larger than $L$. Therefore, we can dimensionally reduce\footnote{We want to emphasize that the dimensional reduction from  $\mathbb R^3\times \mathbb S^1$ to $\mathbb R^3$ still remembers about the four-dimensional nature of the theory.} our theory from $\mathbb R^3\times \mathbb S^1$ to $\mathbb R^3$. Upon dimensional reduction, the quantum fluctuations will generate a vacuum expectation value (vev) of the gauge field component along the $x^3$ direction.  The vev can always be chosen to lie along the Cartan generators by means of a similarity transformation, i.e., we have $\langle A_3^at^a \rangle=\langle \bm A_3\cdot \bm H\rangle $.   Writing $\bm A_3=\frac{\bm \Phi}{L}$, where $L=2\pi R$ is the circumference of the $\mathbb S^1$ circle, then the bosonic part of the 3-dimensional action reads
\begin{eqnarray}
S_{3d\,\scriptsize\mbox{bos}}=\frac{L}{g^2}\int_{\mathbb R^3}\mbox{tr}\left[ -\frac{1}{2}F_{\mu\nu}F^{\mu\nu}+\frac{D_\mu \bm\Phi D^\mu \bm\Phi}{L^2}\right] +V_{\scriptsize\mbox{eff}}(\bm\Phi)\,,
\label{3d action}
\end{eqnarray}
where $\mu,\nu=0,1,2$ and the potential $V(\bm\Phi)$ results from integrating out the Kaluza Klein excitations of the gauge and fermionic fields along the $x^3$ direction. In fact, the effective action (\ref{3d action}) is the three-dimensional Georgi-Glashow model where the gauge field component along the $x^3$ direction $\bm \Phi$ is an adjoint scalar. The reduced theory, however, still remembers its four dimensional origin which manifests in the fact that $V_{\scriptsize\mbox{eff}}(\bm\Phi)$ is a periodic function of $\bm \Phi$ with periodicities $\bm \Phi\equiv \bm \Phi+2\pi\bm \alpha^*_a$, $a=1,2,...,N_c-1$, where $\bm\alpha^*_a$ are the co-roots, see e.g. \cite{Anber:2014lba} for details.  Now, if the vev $\bm \Phi$ lies inside the affine Weyl chamber (we define the affine Weyl chamber in Section \ref{Numerical investigation}), then the theory abelianizes, i.e., the group $su(N_c)$  spontaneously breaks into its maximal abelian subgroup $u(1)^{N_c-1}$. Then, the gauge and fermionic fields along $\{E_{\bm \beta}\}$ acquire masses $\sim n/N_cL$ (these are the W-bosons and the fermionic counterparts), while the adjoint scalar acquires a mass $\sim gn/N_cL$, where $n=1,2,...$ accounts for the Kaluza Klein tower. Therefore,  the long-distance bosonic three dimensional action reads
\begin{eqnarray}
S_{3d\,\scriptsize\mbox{eff bos}}=\frac{L}{g^2}\int_{\mathbb R^3} -\frac{1}{2}\bm F_{\mu\nu}\bm F^{\mu\nu}\,,
\label{3D action in terms of F}
\end{eqnarray}
and we have used $\mbox{tr}\left[H^iH^j \right]=\delta^{ij}$, $i,j=1,2,...,N_c^2-1$. Thus, the effective bosonic theory is a collection of non-interacting three dimensional photons. Since photons in $3$-D have only one degree of freedom, one can use a dual description such that
$
\bm F^{\mu\nu}=\frac{g^2}{4\pi L}\epsilon^{\mu\nu\alpha}\partial_{\alpha}\bm \sigma\,,
$
where $\bm \sigma$ is the dual photon field. Thus, the action (\ref{3D action in terms of F}) becomes
\begin{eqnarray}
S_{3d\,\scriptsize\mbox{eff bos}}=\frac{g^2}{16\pi^2 L}\int_{\mathbb R^3} \left(\partial_\mu\bm\sigma\right)^2\,.
\label{3D action in terms of sigma}
\end{eqnarray}

Fortunately enough, the story does not stop here since there is a non-perturbative sector that has to be taken into account. In fact, the theory also admits composite instantons (e.g. bions) that condense in the vacuum and causes the theory to confine. This mechanism has been elucidated  in previous publications \cite{Unsal:2007jx,Poppitz:2009tw,Poppitz:2009uq, Anber:2011de} for various theories with various fermion contents, and we only mention it briefly in the conclusion. One of the main purposes of the present work, however, is to perform a systematic study of the potential $V(\bm \Phi)$ to determine the class of theories that contain fermions in various representations and yet enjoy stable vevs, and hence, may provide an excellent laboratory to understand the effect of fermions on the confinement phenomenon. The study of the composite instantons in these theories will be pursued in great details in a future work. The validity of our analysis hings on the validity of perturbation analysis at small circle radius, and hence, we first have to examine whether a theory, with a specific number of $n_{G}$ and $n_{\cal R}$ fermions, is weakly coupled in the IR. An abelian weakly coupled theory in the IR demands that (i) the UV theory is asymptotically free and (ii) the absence of light charged particles in the IR. The next section is devoted to study the theories that contain an arbitrary number of fermions in a general representation, and yet, are asymptotically free.

\section{Asymptotically and anomaly free theories}
\label{Asymptotically and anomaly free theories}

\subsection{Asymptotically free theories}
\label{The asymptotic free theories}

In this section we perform a systematic study to determine the asymptotically free $su(N_c)$ Yang-Mills theory with  $n_{G}$  and  $n_{\cal R}$  massless Weyl fermions. The two-loop $\beta$ function of this theory is given by \cite{Caswell:1974gg,Dietrich:2006cm}
\begin{eqnarray}
\nonumber
\beta(g)&=&-\beta_0\frac{g^3}{(4\pi)^2}-\beta_1\frac{g^5}{(4\pi)^4}\,,\\
\nonumber
\beta_0&=&\frac{11}{6}C_2(G)-\frac{1}{3}T(G)n_{G}-\frac{1}{3}T({\cal R})n_{\cal R}\,,\\
\nonumber
\beta_1&=&\frac{34}{12}C_2^2(G)-\frac{5}{6}n_G C_2(G)T(G)-\frac{n_G}{2}C_2(G)T(G)-\frac{5}{6}n_{\cal R} C_2(G)T({\cal R})-\frac{n_{\cal R}}{2}C_2(R)T({\cal R})\,,\\
\label{beta function}
\end{eqnarray} 
where $C_2({\cal R})$ and $T({\cal R})$ are respectively the Casimir and trace operators of representation ${\cal R}$, see Appendix  (\ref{The Casimir and trace operators, and the dimension of representation}) for the expressions of these operators.

Using this information, we find the condition that the theory remains asymptotically free, i.e., $\beta_0>0$, is given by 
\begin{eqnarray}
n_{G}+\frac{C_2({\cal R})d({\cal R})}{2N_c(N_c^2-1)}n_{{\cal R}}<\frac{11}{2}\,,
\label{the main condition}
\end{eqnarray}
where $d({\cal R})$ is the dimension of the representation.  
This equation tells us that for a large number of fermions or for representations with large dimensions the screening effect of the fermions overcomes the anti-screening of the gluons and the theory looses its asymptotic freedom. 

We are interested in weakly coupled theories in the IR, and therefore, we first check that our theories are asymptotically free in the UV, i.e., they respect the inequality (\ref{the main condition}). In the second column of Tables \ref{su2 table} to \ref{su8 table} we list all asymptotically free representations of $su(N_c)$ with $2\leq N_c\leq 8$. It is trivial to see that the number of allowed representations decreases as we use more adjoint fermions. Also, theories with a larger number of colors admit more asymptotically free representations, which can be easily found from (\ref{the main condition}). However, it is extremely challenging  to numerically study the effective potential beyond $su(8)$, see Section \ref{Numerical investigation}, and hence, we limit our analysis to $2\leq N_c\leq 8$. Once the asymptotically free theories are identified, the next step is to  make sure that the theory does not have an anomaly. 

\subsection{Anomalies}
\label{Anomalies}

\subsection*{Gauge anomaly}
\label{Gauge anomaly}

Theories with complex representations suffer from gauge anomalies. The anomaly is given by
\begin{eqnarray}
\mbox{tr}_{\cal R}\left[\left\{t_a,t_b \right\}t_c\right]=d_{abc}A({\cal R})\,,
\end{eqnarray}
where $A({\cal R})$ is an integer called the cubic Dynkin index or the anomaly of the representation and $d_{abc}$ is a symmetric third-rank tensor made out of the structure constants of the algebra $f_{abc}$. The cubic Dynkin index vanishes for all simple Lie algebras except $su(N_c)$ for $N_c \geq 3$ (and $so(6)$ which is isomorphic to $su(4)$). The values of $A({\cal R})$ are given in Appendix \ref{Cubic Dynkin index} for the asymptotically free representations we encounter in this work. There we also show that real representations (the ones with Dynkin labels that satisfy $(m_1,m_2,...,m_r)=(m_r,m_{r-1},...,m_1)$) have vanishing cubic Dynkin indices. Thus, in order for a theory to be anomaly free the Weyl fermions have to belong to a real representation, or otherwise we need to consider Dirac fermions (i.e. an even number of Weyl fermions).

\subsection*{Witten anomaly}
\label{Terminal anomaly}
Some of the theories with fermions in a particular representation are not mathematically defined since they have Witten (terminal) anomaly \cite{Witten:1982fp,AlvarezGaume:1983ig}. A good diagnose of such theories is that they have an odd number of fermionic zero modes in the background of a  Belavin-Polyakov-Schwarz-Tyupkin (BPST) instanton \cite{Belavin:1975fg}. Atiyah-Singer index theorem \cite{Atiyah:1968mp} gives the number of fermionic zero modes in the background of a BPST instanton as
\begin{eqnarray}
{\cal I}_f({\cal R})=n_{\cal R}T({\cal R}){\cal K}\,,
\end{eqnarray}
where 
\begin{eqnarray}
{\cal K}=\frac{1}{16\pi^2}\int d^4x F_{MN}^a\tilde F_{MN}^a\,,
\end{eqnarray}
is the instanton number. 

In Tables \ref{su2 table} to \ref{su8 table} we compute Atiyah-Singer index in the background of a single BPST instanton for the asymptotically free $su(N_c)$ theories with $2\leq N_c\leq 8$ and fermions in representations $n_G\oplus n_{\cal R}$. Theories with an odd index have Witten anomalies, and hence,  are ill-defined. For example, $su(2)$ with $n_F=1$ is ill defined despite the fact that it is gauge anomaly free. 

Having found all asymptotically and anomaly free theories, the next task will be summing up contributions from the Kaluza-Klein tower. This is discussed in the next section. 

\section{Integrating out the Kaluza Klein tower: the effective potential}
\label{Integrating out the Kaluza Klein tower: the effective potential}

In this section we apply dimensional reduction to our theory, Eq. (\ref{the main Lagrangian}), assuming that condition (\ref{the main condition}) holds, i.e., the theory is asymptotically free, and hence, perturbation theory is at work. Upon dimensionally reducing the theory from four to three dimensions we obtain a tower of heavy Kaluza-Klein excitations that can be integrated out. The effect of the tower on the low energy phenomena can be taken into account by calculating the effective potential $V(\bm \Phi)$. The purpose of this section is to elucidate this calculation which originally appeared in the pioneering work of Gross, Pisarski, and Yafee \cite{Gross:1980br}. In the following, we calculate $V(\bm \Phi)$ for fermions in a generic representation ${\cal R}$. The calculations for the gauge fields follow the exact same steps. The reader can refer to \cite{Gross:1980br} for details.  

The calculations start by assuming that the potential develops a holonomy $\bm \Phi$ along the $x^3$-direction, and therefore, we seek a  self-consistent solution in the sense that at the end of the calculations one should check that the potential yields a minimum at $\bm \Phi$. To this end, we write (from here on we remove the subscript ${\cal R}$ to reduce notational clutter)
\begin{eqnarray}
A_M(\vec x, x^3)=\frac{\Phi^iH^i}{L}\delta_{M,3}+{\cal A}_M(\vec x, x^3)\,,
\end{eqnarray}
 where ${\cal A}_M$ are the field fluctuations, and expand the fields $\chi$ and $A_M$ as in (\ref{expansion in Cartan basis}).  Explicitly, we have
\begin{eqnarray}
\nonumber
A_M&=&A_M^a t^a=A_M^i H^i+\sum_{\bm\beta_+}A_M^{\bm \beta}E_{\bm \beta}+\sum_{\bm\beta_+}A_M^{*\bm \beta}E_{-\bm \beta} \,, \\
\chi&=& \chi^a t^a=\chi^i H^i+\sum_{\bm \beta_+}\chi^{\bm\beta} E_{\bm \beta}+\sum_{\bm \beta_+} \chi^{-\bm \beta}E_{-\bm \beta} \,.
\label{the expansion for fermion and gauge boson}
\end{eqnarray}
Substituting Eq. (\ref{the expansion for fermion and gauge boson}) into Eq. (\ref{the main Lagrangian}) and using $\left[H^i, H^j\right]=0$ ,$\left[H^i, E_{\pm \bm \beta}\right]=\pm \beta^iE_{\pm \bm \beta}$, and $\mbox{tr}\left[t^at^b\right]=\delta^{ab}$, we obtain
\begin{eqnarray}
S_{\chi}=\int_{\mathbb R^3 \times \mathbb S^1}i\left[\bar \chi \bar\sigma_\mu^m \partial^\mu\chi^m+ \sum_{n=1}^{d({\cal R})}  \bar \chi^m\bar \sigma_3 \frac{\bm \Phi\cdot \bm\mu_n}{L}\chi_m+\mbox{interaction terms} \right]\,,
\label{simple fermion action}
\end{eqnarray}
where $\{\bm \mu_n\}$ is the set of all weights of the ${\cal R}$ representation and $d({\cal R})$ is its dimension. Since the action (\ref{simple fermion action}) is quadratic in $\chi$, we can readily integrate out the $\chi$ field. Assuming periodic boundary conditions for the fermions along the spatial circle, and performing continuous and discrete Fourier transforms along $x^{0,1,2}$ and $x^3$, respectively, we obtain the one-loop effective potential
\begin{eqnarray}
\nonumber
V_{\chi}(\bm \phi)&=&-\sum_{m=1}^{d({\cal R})}\sum_{p \in \mathbb Z}\int \frac{d^3k}{(2\pi)^3}\log\left[k^2+\left(\frac{2\pi p}{L}+\frac{\bm \Phi\cdot \bm \mu_m}{L}\right)^2 \right]\\
&\equiv& -\sum_{p \in \mathbb Z}\mbox{tr}_{\cal R}\left\{\int \frac{d^3k}{(2\pi)^3}\log\left[k^2+\left(\frac{2\pi p}{L}+\frac{\bm \Phi\cdot \bm H}{L}\right)^2 \right]\right\}\,.
\label{unregulaized effective potential}
\end{eqnarray}
The potential in Eq. (\ref{unregulaized effective potential}) is UV divergent and it needs to be regularized before we can make use of it. The regularization can be performed using the zeta-function technique.\footnote{Basically, this can be done by writing 
\begin{eqnarray}
\sum_{p \in \mathbb Z}\int \frac{d^3k}{(2\pi)^3}\log \left[ k^2+ (p+a)^2 \right]=-\mbox{lim}_{s\rightarrow 0}\frac{d}{ds}\left\{\sum_{p \in \mathbb Z}\int \frac{d^3k}{(2\pi)^3} \left[ k^2+ (p+a)^2 \right]^{-s}\right\}\,,
\end{eqnarray}
and then using the identity \cite{Ponton:2001hq,Elizalde:1994gf}
\begin{eqnarray}
\sum_{p \in \mathbb Z}\left[c+(p+a)^2  \right]^{-s}=\frac{\sqrt{\pi}}{\Gamma[s]}|c|^{1-2s}\left[\Gamma\left(s-\frac{1}{2}\right)+4\sum_{p=1}^\infty \left(\pi p |c|\right)^{s-\frac{1}{2}}\cos\left(2\pi p a\right)K_{s-\frac{1}{2}}\left(2\pi p|c|\right) \right]\,,
\end{eqnarray}
where $K_{n}$ are the Bessel functions of the second kind. Finally, one can extract the physically relevant information by carefully taking the limit $s\rightarrow 0$.}

One can also follow the same steps to calculate the effective potential that results from integrating out the Kaluza-Klein tower of the gauge field. The final form of the effective potential reads
\begin{eqnarray}
V_{\scriptsize\mbox{eff}}\left(\bm \Phi\right)=\frac{2}{\pi^2L^3}\sum_{p=1}^\infty\frac{1}{p^4}\left\{n_{{\cal R}}\mbox{tr}_{{\cal R}}\left[\cos\left(p \bm \Phi \cdot \bm H\right)\right]+(n_{G}-1)\mbox{tr}_{G}\left[\cos\left(p \bm \Phi \cdot \bm H\right)\right]  \right\}\,.
\label{final form of the potential}
\end{eqnarray}
One of the main purposes of this work is to determine whether the potential (\ref{final form of the potential}) has non-trivial minima. As we emphasized above, the presence of non-trivial minima means that the gauge group $su(N_c)$ breaks spontaneously down to its maximum abelian subgroup $u(1)^{N_c-1}$. This in turn makes it possible to tackle the theory, e.g., understand the confinement dynamics, by analytical means.

The potential (\ref{final form of the potential}) is expressed as a sum over the weight vectors $\{\bm \mu\}$ of representation ${\cal R}$. At this point, we can proceed either by explicitly constructing the weights of ${\cal R}$ or using Frobenius formula. In the next section we elaborate on these two different methods.

\section{Computation of traces}
\label{Computation of the traces}

A direct computation  of the traces in (\ref{final form of the potential}) for a general representation can be circumvented by using the Frobenius formula, which will be discussed in this section. However, in this work we will also need to determine the number of fermionic zero modes in the background of instantons and  check whether any of the charged fermions in the IR are massless. These two pieces of computations require the explicit knowledge of the weights in every representation. In fact, computing the traces using both the explicit weights and Frobenius formula  works as a cross check on our computations. In the rest of this section we discuss both methods in some detail.  

\subsection{Constructing the weights using Verma bases}

Obtaining the weights of a particular representation can be a formidable task. Fortunately enough, this task is made possible by exploiting  a method that was originally due to  Verma and then further developed by Li, Moody, Nicolescu, and Patera \cite{:/content/aip/journal/jmp/27/3/10.1063/1.527222}, and can be applied to construct all finite dimensional irreducible representations\footnote{In fact, this method can also be applied to construct all finite dimensional irreducible representations of the simple Lie algebras of types $B_n$ and $C_n$ ($2\leq n\leq 6$), $D_n$ ($4\leq n \leq 6$), and $G_2$; see \cite{:/content/aip/journal/jmp/27/3/10.1063/1.527222} for details.} of $su(N_c)$. This method is a set of basis-defining inequalities which satisfy the following criteria. (1) The inequalities define a set of linearly independent vectors that span the whole representation space. (2) The number of inequalities is equal to the number of the positive roots. (3) The bases provided by this method consist of eigenvectors of the Cartan subalgebra and thus each basis vector is labeled by additive quantum numbers that are the components of a weight of the representation.    

For a given  representation ${\cal R}$ of $su(N_c)$, which we denote by ${\cal R}=(m_1,m_2,...,m_{N_c-1})$ (see Appendix \ref{Lie Algebra and Conventions} for a review of Lie Algebra and the conventions used in this work), the basis vectors are
\begin{eqnarray}
\nonumber
&&\left[\left(E_{-\bm \alpha_1} \right)^{a_N}\left(E_{-\bm \alpha_2} \right)^{a_{N-1}}...\left(E_{-\bm \alpha_{N_c-1}} \right)^{a_{N-N_c+2}} \right]\left[\left(E_{-\bm \alpha_1} \right)^{a_{N-N_c+1}}...\left(E_{-\bm \alpha_{N_c-2}} \right)^{a_{N-2N_c+4}} \right]\\
&&\times...\left[\left(E_{-\bm \alpha_1} \right)^{a_3}\left(E_{-\bm \alpha_2} \right)^{a_2}\right]\left(E_{-\bm \alpha_1} \right)^{a_1}|{\cal R}\rangle\,,
\label{verma labels}
\end{eqnarray}
where $N=N_c(N_c-1)/2$ and $\{E_{-\bm \alpha_a} \}$, $a=1,2,...,N_c-1$ is the set of the simple-root generators. The coefficients $\{a_i\}$ satisfy a set of inequalities given in Table \ref{verma basis table} in Appendix (\ref{Constructing the wights using Verma bases}), where  we also give an example of using the Verma bases to construct the weights of $su(3)$.

The set of rules used to construct the weights in any representation can be easily coded and implemented in a numerical scheme to obtain the global minimum of the effective potential. Before doing that, we will also discuss the Frobenius formula that can be used to circumvent the explicit construction of the weights. 

\subsection{The Frobenius formula}

In the rest of this section we apply the Frobenius formula in order to compute the traces in Eq. (\ref{final form of the potential}). We first note that one can write $\mbox{tr}_{{\cal R}}\left[\cos\left(p \bm \Phi \cdot \bm H\right)\right]$ as $\mbox{Re}\left\{\mbox{tr}_{{\cal R}}\left[e^{ip \bm \Phi \cdot \bm H}\right]\right\}$. Defining the Polyakov loop along the compact circle as 
\begin{eqnarray}
\Omega\equiv e^{i \bm \Phi \cdot \bm H}\,,
\end{eqnarray}
we find that the effective potential can be expressed as 
\begin{eqnarray}
V_{\scriptsize\mbox{eff}}\left(\bm \Phi\right)=\frac{2}{\pi^2L^3}\sum_{p=1}^\infty\frac{\mbox{Re}\left\{n_{{\cal R}}\mbox{tr}_{{\cal R}}\left[\Omega^p\right]+(n_{G}-1)\mbox{tr}_{G}\left[\Omega^p\right]  \right\}}{p^4}\,.
\label{final form of the potential in terms of the polyakov loop}
\end{eqnarray}

Now, we are in a position to compute the traces in a general representation ${\cal R}$ using the Frobenius formula. For $P\in su(N_c)$, the Frobenius formula gives the trace of $P$ in the representation $(n,\vec 0)\equiv(n,0,0,...,0)$ in terms of the trace of the fundamental (defining) representation as \cite{Myers:2009df,Gross:1993yt}
\begin{eqnarray}
\mbox{tr}_{(n,\vec 0)}P=\sum_{j \in S_n}\frac{1}{\prod_{k=1}^n k^{j_k}j_k!}\left(\mbox{tr}_F P\right)^{j_1}\left(\mbox{tr}_F P^2\right)^{j_2}...\left(\mbox{tr}_F P^n\right)^{j_n}\,,
\label{Frobenius formula}
\end{eqnarray}
where the sum is over all the $n!$ permutations $\{j_1,j_2,...,j_n\}$ of the symmetric group $S_n$. The vector  $\{j_1,j_2,...,j_n\}$ can be obtained as the solution of the equation $1j_1+2j_2+...+nj_n=n$ for all integers $j_i\geq 0$. The traces for a general representation ${\cal R}=(m_1,m_2,...,m_r)$ can be obtained using tensor methods and Young tableau. In Appendix \ref{Using the Frobenius formula} we list $\mbox{tr}_{\cal R}P$ for all the needed representations in this work, i.e., the representations that are asymptotically free.

In this section we introduced all the necessary tools needed to compute the traces. In the next section we study the minimization problem of the effective potential.

\section{The global minima of the effective potential}
\label{Numerical investigation}

 The existence of a minimum (or minima) of the effective potential inside (and not on the boundary of) the affine Weyl chamber (defined below) guarantees that the gauge group $su(N_c)$ breaks down to its maximum abelian subgroup $u(1)^{N_c-1}$. In this section we search for the minima of the effective potential using both analytical and numerical means.  The analytical method is a succinct  way to understand the behavior of the minima in the presence of fermions in mixed representations. In addition, it provides us with valuable information that we can use to check our numerical calculations. Unfortunately, it is  extremely difficult to  analytically  find the minima of the potential for any group beyond $su(3)$. 

\subsection{The affine Weyl chamber}

\begin{figure}[t] 
   \centering
   \includegraphics[width=2in]{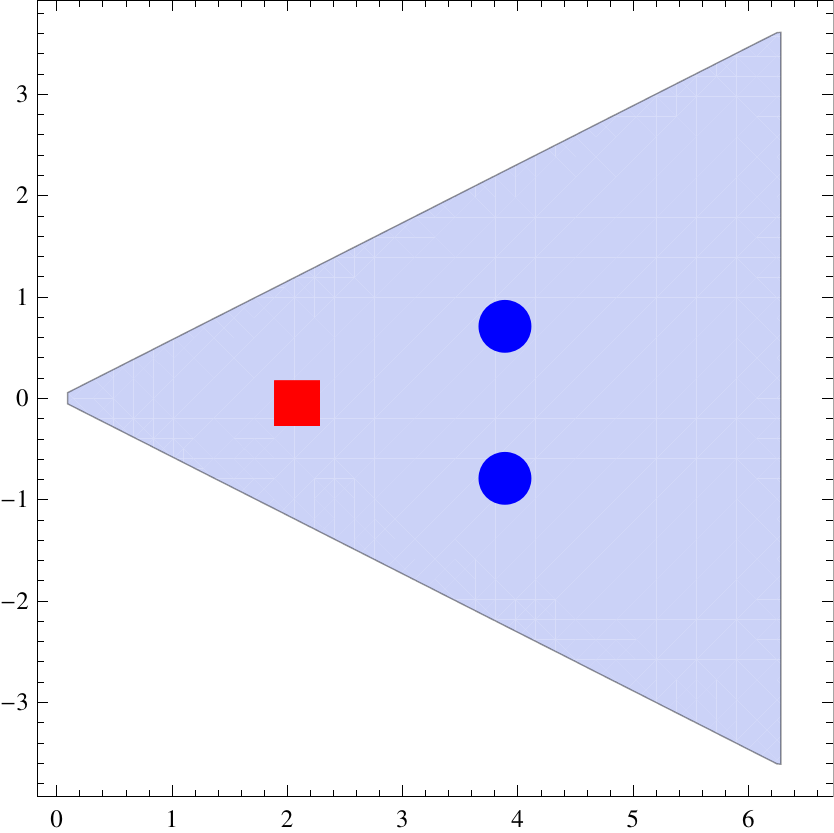} 
   \caption{The shaded region is the affine Weyl chamber of $su(3)$. The horizontal and vertical axes are $\phi_1$ and $\phi_2$, respectively. We take the simple roots to be $\bm \alpha_1=\left(\frac{1}{2},\frac{\sqrt 3}{2}\right)$ and $\bm \alpha_2=\left(\frac{1}{2},-\frac{\sqrt 3}{2}\right)$.  The blue circles indicate the location of the two degenerate vacua  in the case $\left(n_G=2\right)\oplus\left( n_{(20)}=2\right)$. The red square is the center symmetric point, which is the vacuum location for $1\leq n_G \leq 5$ adjoint fermions.}
   \label{su3 Weyl chamber}
\end{figure}

The affine Weyl chamber is the region of physically inequivalent values of $\bm \Phi$. This region is a polyhedron in $N_c-1$ dimensional space defined by the inequalities
\begin{eqnarray}
\bm \alpha_a\cdot\bm \Phi>0 \quad \mbox{for all}\quad a=1,2,3,...,N_c-1\quad \mbox{and} \quad-\bm\alpha_0\cdot \bm \Phi<2\pi\,,
\end{eqnarray}
where $\bm \alpha_a$, $a=1,2,...,N_c-1$, are the simple roots and $\bm\alpha_0$ is the affine root which is given by
\begin{eqnarray}
\bm \alpha_0=-\sum_{a=1}^{N_c-1}\bm \alpha_a\,.
\end{eqnarray} 
The interior of this region (not including the boundary) is the smallest region in the $\bm \Phi$-space with no massless $W$-bosons (including their Kaluza-Klein excitations), see \cite{Argyres:2012ka} for more details. The existence of the global minimum on the boundary means that one (or several)  $W$-bosons become(s) massless, which makes the theory strongly coupled and invalidates its semi-classical description. As an example,  in Figure \ref{su3 Weyl chamber}  we plot the affine Weyl chamber of $su(3)$ and indicate the locations of the minima  in the two cases $1\leq n_G \leq 5$ and $\left(n_G=2\right)\oplus \left(n_{(20)}=2\right)$ (two fermions in the adjoint and two fermions in the two-index symmetric representations). In the latter case we see that the potential admits two degenerate vacua, as we discuss extensively in Section \ref{Perturbative vacua and the role of discrete symmetries}. In both cases the minima are located inside the affine Weyl chamber. 

In addition, we also define the quantity
\begin{eqnarray}
{\cal Z}=\frac{\left|\mbox{tr}_{F}\left[e^{i\bm \Phi_{\scriptsize \mbox{min}}\cdot \bm H}\right]\right|}{N_c}\,.
\label{definition of Z}
\end{eqnarray}
Theories that respect the center symmetry of $su(N_c)$ have ${\cal Z}=0$ at $\bm \Phi_{\scriptsize \mbox{min}}$. On the other hand, theories that break the center of the group badly, for example theories in the deconfined phase, have ${\cal Z}\cong1$. Since $0\leq \left|\mbox{tr}_{F}\left[e^{i\bm \Phi_{\scriptsize \mbox{min}}\cdot \bm H}\right]\right|\leq  N_c$, we find that $0\leq {\cal Z}\leq 1$ is a gauge invariant  measure of the distance between the global minimum and center symmetric point.

\subsection{Analytical solutions}

To this end, we try to minimize $V_{\scriptsize \mbox{eff}}$ analytically. In fact, the sum over $p$ in (\ref{final form of the potential}) can be performed exactly:
\begin{eqnarray}
B(z)\equiv\sum_{p=1}^\infty\frac{\cos pz}{p^4}=\frac{-15z^4+60\pi z^3-60 \pi^2 z^2+8\pi^4}{720}\,,\quad 0 \leq z\leq 2\pi\,.
\label{analytic sum}
\end{eqnarray}
Next, with the aid of the Frobenius formula we find that the effective potential of $su(2)$ is given by \footnote{We find it more convenient to use the normalization $\bm\alpha^2=1$ for $su(2)$.}
\begin{eqnarray}
V_{\scriptsize\mbox{eff}}(\Phi)=\sum_{p=1}^\infty\frac{2n_F \cos\left(\frac{p\Phi}{2}\right)+4(n_G-1)\cos^2\left(\frac{p\Phi}{2}\right)}{p^4}\,,
\end{eqnarray} 
for $n_G$ and $n_F$ fermions. Then, using Eq. (\ref{analytic sum}) we obtain, apart from irrelevant additive and multiplicative constants,
\begin{eqnarray}
\nonumber
V_{\scriptsize\mbox{eff}}(\Phi)&\sim&-15(16(n_G-1)+n_F)\Phi^4+120\pi(8(n_G-1)+n_F)\Phi^3-240\pi^2(4(n_G-1)+n_F)\Phi^2\,,\\
\end{eqnarray}
which admits a minimum at
\begin{eqnarray}
\Phi_{\scriptsize\mbox{min}}=\frac{4\pi(4n_G+n_F-4)}{16n_G+n_F-16}\,.
\label{analytic expression for su2}
\end{eqnarray}
Expression (\ref{analytic expression for su2}) shows that there cannot be a minimum inside the affine Weyl chamber, $0< \Phi < 2\pi$, for a single flavor of adjoint fermion and any number of fundamentals. It also shows that the maximum number of fundamental fermions that can yield a minimum inside the affine Weyl chamber increases as we increase the number of adjoint fermions. Of course,  one should not trust (\ref{analytic expression for su2}) when the theory ceases to be asymptotically free.  

\begin{figure}[t] 
   \centering
   \includegraphics[width=2in]{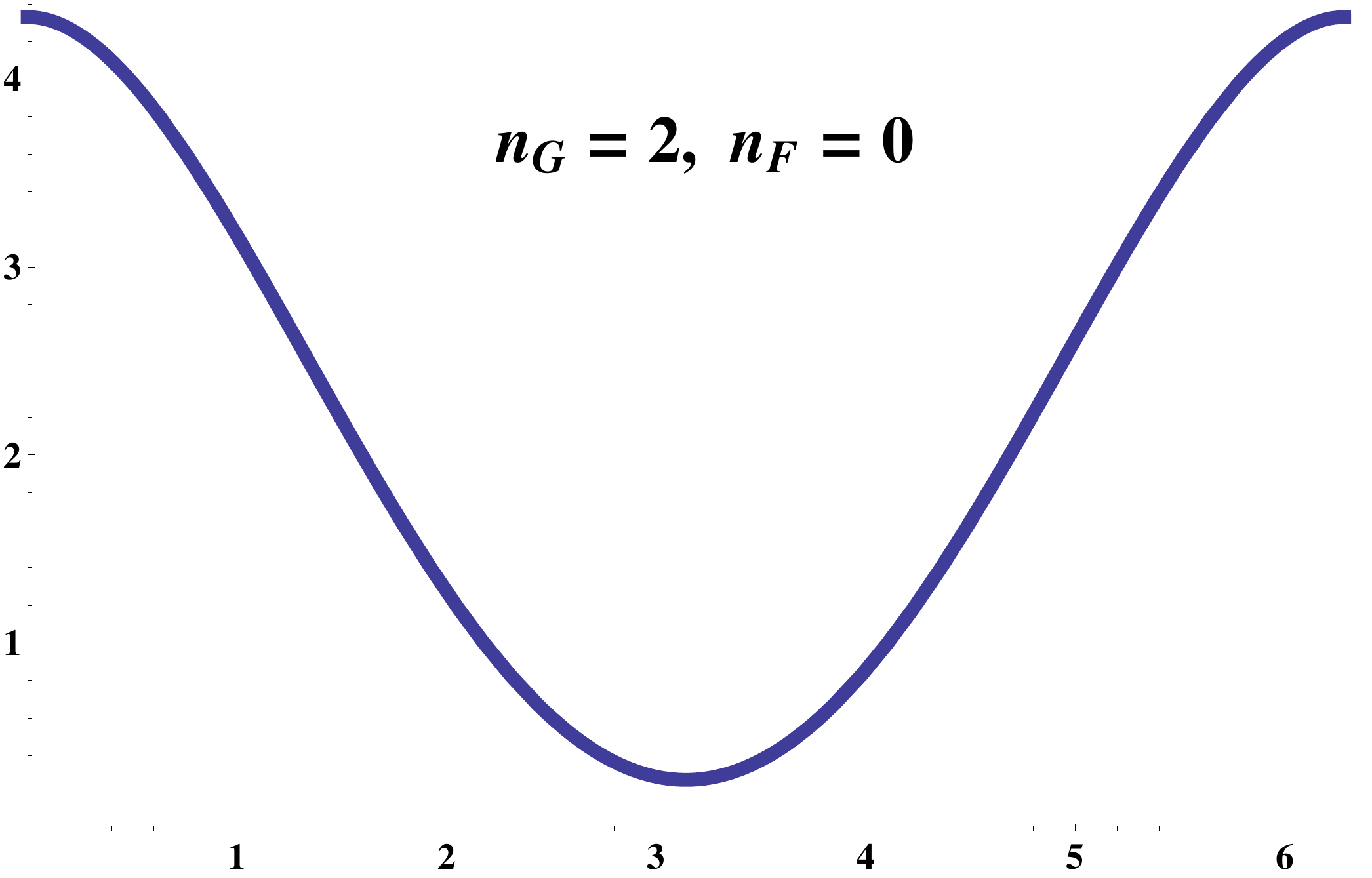} 
	\includegraphics[width=2in]{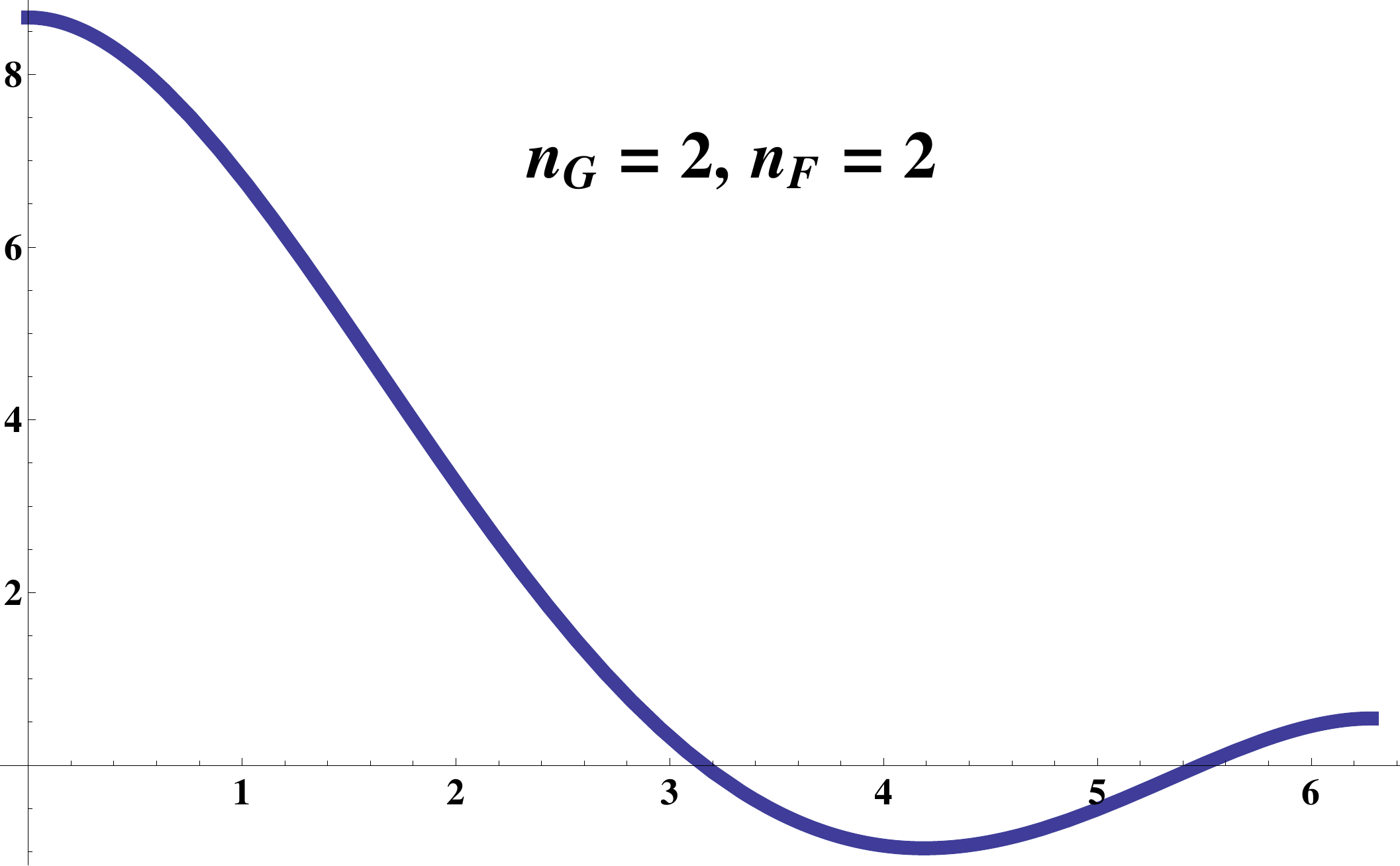} 
	\includegraphics[width=2in]{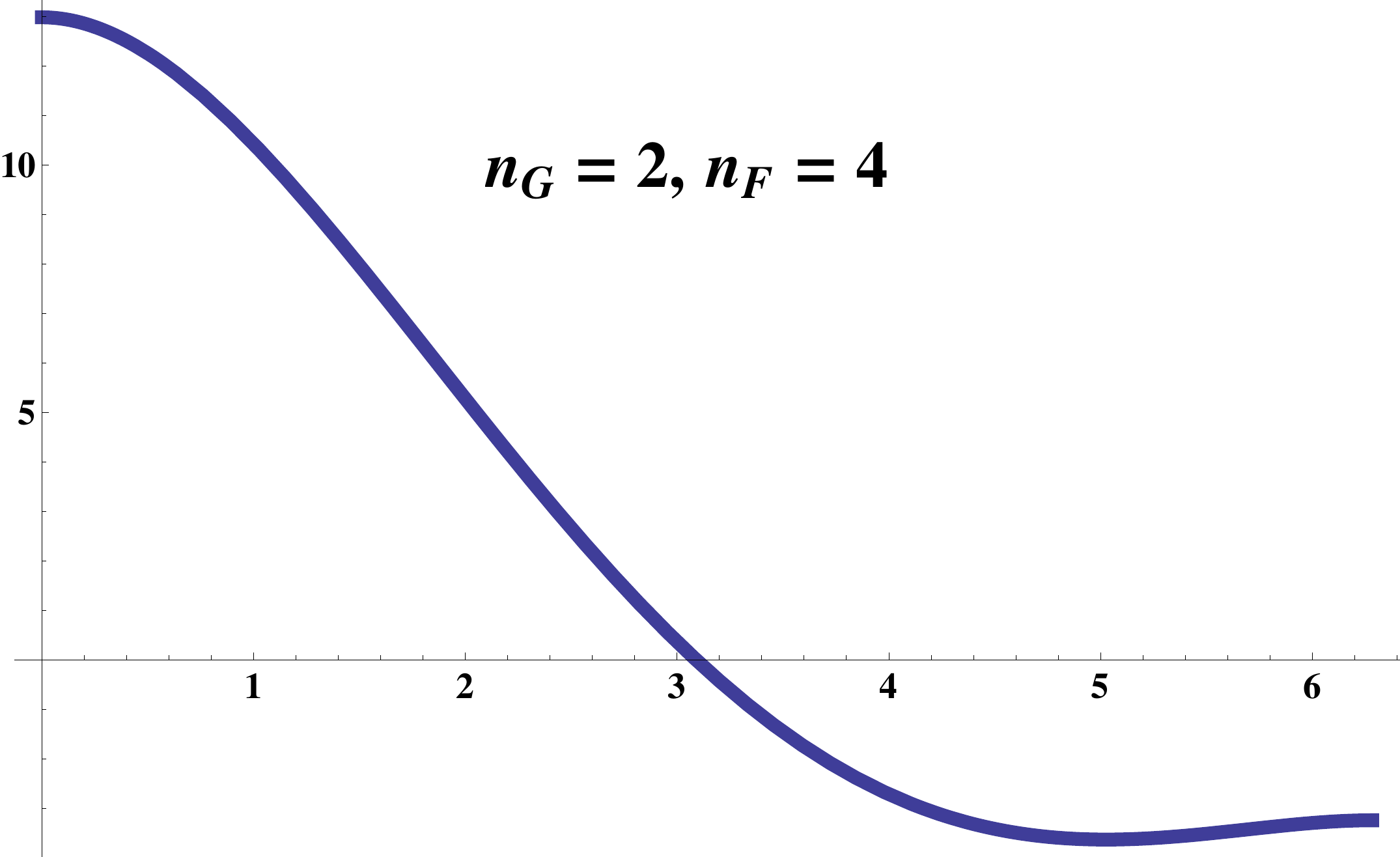} 
   \caption{The effective potential, $V_{\scriptsize \mbox{eff}}(\Phi)$, for $su(2)$ with $n_G=2$ and $n_{F}=0,2,4$.}
   \label{plots for ng2 and various fundamentals}
\end{figure}

 In Figure \ref{plots for ng2 and various fundamentals} we plot the $su(2)$ effective potential for $n_G=2$ and $n_F=0,2,4$ in the range $0<\Phi<2\pi$. For $n_F=0$ the potential develops a minimum at $\phi=\pi$, which is a center symmetric point since ${\cal Z}=0$. As we increase the number of fundamentals, the minimum shifts closer to the boundary of the affine Weyl chamber $\Phi=2\pi$. 

Similarly, one can also use analytical means to find the global minimum in the $su(3)$ case. Here, it is more convenient to use the $\mathbb R^{N_c}$ root and weight bases given in Appendix  \ref{Lie Algebra and Conventions}. Modulo multiplicative and additive constants, the effective potential for $n_G$ adjoint and $n_F$ fundamental fermions reads 
\begin{eqnarray}
\nonumber
V(\Phi_1,\Phi_2)&\sim& n_F \left(B(\Phi_1)+B(\Phi_2)+B(\Phi_1+\phi_2)  \right)\\
&&+2(n_G-1)\left(B(\Phi_1-\Phi_2)+B(2\Phi_1+\Phi_2)+B(\Phi_1+2\Phi_2)\right)\,.
\end{eqnarray}
The global minimum of the potential is given by
\begin{eqnarray}
\Phi_{1\,\scriptsize\mbox{min}}=\frac{2\pi\left[ 6(n_G-1)+n_F\right]}{n_F+18(n_G-1)}\,, \quad \Phi_{2\,\scriptsize\mbox{min}}=0\,,
\label{minima su3}
\end{eqnarray}
and the eigenvalues of the Hessian are
\begin{eqnarray}
\left\{\frac{360\pi^2 (36 (-1 + n_G)^2-n_F^2)} {
  n_F + 18 (-1 + n_G)},\frac{
  120\pi^2 (36 (-1 + n_G)^2-n_F^2)}{
  n_F + 18 (-1 + n_G)}\right\}\,.
	\label{hessian su3}
\end{eqnarray}
We can also check easily that the maximum number of the allowed fundamentals increases with increasing $n_G$, as in the $su(2)$ case. 
However, we show in Section \ref{The 3D coupling constant}  that $su(3)$ with fundamental fermions have massless modes charged under $u(1)^2$ rendering the theory strongly coupled in the IR. Yet, this case is important since it provides us with information that we can use to check our numerical computations.  

\subsection{Numerical investigation}

It is extremely difficult to use analytical expressions to find the global minima of the potential for any group beyond $su(3)$. In this section we numerically search for the minima of the effective potential (\ref{final form of the potential in terms of the polyakov loop}) for all  asymptotically free theories.

Since the sum in (\ref{final form of the potential in terms of the polyakov loop}) is rapidly convergent, we could try to find the minimum for fixed small values of $p$ and then check whether the obtained minimum is the true global minimum of the full potential. Unfortunately, this method does not work when the global minumum is far from the center symmetric point, specially when it is close to the boundary of the affine Weyl chamber. Therefore, we prefer to use a full numerical approach. Apart from additive and multiplicative constants, the  final form of the effective potential is given by
\begin{eqnarray}
V_{\scriptsize\mbox{eff}}\sim n_{\cal R}\mbox{tr}_{\cal R} B\left[\bm \Phi\cdot \bm H\right] +(n_G-1)\mbox{tr}_{G} B\left[\bm \Phi\cdot \bm H\right]\,,
\end{eqnarray}
where $B(x)$ is given by (\ref{analytic sum}). As we mentioned before, we use both Frobenius method and Verma bases to double check our computations. 

 In our numerical scheme, we seek the global minimum of the effective potential  (\ref{final form of the potential in terms of the polyakov loop}) with 16-digit precision. However, we are conservative in deciding whether the minimum is located inside or on the boundaries of the affine Weyl chamber. Therefore, we exclude all theories that have global minima  within a distance less than $10^{-3}$ from the boundary, i.e., the conditions for accepting a theory is that its global minimum (or minima) satisfies the conditions
\begin{eqnarray}
\bm \alpha_a\cdot\bm \Phi_{\scriptsize \mbox{min}}>10^{-3} \quad \mbox{for all}\quad a=1,2,3,...,N_c-1\quad \mbox{and} \quad-\bm\alpha_0\cdot \bm \Phi_{\scriptsize \mbox{min}}<2\pi-10^{-3}\,.
\label{numerical condition}
\end{eqnarray}
The cutoff window is consistent with the analytical results of $su(2)$ and $su(3)$ that we discussed in the previous section. In particular, increasing the size of  the cutoff window,  to allow for the minima to reside within a distance smaller than $10^{-3}$ from the boundaries of the Weyl chamber, overestimates the  number of the allowed fermions compared to the number we obtain from the analytical expressions of $su(2)$ and $su(3)$.  Therefore, we use the empirical affine Weyl chamber as given by (\ref{numerical condition}) throughout our numerical work.

 Indeed, we can also have situations of degenerate global minima. We study these cases in great details in Section \ref{Perturbative vacua and the role of discrete symmetries}.

Our results are discussed in the following subsections.

\subsection*{Theories with fermions in  pure representations}
%
\begin{enumerate}[(a)]

\item Theories with $1\leq n_G<5$ flavors of fermions in the adjoint representation of $su(N_c)$ have global minima inside the affine Weyl chamber\footnote{The special case $n_G=1$ is ${\cal N}=1$ super Yang-Mills.}. In fact, these global minima preserve the center symmetry of the gauge group, i.e., ${\cal Z}=0$. This class of theories is well studied in the literature, see e.g., \cite{Unsal:2007jx,Anber:2011de}, and we refrain from giving further comments. 

\item $su(2)$ with  a single fermion flavor in the representation ${\cal R}=(4)$ has global minima (two minima, see Section \ref{Perturbative vacua and the role of discrete symmetries}) inside the affine Weyl chamber. We also find ${\cal Z}=\frac{1}{\sqrt{2}}$, which means that the global minima of the theory breaks the center symmetry.  

\item  All other theories with fermions in all other representations of $su(N_c)$ fail to have global minima inside the affine Weyl chamber. The physical reason behind this observation will be discussed below.

\end{enumerate}

Therefore, beyond $su(2)$, only theories with adjoint fermions abelianize, and hence, have semi-classical description.

\subsection*{Theories with fermions in mixed representations $n_G\oplus n_{\cal R}$}

Now, we consider the mixed representations of $1\leq n_G\leq 5$ flavors of massless adjoint fermions and $n_{\cal R}$ massless fermions in representation ${\cal R}$. Our results for the gauge groups $su(2)$ to $su(8)$ are displayed in Tables \ref{su2 table} to \ref{su8 table}. The first column, $n_G$,  is the number of the adjoint fermions. In the second column we list the asymptotically free theories with fermions in representation ${\cal R}$. In the third column we give the maximum number of  ${\cal R}$ fermions,  $n_{\cal R}^{\scriptsize\mbox{max}}$, in the presence of $n_G$ fermions such that the theory remains asymptotically free. The fourth column gives the range of fermions that lead to a global minimum (or minima) of the effective potential inside the affine Weyl chamber. This range is represented either as an interval $[n_{\cal R}^1, n_{\cal R}^2]$ (keeping in mind that we really mean the integer values of the fermion number), or as a list of numbers $\{n_{\cal R}^1, n_{\cal R}^2,...\}$. The abbreviation NON indicates the absence of a global minimum inside the affine Weyl chamber. The fifth column gives the deviation of the global minimum of the potential from the center symmetric point as defined in (\ref{definition of Z}). In particular, we give the range of ${\cal Z}$ as an interval $[{\cal Z}^{1}, {\cal Z}^{2}]$, where ${\cal Z}^{1}$ corresponds to $n_{\cal R}^1$ and ${\cal Z}^{2}$ corresponds to $n_{\cal R}^2$. Notice that we do not compute ${\cal Z}$ for theories with no global minima inside the affine Weyl chamber. The final column gives the Atiyah-Singer index in the background of a single Belavin-Polyakov-Schwarz-Tyupkin (BPST) instanton. An odd index would indicate that the theory has a Witten anomaly, see section \ref{Terminal anomaly}. Finally, theories that are asymptotically free, have global minima inside the affine Weyl chamber, anomaly free, and do not admit massless charged fermions, under $u(1)^{N_c-1}$, in the IR are indicated by blue bold face, see Section \ref{The admissible class of theories} for more details.

Interestingly, we find that the same pattern we observed for $su(2)$ and $su(3)$ continues to hold for higher groups and higher representations. In particular, the number of allowed fundamental fermions increases with increasing $n_G$. One can understand this fact by analyzing the fields that compete among each other to break or restore the center symmetry. In the absence of fermions, gauge fluctuations will always break the center of the group badly by pushing the minimum of the potential to the boundaries of the affine Weyl chamber. Fundamental fermions also tend to break the center symmetry. In fact, our analysis shows that fermions in all pure representations, except the adjoint, either push the minimum of the potential to the boundary of the affine Weyl chamber or, at least, cannot fight against the gauge fields.  Adjoint fermions, on the contrary, prefer a stable center and adding them is the only hope to have a theory with a non-trivial global minimum. Adding a single Weyl adjoint fermion to pure Yang-Mills renders the theory supersymmetric, which has a vanishing perturbative potential to all orders in perturbation theory. Nonperturbative contributions (neutral bions), however, restore the center symmetry \cite{Poppitz:2012sw}. The contribution from neutral bions is exponentially suppressed and it cannot fight against any additional non-adjoint fermion field. Hence, in general we find no nontrivial global minimum for a theory with zero or one adjoint fermion in the presence of any number of additional fermions in a representation ${\cal R}\neq\mbox{adj}$. An exception to this generality is the $(020)$ representation in $su(4)$ with a single adjoint fermion. Although the contributions from gauge and adjoint fermion cancel each other, the representation $(020)$ is capable of generating an effective potential with a nontrivial minimum. Notice, however, that the potential generated from this representation is not capable by itself in fighting against the gauge field; it needs at least a single adjoint fermion to join the battle (the maximum number of $(020)$ fermions allowed by asymptotic freedom, which is $2$ in this case, is not enough to fight against the gauge field). Adding more than one adjoint fermion empowers the theory in its fight against the gauge fluctuations and other additional fermion fields. Thus, for $n_G\geq 2$ we expect that more fermions will be allowed, which is limited only by the requirement of asymptotic freedom of the theory, as can be seen in the tables.

Of course, the location of the global minimum (or minima) will be determined as a compromise between the different components of the theory.   In general, the minimum will not respect the center symmetry of the gauge group. The deviation of the minimum from the center-symmetric point is measured using ${\cal Z}$ defined in (\ref{definition of Z}). In Tables  (\ref{su2 table}) to (\ref{su8 table}) we list the range of ${\cal Z}$ for the range of the allowed fermions in each representation. We have checked explicitly that for a given number of $n_{G}$ adjoint fermions, the value of ${\cal Z}$ increases monotonically with the number of additional fermions in a representation ${\cal R}$ . In particular, ${\cal Z}\ll 1$ for very small number of fermions (not in the adjoint) and it reaches its peak for the maximum number of allowed fermions.  Such behavior is consistent with our discussion above since fermions in a representation ${\cal R}\neq \mbox{adj}$ favor destabilizing the center.  
We also note that the global minima of the adjoint-antisymmetric mixed representations $n_G\oplus(0,1,0,..,0)$, for $N_c$ even, are very close to the center symmetric point as these theories have ${\cal Z}\cong0$. However, as we will discuss in Section(\ref{The admissible class of theories}), all these theories are strongly coupled in the IR. It is also important to emphasize that although Tables  (\ref{su2 table}) to (\ref{su8 table}) show a few cases with ${\cal Z}=0$, this is meant to be true only within our numerical accuracy.

\begin {table}
\begin{center}
\tabcolsep=0.11cm
\footnotesize
\begin{tabular}{|c|c|c|c|c|c|}
\hline
$n_G$ & ${\cal R}$ & $n_{\cal R}^{\scriptsize\mbox{max}}$ & Range of $n_{\cal R}$ & Range of ${\cal Z}$  & ${\cal I}_f$\\
\hline
$0$&  $(1)$=F &  $22$ & NON & NA & $n_{\cal R}$\\
&  $\color{blue}\bm{(2)=\mbox{adj}}$ & $5\frac{1}{2}$ & $\color{blue}{\bm{[1,5]}}$ & $\{0\}$ & 4$n_{\cal R}$\\
&  $(3)$ & $2\frac{1}{5}$ & NON &  NA& 10$n_{\cal R}$\\
&  $\color{blue}\bm{(4)}$& $1\frac{1}{10}$ & $\color{blue}{\bm{ \{1\}}}$ & $\{0.707\}$ & 20$n_{\cal R}$\\
\hline
$1$&  $(1)$ & $18$ & NON &  NA& $4+n_{\cal R}$\\
&  $(3)$ & $1\frac{4}{5}$ & NON &  NA& $4+10n_{\cal R}$\\
\hline
$2$&  $\color{blue}\bm{(1)}$ & $14$ & $\color{blue}\bm{[1,7]_{\scriptsize \mbox{even}}}$ & $[0.274,0.991]$  &  8+$n_{\cal R}$\\
&  $\color{blue}\bm{(3)}$ & $1\frac{2}{5}$ & $\color{blue}{\bm{ \{1\}}}$ & $\{0.257\}$ &  $8+10n_{\cal R}$\\
\hline
$3$&  $\color{blue}\bm{(1)}$ & $10$ & $\color{blue}\bm{[1,10]_{\scriptsize \mbox{even}}}$ & $[0.142,0.901]$ &  12+$n_{\cal R}$\\
&  $\color{blue}\bm{(3)}$ & $1$ & $\color{blue}{\bm{ \{1\}}}$ & $\{0.188\}$ &  $12+10n_{\cal R}$\\
\hline
$4$&  $\color{blue}\bm{(1)}$ & $6$ & $\color{blue}\bm{[1,6]_{\scriptsize \mbox{even}}}$ & $[0.096,0.500]$  & $16+n_{\cal R}$\\
\hline
$5$&  $\color{blue}\bm{(1)}$ & $2$ & $\{1,{\color{blue}\bm{2}}\}$ & $\{0.072,0.142\}$  &  $20+n_{\cal R}$\\
\hline
\end{tabular}
\caption{ \label{su2 table} $su(2)$: Classification of all  asymptotically free theories. The first column, $n_G$,  is the number of the adjoint fermions. In the second column we list the asymptotically free theories with fermions in representation ${\cal R}$. In the third column we give the maximum number of  ${\cal R}$ fermions,  $n_{\cal R}^{\scriptsize\mbox{max}}$, in the presence of $n_G$ fermions such that the theory remains asymptotically free. The fourth column gives the range of fermions that lead to a global minimum (or minima) of the effective potential inside the affine Weyl chamber. This range is represented either as an interval $[n_{\cal R}^1, n_{\cal R}^2]$ (keeping in mind that we really mean the integer values of the fermion number), or as a list of numbers $\{n_{\cal R}^1, n_{\cal R}^2,...\}$. The abbreviation NON indicates the absence of a global minimum inside the affine Weyl chamber. The fifth column gives the deviation of the global minimum of the potential from the center symmetric point as defined in (\ref{definition of Z}). In particular, we give the range of ${\cal Z}$ as an interval $[{\cal Z}^{1}, {\cal Z}^{2}]$, where ${\cal Z}^{1}$ corresponds to $n_{\cal R}^1$ and ${\cal Z}^{2}$ corresponds to $n_{\cal R}^2$. Notice that we do not compute ${\cal Z}$ for theories with no global minima inside the affine Weyl chamber. The final column gives the Atiyah-Singer index in the background of a Belavin-Polyakov-Schwarz-Tyupkin (BPST) instanton.}
\end{center}
\end{table}

\begin {table}
\caption{$su(3)$}
\begin{center}
\tabcolsep=0.11cm
\footnotesize
\begin{tabular}{|*{7}{c|}}
\hline
$n_G$ & ${\cal R}$ & $n_{\cal R}^{\scriptsize\mbox{max}}$ & Range of $n_{\cal R}$ & Range of ${\cal Z}$& ${\cal I}_f$\\
\hline
$0$& $(10)$ & $33$ & NON & 	NA    & $n_{\cal R}$ \\ 
& ($20)$ & $6\frac{3}{5}$ & NON &  NA  & $5n_{\cal R}$  \\
& $(30)$ & $2\frac{1}{5}$ & NON  &  NA   & $15n_{\cal R}$\\
& $\color{blue}\bm{(11)}$ & $5\frac{1}{2}$ & $\color{blue}{\bm{[1,5]}}$  & $\{0\}$  & $6n_{\cal R}$ \\
& $(21)$ & $1\frac{13}{20}$ & NON  & NA   & $20n_{\cal R}$\\
\hline
$1$& $(10$) & $27$ & NON & NA &  $6+n_{\cal R}$ \\
& $(20)$ & $5\frac{2}{5}$ & NON & NA  & $6+5n_{\cal R}$ \\
& $(30)$ & $1\frac{4}{5}$ & NON  & NA   & $6+15n_{\cal R}$\\
& $(21)$ & $1\frac{7}{20}$ & NON  & NA  & $6+20n_{\cal R}$ \\
\hline
$2$& $(10)$ & $21$ & $[1,5]$ & $[0.118,0.327]$   & $12+n_{\cal R}$ \\
& $(20)$ & $4\frac{1}{5}$ & $\{1\}$ &  $\{0.202\}$  &  $12+5n_{\cal R}$\\
& $(30)$ & $1\frac{2}{5}$ & $\{1\}$  & $\{0.314\}$   & $12+15n_{\cal R}$ \\
& $(21)$ & $1\frac{1}{20}$ & $\{1\}$  & $\{0.215\}$   &$12+20n_{\cal R}$ \\
\hline
$3$& $(10)$ & $15$ & $[1,11]$ & $[0.063,0.331]$   &  $18+n_{\cal R}$ \\
& $\color{blue}\bm{(20)}$ & $3$ & $\{1,{\color{blue}{\bm 2}},3\}$ &  $[0.047,0.202, 0.337]$  & $18+5n_{\cal R}$ \\
& $(30)$ & $1$ & $\{1\}$  & $\{0.206\}$  &  $18+15n_{\cal R}$\\
\hline
$4$& $(10)$ & $9$ & $[1,9]$ & $[0.042,0.267]$   & $24+n_{\cal R}$\\
& $(20)$ & $1\frac{4}{5}$ & $\{1\}$ & $\{0.035\}$   & $24+5n_{\cal R}$ \\
\hline
$5$& $(10)$ & $3$ & $[1,3]$ & $[0.033,0.092]$  & $30+n_{\cal R}$ \\
\hline
\end{tabular}
\label{su3 table}
\end{center}
\end{table}

\begin {table}
\caption{$su(4)$}
\begin{center}
\tabcolsep=0.11cm
\footnotesize
\begin{tabular}{|*{7}{c|}}
\hline
$n_G$ & ${\cal R}$ & $n_{\cal R}^{\scriptsize\mbox{max}}$ & Range of $n_{\cal R}$ & Range of ${\cal Z}$ & ${\cal I}_f$\\
\hline
$0$ & $(100)$ & $44$ & NON & NA  & $n_{\cal R}$ \\
& $(200)$ & $7\frac{1}{3}$ & NON   & NA & $6n_{\cal R}$ \\
& $(300)$ & $2\frac{2}{21}$ & NON   & NA & $21n_{\cal R}$ \\
& $(010)$ & $22$ & NON  & NA   & $2n_{\cal R}$ \\
& $(020)$ & $2\frac{3}{4}$ & NON  & NA  & $16n_{\cal R}$  \\
& $\color{blue}{\bm{(101)}}$ & $5\frac{1}{5}$ & $\color{blue}{\bm{[1,5]}}$  & $\{0\}$  & $8n_{\cal R}$ \\
& $(110)$ & $3\frac{5}{13}$ & NON  & NA  & $13n_{\cal R}$  \\
& $(201)$ & $2\frac{1}{3}$ & NON   & NA & $33n_{\cal R}$ \\
\hline 
$1$& (100) & $36$ & NON & NA   & $8+n_{\cal R}$ \\
& $(200)$ & $6$ & NON &  NA  & $8+6n_{\cal R}$ \\
& $(300)$ & $1\frac{5}{7}$ & NON   & NA & $8+21n_{\cal R}$ \\
& $(010)$ & $18$ & NON  & NA  &  $8+2n_{\cal R}$ \\
& $\color{blue}{\bm{(020)}}$ & $2\frac{1}{4}$ & $\color{blue}{\bm{\{1,2\}}}$  & $\{0.485,0.485\}$   & $8+16n_{\cal R}$  \\
& $(110)$ & $2\frac{10}{13}$ & NON  & NA  & $8+13n_{\cal R}$  \\
& $(201)$ & $1\frac{1}{11}$ & NON    & NA & $8+33n_{\cal R}$ \\
\hline
$2$& $\color{blue}{\bm{(100)}}$ & $28$ & $\color{blue}{\bm{[1,15]_{\scriptsize\mbox{even}}}}$ & $[0.126,0.998]$   & $16+n_{\cal R}$ \\
& $\color{blue}{\bm{(200)}}$ & $4\frac{2}{3}$  & $\{1,{\color{blue}{\bm 2}}\}$ & $\{0,0.95\}$  & $16+6n_{\cal R}$ \\
& $(300)$ & $1\frac{1}{3}$ & NON   & NA & $16+21n_{\cal R}$  \\
& $(010)$ & $14$ & $\{1\}$  & $\{0\}$    & $16+2n_{\cal R}$ \\
& $\color{blue}{\bm{(020)}}$ & $1\frac{3}{4}$ & ${\color{blue}{\bm{ \{1\}}}}$  & $\{0.25\}$  & $16+16n_{\cal R}$ \\
& $(110)$ & $2\frac{2}{13}$ & NON  & NA  & $16+13n_{\cal R}$ \\
\hline
$3$& $\color{blue}{\bm{(100)}}$ & $20$ & $\color{blue}{\bm{[1,20]{\scriptsize\mbox{even}}}}$ & $[0.063,0.900]$   & $24+n_{\cal R}$  \\
& $(200)$ & $3\frac{1}{3}$ & $\{1,2,3\}$  & $\{0,0,8.85\times 10^{-4}\}$  & $24+6n_{\cal R}$ \\
& $(010)$ & $10$ & $[1,3]$  & $\{0\}$   & $24+2n_{\cal R}$ \\
& $\color{blue}{\bm{(020)}}$ & $1\frac{1}{4}$ & ${\color{blue}{\bm {\{1\}}}}$  & $\{0.34\}$   & $24+16n_{\cal R}$ \\
& $(110)$ & $1\frac{7}{13}$ & $\{1\}$   & $\{0.057\}$  & $24+13n_{\cal R}$ \\
\hline
$4$& $\color{blue}{\bm{(100)}}$ & $12$ & $\color{blue}{\bm{[1,12]_{\scriptsize\mbox{even}}}}$  & $[0.042,0.479]$   &  $32+n_{\cal R}$  \\
& $(200)$ & $2$ & $\{1,2\}$  &   $\{0,0\}$  & $32+6n_{\cal R}$  \\
& $(010)$ & $6$ & $[1,5]$  & $\{0\}$   & $32+16n_{\cal R}$ \\
\hline
$5$& $\color{blue}{\bm{ (100)}}$ & $4$ & $\color{blue}{\bm{[1,4]_{\scriptsize\mbox{even}}}}$  & $[0.031,0.126]$   &  $40+n_{\cal R}$  \\
& $(010)$ & $2$ & $\{1,2\}$   & $\{0\}$   & $40+16n_{\cal R}$ \\
\hline
\end{tabular}
\label{su4 table}
\end{center}
\end{table}


\begin {table}
\caption{$su(5)$}
\begin{center}
\tabcolsep=0.11cm
\footnotesize
\begin{tabular}{|*{7}{c|}}
\hline
$n_G$ & ${\cal R}$ & $n_{\cal R}^{\scriptsize\mbox{max}}$ & Range of $n_{\cal R}$ & Range of ${\cal Z}$  & ${\cal I}_f$\\
\hline 
$0$& $(1000)$ & $55$ & NON &  NA  &$n_{\cal R}$\\
& $(2000)$ & $7\frac{6}{7}$ & NON & NA   & $7n_{\cal R}$\\
& $(3000)$ & $1\frac{27}{28}$ & NON    & NA & $28n_{\cal R}$ \\
& $(0100)$ & $18\frac{1}{3}$ & NON  & NA    & $3n_{\cal R}$\\
& $(0200)$ & $1\frac{4}{7}$ & NON    & NA & $35n_{\cal R}$\\
& $\color{blue}{\bm{(1001)}}$ & $5\frac{1}{2}$ & $\color{blue}{\bm{[1,5]}}$  & $\{0\}$   & $10n_{\cal R}$ \\
& $(2001)$ & $1\frac{6}{49}$ & NON   & NA & $49n_{\cal R}$\\
& $(1100)$ & $2\frac{1}{2}$ & NON  & NA  &$22n_{\cal R  }$ \\
& $(1010)$ & $2\frac{7}{24}$ & NON  & NA   & $24 n_{\cal R}$ \\
& $(0110)$ & $1\frac{1}{10}$ & NON   & NA &$50n_{\cal R}$ \\
\hline
$1$& $(1000)$ & $45$ & NON & NA   & $10+n_{\cal R}$ \\
& $(2000)$ & $6\frac{3}{7}$ & NON & NA  &$10+7n_{\cal R}$ \\
& $(3000)$ & $1\frac{17}{28}$ & NON    & NA & $10+28n_{\cal R}$\\
& $(0100)$ & $15$ & NON  & NA  & $10+3n_{\cal R}$ \\
& $(0200)$ & $1\frac{2}{7}$ & NON    & NA & $10+35n_{\cal R}$\\
& $(1100)$ & $2\frac{1}{22}$ & NON    & NA &$10+22n_{\cal R  }$ \\
& $(1010)$ & $1\frac{7}{8}$ & NON    & NA & $10+24 n_{\cal R}$\\
\hline
$2$& $\color{blue}{\bm{(1000)}}$ & $35$ & $\{1,2,3,4,5,{\color{blue}{\bm 6}},7,{\color{blue}{\bm 8}},9\}$ & $[0.095,0.602]$   & $20+n_{\cal R}$ \\
& $\color{blue}{\bm{(2000)}}$ & $5$ & $\{1,{\color{blue}{\bm 2}}\}$ & $\{0.060,0.193\}$  & $20+7n_{\cal R}$   \\
& $(3000)$ & $1\frac{1}{4}$ & NON   & NA & $20+28n_{\cal R}$\\
& $(0100)$ & $11\frac{2}{3}$ & $\{1\}$  &  $\{0.0516\}$  & $20+3n_{\cal R}$ \\
& $(0200)$ & $1$ & NON  &  NA & $20+35n_{\cal R}$ \\
& $(1100)$ & $1\frac{13}{22}$ & NON  & NA   & $20+22n_{\cal R  }$ \\
& $(1010)$ & $1\frac{11}{24}$ & NON   & NA & $20+24 n_{\cal R}$ \\
\hline 
$3$& $\color{blue}{\bm{(1000)}}$ & $25$ & $[1,11]\bigcup\color{blue}{\bm{[12,19]_{\scriptsize\mbox{even}}}}$ & $[0.048,0.631]$  &  $30+n_{\cal R}$\\
& $\color{blue}{\bm{(2000)}}$ & $3\frac{4}{7}$ & $\{1,{\color{blue}{\bm 2}},3\}$  &  $\{0.007,0.060,0.143\}$   & $30+7n_{\cal R}$ \\
& $(0100)$ & $8\frac{1}{3}$ & $\{1,2\}$  & $\{0.003,0.016\}$    & $30+3n_{\cal R}$ \\
& $(1100)$ & $1\frac{3}{22}$ & $\{1\}$  & $\{0.015\}$   & $30+22n_{\cal R  }$ \\
& $(1010)$ & $1\frac{1}{24}$ & $\{1\}$  &  $\{0.059\}$   & $20+24 n_{\cal R}$\\
\hline 
$4$& $(1000)$ & $15$ & $[1,15]$  & $[0.032,0.392]$    &  $40+n_{\cal R}$ \\
& $\color{blue}{\bm{(2000)}}$ & $2\frac{1}{7}$ & $\{1,{\color{blue}{\bm 2}}\}$  & $\{0.004,0.009\}$ & $40+7n_{\cal R}$  \\
& $(0100)$ & $5$ & $[1,3]$  & $[3.7\times 10^{-4},0.016]$   & $40+3n_{\cal R}$\\
\hline
$5$& $(1000)$ & $5$ & $[1,5]$  & $[0.024,0.118]$  &  $50+n_{\cal R}$ \\
& $(0100)$ & $1\frac{2}{3}$ & $\{1\}$   & $\{2.9\times 10^{-4}\}$    & $50+3n_{\cal R}$\\
\hline
\end{tabular}
\label{su5 table}
\end{center}
\end{table}

\begin {table}
\caption{$su(6)$}
\begin{center}
\tabcolsep=0.11cm
\footnotesize
\begin{tabular}{|*{7}{c|}}
\hline
$n_G$ & ${\cal R}$ & $n_{\cal R}^{\scriptsize\mbox{max}}$ & Range of $n_{\cal R}$ & Range of ${\cal Z}$ & ${\cal I}_f$\\
\hline
$0$& $(10000)$ & $66$ & NON & NA  & $n_{\cal R}$ \\
& $(20000)$ & $8\frac{1}{4}$ & NON  & NA & $8n_{\cal R}$\\
& $(30000)$ & $1\frac{5}{6}$ & NON   & NA & $36n_{\cal R}$ \\
& $(01000)$ & $16\frac{1}{2}$ & NON   & NA & $4n_{\cal R}$  \\
& $(02000)$ & $1\frac{1}{32}$ & NON   & NA & $64n_{\cal R}$ \\
& $(00100)$ & $11$ & NON & NA  & $6n_{\cal R}$ \\
& $\color{blue}{\bm{(10001)}}$ & $5\frac{1}{2}$ & $\color{blue}{\bm{[1,5]}}$  & $\{0\}$   & $12n_{\cal R}$ \\
& $(11000)$ & $2$ & NON   & NA & $33n_{\cal R}$ \\
& $(10100)$ & $1\frac{7}{26}$ & NON   & NA & $52n_{\cal R}$ \\
& $(10010)$ & $1\frac{14}{19}$ & NON   & NA  & $38n_{\cal R}$ \\
\hline
$1$& $(10000)$ & $54$ & NON & NA  & $12+n_{\cal R}$ \\
& $(20000)$ & $6\frac{3}{4}$ & NON  & NA & $12+8n_{\cal R}$  \\
& $(30000)$ & $1\frac{1}{2}$ & NON  & NA & $12+36n_{\cal R}$ \\
& $(01000)$ & $13\frac{1}{2}$ & NON   & NA & $12+4n_{\cal R}$  \\
& $(00100)$ & $9$ & NON & NA  & $12+6n_{\cal R}$  \\
& $(11000)$ & $1\frac{7}{11}$ & NON   & NA & $12+33n_{\cal R}$ \\
& $(10100)$ & $1\frac{1}{26}$ & NON   & NA & $12+52n_{\cal R}$ \\
& $(10010)$ & $1\frac{8}{19}$ & NON   & NA  & $12+38n_{\cal R}$ \\
\hline
$2$& $\color{blue}{\bm{(10000)}}$ & $42$ & $\color{blue}{\bm{[1,14]_{\scriptsize \mbox{even}}}}$ & $[0.083,0.872]$  & $24+n_{\cal R}$ \\
& $\color{blue}{\bm{(20000)}}$ & $5\frac{1}{4}$ & $\{1,{\color{blue}{\bm 2}}\}$ & $\{0,6\times 10^{-7}\}$  & $24+8n_{\cal R}$  \\
& $(30000)$ & $1\frac{1}{6}$ & NON   & NA & $24+36n_{\cal R}$ \\
& $(01000)$ & $10\frac{1}{2}$ & $\{1\}$  & $\{0\}$   & $24+4n_{\cal R}$ \\
& $(00100)$ & $7$ & NON & NA  & $24+6n_{\cal R}$ \\
& $(11000)$ & $1\frac{3}{11}$ & NON  & NA & $24+33n_{\cal R}$ \\
& $(10010)$ & $1\frac{2}{19}$ & NON  & NA & $24+38n_{\cal R}$ \\
\hline 
$3$& $\color{blue}{\bm{(10000)}}$ & $30$ & $\color{blue}{\bm{[1,28]}_{\scriptsize \mbox{even}}}$ & $[0.042,0.872]$ & $36+n_{\cal R}$  \\
& $(20000)$ & $3\frac{3}{4}$ & $\{1,2,3\}$  & $\{0\}$   & $36+8n_{\cal R}$  \\
& $(01000)$ & $7\frac{1}{2}$ & $[1,2]$  & $\{0\}$  &  $36+4n_{\cal R}$  \\
& $\color{blue}{\bf{(00100)}}$ & $5$ & $\color{blue}{\bm {\{1\}}}$ & $\{0\}$  & $36+6n_{\cal R}$  \\
\hline
$4$& $\color{blue}{\bf{(10000)}}$ & $18$ & $\color{blue}{\bm{[1,18]_{\scriptsize \mbox{even}}}}$  & $[0.028,0.478]$  & $48+n_{\cal R}$   \\
& $(20000)$ & $2\frac{1}{4}$ & $[1,2]$  & $\{0\}$  & $48+8n_{\cal R}$  \\
& $(01000)$ & $4\frac{1}{2}$ & $[1,3]$  & $\{0\}$   & $48+4n_{\cal R}$  \\
& $\color{blue}{\bm{(00100)}}$ & $3$ & $\color{blue}{\bm{[1,2]}}$ & $\{0\}$  & $48+6n_{\cal R}$\\
\hline 
$5$& $\color{blue}{\bm{(10000)}}$ & $6$ & $\color{blue}{\bm{[1,6]_{\scriptsize \mbox{even}}}}$  & $[0.021,0.125]$ & $60+n_{\cal R}$  \\
& $(01000)$ & $1\frac{1}{2}$ & $\{1\}$  & $\{0\}$   & $60+4n_{\cal R}$  \\
& $\color{blue}{\bm{(00100)}}$ & $1$ & $\color{blue}{\bm {\{1\}}}$ & $\{0\}$  & $60+6n_{\cal R}$ \\
\hline
\end{tabular}
\label{su6 table}
\end{center}
\end{table}


\begin {table}
\caption{$su(7)$}
\begin{center}
\tabcolsep=0.11cm
\footnotesize
\begin{tabular}{|*{7}{c|}}
\hline
$n_G$ & ${\cal R}$ & $n_{\cal R}^{\scriptsize\mbox{max}}$ & Range of $n_{\cal R}$ & Range of ${\cal Z}$  & ${\cal I}_f$\\
\hline
$0$& $(100000)$ & $77$ & NON & NA & $n_{\cal R}$\\
& $(200000)$ & $8\frac{5}{9}$ & NON & NA & $9n_{\cal R}$\\
& $(300000)$ & $1\frac{32}{45}$ & NON   & NA & $45n_{\cal R}$ \\
& $(010000)$ & $15\frac{2}{5}$ & NON  &NA  & $5n_{\cal R}$ \\
& $(001000)$ & $7\frac{7}{10}$ & NON  & NA &  $10 n_{\cal R}$  \\
& $\color{blue}{\bm{(100001)}}$ & $5\frac{1}{2}$ & $\color{blue}{\bm{[1,5]}}$  & $\{0\}$  & $14n_{\cal R}$\\
& $(110000)$ & $1\frac{31}{46}$ & NON   & NA & $46n_{\cal R}$ \\
& $(100010)$ & $1\frac{2}{5}$ & NON  & NA & $55n_{\cal R}$\\
\hline
$1$& (100000) & $63$ & NON &  NA &   $14+n_{\cal R}$ \\
& $(200000)$ & $7$ & NON & NA  & $14+9n_{\cal R}$ \\
& $(300000)$ & $1\frac{2}{5}$ & NON   & NA & $14+45n_{\cal R}$ \\
& $(010000)$ & $12\frac{3}{5}$ & NON   & NA & $14+5n_{\cal R}$ \\
& $(001000)$ & $6\frac{3}{10}$ & NON   & NA & $14+10 n_{\cal R}$ \\
& $(110000)$ & $1\frac{17}{46}$ & NON   & NA& $14+46n_{\cal R}$\\
& $(100010)$ & $1\frac{8}{55}$ & NON   & NA & $14+55n_{\cal R}$\\
\hline
$2$& $\color{blue}{\bm{(100000)}}$ & $49$ & $\{1,2,3,4,5,{\color{blue}{\bm 6}},7,{\color{blue}{\bm 8}},9\}$ &$[0.070,0.527]$  & $28+n_{\cal R}$  \\
& $\color{blue}{\bm{(200000)}}$ & $5\frac{4}{9}$ & $\{1,{\color{blue}{\bm 2}}\}$ & $\{0.0226,0.666\}$  & $28+9n_{\cal R}$\\
& $(300000)$ & $1\frac{4}{45}$ & NON  & NA& $28+38n_{\cal R}$ \\
& $(010000)$ & $9\frac{4}{5}$ & $\{1\}$  & $\{4.4\times 10^{-4}\}$ & $28+5n_{\cal R}$ \\
& $(001000)$ & $4\frac{9}{10}$ & NON  & NA  & $28+10 n_{\cal R}$ \\
& $(110000)$ & $1\frac{3}{46}$ & NON   & NA & $28+46n_{\cal R}$\\
\hline
$3$& $\color{blue}{\bm{(100000)}}$ & $35$ & $[1,11]\bigcup{\color{blue}\bm{[12,19]_{\scriptsize\mbox{even}}}}$ & $[0.035,0.551]$  & $42+n_{\cal R}$ \\
& $\color{blue}{\bm{(200000)}}$ & $3\frac{8}{9}$ & $\{1,{\color{blue}{\bm 2}},3\}$ & $[0.002,0.111]$   & $42+9n_{\cal R}$\\
& $(010000)$ & $7$ & $\{1,2\}$  & $\{3.09\times 10^{-4},4.44\times 10^{-4}\}$  & $42+5n_{\cal R}$ \\
& $(001000)$ & $3\frac{1}{2}$ & $\{1\}$  & $\{0.024\}$  & $42+10 n_{\cal R}$ \\
\hline
$4$& $\color{blue}{\bm{(100000)}}$ & $21$ & $[1,18]\bigcup\{19,21\}\bigcup\color{blue}{\bm{\{20\}}}$ & $[0.024,0.428]$ & $56+n_{\cal R}$ \\
& $\color{blue}{\bm{(200000)}}$ & $2\frac{1}{3}$ & $\{1,{\color{blue}{\bm{2}}}\}$ & $[0.001,0.003]$   & $56+9n_{\cal R}$ \\
& $(010000)$ & $4\frac{1}{5}$ & $[1,3]$  & $[8\times 10^{-5},4.44\times 10^{-4}]$ &  $56+5n_{\cal R}$ \\
& $(001000)$ & $2\frac{1}{10}$ & $\{1\}$  & $\{0.013\}$  &  $56+10n_{\cal R}$\\
\hline
$5$& $(100000)$ & 7 & $[1,7]$ & $[0.018,0.122]$  & $70+n_{\cal R}$ \\
& $(010000)$ & $1\frac{2}{5}$ & $\{1\}$  &  $\{1.58\times 10^{-5}\}$ & $70+5n_{\cal R}$ \\
\hline
\end{tabular}
\label{su7 table}
\end{center}
\end{table}

\begin {table}
\caption{$su(8)$}
\begin{center}
\tabcolsep=0.11cm
\footnotesize
\begin{tabular}{|*{7}{c|}}
\hline
$n_G$ & ${\cal R}$ & $n_{\cal R}^{\scriptsize\mbox{max}}$ & Range of $n_{\cal R}$ & Range of ${\cal Z}$  & ${\cal I}_f$\\
\hline
$0$& $(1000000)$ & $88$ & NON & NA  & $n_{\cal R}$ \\
& $(2000000)$ & $8\frac{4}{5}$ & NON  & NA  & $10n_{\cal R}$ \\
& $(3000000)$ & $1\frac{3}{5}$ & NON   &  NA & $55n_{\cal R}$ \\
& $(0100000)$ & $14\frac{2}{3}$ & NON   & NA  & $6n_{\cal R}$\\
& $(0010000)$ & $5\frac{13}{15}$ & NON   & NA  &  $15n_{\cal R}$\\
& $(0001000)$ & $4\frac{2}{5}$ & NON & NA   & $20n_{\cal R}$\\
& $\color{blue}{\bm{(1000001)}}$ & $5\frac{1}{2}$ & $\color{blue}{\bm{[1,5]}}$  & $\{0\}$   &  $16n_{\cal R}$ \\
& $(1100000)$ & $1\frac{27}{61}$ & NON   & NA  & $61n_{\cal R}$\\
& $(1000010)$ & $1\frac{13}{75}$ & NON   &  NA & $75n_{\cal R}$\\
\hline
$1$& (1000000) & $72$ & NON & NA  & $16+n_{\cal R}$ \\
& $(2000000)$ & $7\frac{1}{5}$ & NON  &  NA & $16+10n_{\cal R}$ \\
& $(3000000)$ & $1\frac{17}{55}$ & NON   &  NA & $16+55n_{\cal R}$ \\
& $(0100000)$ & $12$ & NON  & NA  &  $16+6n_{\cal R}$\\
& $(0010000)$ & $4\frac{4}{5}$ & NON  & NA   & $16+15n_{\cal R}$ \\
& $(0001000)$ & $3\frac{3}{5}$ & NON & NA   & $16+20n_{\cal R}$\\
& $(1100000)$ & $1\frac{11}{61}$ & NON  & NA   & $61n_{\cal R}$ \\
\hline
$2$& $\color{blue}{\bm{(1000000)}}$ & $56$ & $\color{blue}{\bm{[1,14]_{\scriptsize \mbox{even}}}}$ & $[0.063,0.740]$    & $32+n_{\cal R}$ \\
& $\color{blue}{\bm{(2000000)}}$ & $5\frac{3}{5}$ & $\{1,{\color{blue}{\bm 2}}\}$ & $\{0, 0.106\}$   & $32+10n_{\cal R}$\\
& $(3000000)$ & $1\frac{1}{55}$ & NON   &  NA &  $32+55n_{\cal R}$\\
& $(0100000)$ & $9\frac{1}{3}$ & $\{1\}$  & $\{0\}$  & $32+6n_{\cal R}$ \\
& $(0010000)$ & $3\frac{11}{15}$ & NON  & NA   & $32+15n_{\cal R}$ \\
& $(0001000)$ & $2\frac{4}{5}$ & NON & NA   & $32+20n_{\cal R}$ \\
\hline
$3$& $\color{blue}{\bm{(1000000)}}$ & $40$ & $\color{blue}{\bm{[1,28]_{\scriptsize \mbox{even}}}}$ & $[0.031,0.740]$    & $48+n_{\cal R}$ \\
& $\color{blue}{\bm{(2000000)}}$ & $4$ & $\{1,2,3,{\color{blue}{\bm 4}}\}$& $\{0,0,0.001,0.230\}$   & $48+10n_{\cal R}$ \\
& $(0100000)$ & $6\frac{2}{3}$ & $\{1,2\}$  & $\{0\}$   & $48+6n_{\cal R}$ \\
& $(0010000)$ & $2\frac{2}{3}$ & $\{1\}$  & $\{0.002\}$  & $48+15n_{\cal R}$ \\
& $(0001000)$ & $2$ & $\{1\}$ & $\{0\}$   & $48+20n_{\cal R}$ \\
\hline
$4$& $\color{blue}{\bm{(1000000)}}$ & $24$ & $\color{blue}{\bm{[1,24]_{\scriptsize \mbox{even}}}}$  & $[0.021,0.478]$  & $64+n_{\cal R}$ \\
& $(2000000)$ & $2\frac{2}{5}$ & $\{1,2\}$  & $\{0\}$   &$64+10n_{\cal R}$\\
& $(0100000)$ & $4$ & $[1,3]$  & $\{0\}$   & $64+6n_{\cal R}$ \\
& $(0010000)$ & $1\frac{3}{5}$ & $\{1\}$  & $\{0.003\}$  & $64+15n_{\cal R}$\\
& $(0001000)$ & $1\frac{1}{5}$ & $\{1\}$  & $\{0\}$ &  $64+20n_{\cal R}$ \\
\hline
$5$& $\color{blue}{\bm{(1000000)}}$ & $8$ & $\color{blue}{\bm{[1,8]_{\scriptsize \mbox{even}}}}$  & $[0.016,0.125]$   & $80+n_{\cal R}$ \\
& $(0100000)$ & $1\frac{1}{3}$ & $\{1\}$ & $\{0\}$   & $80+6n_{\cal R}$\\
\hline
\end{tabular}
\label{su8 table}
\end{center}
\end{table}


\section{The flow of the $3$-D coupling constant}
\label{The 3D coupling constant}

\begin{figure}[t] 
   \centering
   \includegraphics[width=3in]{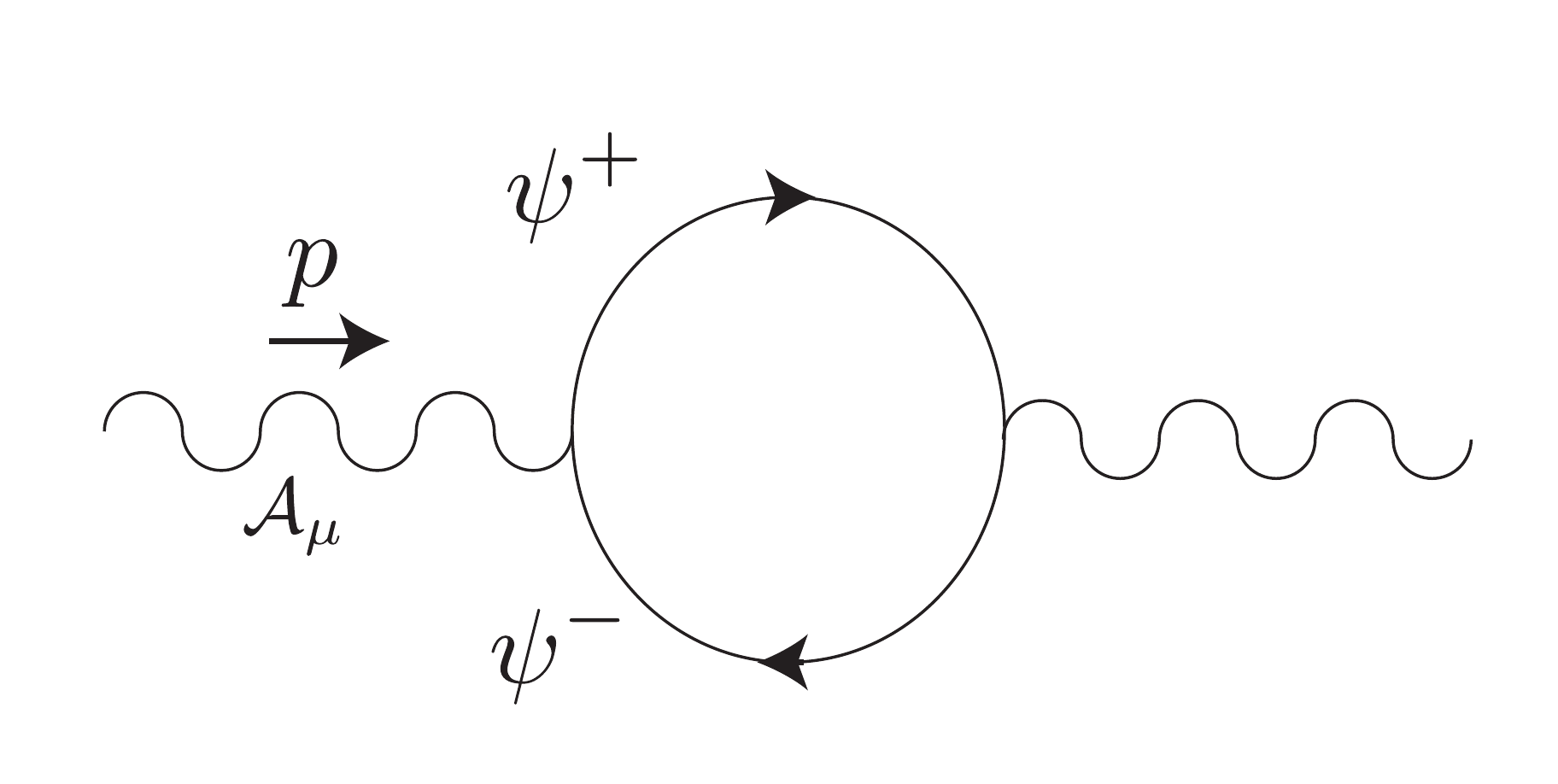} 
   \caption{Fermion contribution to the vacuum polarization.}
   \label{fig:vac_polarization_ferm}
\end{figure}

Having found all theories that are asymptotically and anomaly free and have global minima inside the affine Weyl chamber, now we turn to the question whether the $3$-D effective coupling constant is small in the IR. The weak coupling is necessary in order to trust the semi-classical treatment of such theories. We will find that the necessary condition that a theory stays in the weakly coupled regime is that
\begin{eqnarray}
\frac{\bm \mu_m\cdot\bm \Phi_{\scriptsize \mbox{min}}}{2\pi} \notin \mathbb Z \quad \mbox{for all non-zero}~\bm\mu_m\,, m=\{1,2,...,d({\cal R})\}\,,
\label{strong coupling condition}
\end{eqnarray}
where $\mathbb Z$ is the set of integer numbers. 

In order to obtain the effective $3$-D coupling constant starting from the four dimensional theory, we need to compute the contribution from both gauge and fermion loops upon compactifying the theory over $\mathbb S^1$. The contribution from the gauge fluctuations was found in \cite{Anber:2014sda}. The contribution from fermions in a general representation ${\cal R}$ can be obtained from the vacuum polarization diagram of Figure \ref{fig:vac_polarization_ferm}.  Using the propagator on $\mathbb R^3 \times \mathbb S^1$ (see \cite{Anber:2014sda} for details), the one-loop fermion contribution to the vacuum polarization reads     
\begin{eqnarray}
\nonumber
\Pi^{ed}_{MN}(p, \omega)=
-n_{\cal R}\frac{g^2(L)}{2L}\sum_{q\in \mathbb Z}\int\frac{d^3 k}{(2\pi)^3}\mbox{tr}_{{\cal R}}\left[t^et^d\gamma_M \frac{1}{\slashed{K}}\gamma_{N}\frac{1}{\slashed{K}+\slashed{P}} \right]\,,
\label{Pi for fermions all groups}
\end{eqnarray}
where $\gamma_M$ are the Dirac matrices\footnote{We use Dirac fermions in this computation, and hence, the result for Weyl fermions is obtained by dividing by 2.}.
Then, employing the Weyl-Cartan basis we find that the gauge fluctuations ${\cal A}_\mu$ couple to the fermions $\chi_m$ as $\bm \mu_m \cdot{\cal\bm A}_\mu$, where $m=1,2,...d({\cal R})$, and hence, the polarization tensor is given by 
\begin{eqnarray}
\nonumber
\Pi^{ij}_{\mu\nu}(\bm p, \omega)=-\frac{2g^2(L)}{L}n_{\cal R}\sum_{m=1}^{d({\cal R })}\sum_{q\in \mathbb Z}\int\frac{d^3 k}{(2\pi)^3}\frac{\mu^i_m \mu^j_m\left[g_{\mu\nu}\left(-K\cdot P-K^2\right)+K_\mu K_\nu +K_\nu P_\mu+2K_\mu K_\nu\right]}{\left[k^2+\left(\frac{2\pi q}{L}+\frac{ \bm \mu_m \cdot \bm \Phi}{ L}\right)^2 \right]\left[(\pmb k+\pmb p)^2+\left(\frac{2\pi q}{L}+\frac{\bm \mu_m\cdot \bm \Phi}{ L}+\omega\right)^2 \right]}\,,
\label{Pi fermion in Cartan}
\end{eqnarray}
where $K_\mu=\left(\frac{2\pi q}{L}+\frac{\bm \mu_m \cdot\bm\Phi}{L},\vec k\right)$ and $P_\mu=\left(\omega, \bm p\right)$. The computation of the integral and sum was detailed in \cite{Anber:2014sda}. After a few manipulations we find that the IR limit, $\omega=0$, of the vacuum polarization is 
\begin{eqnarray}
\nonumber
\frac{\Pi^{ij}_{\mu\mu} (\bm p, \omega=0)}{\bm p^2}&=&\frac{n_{\cal R}g^2(L)}{6\pi^2}\sum_{m=1}^{d({\cal R})}\mu_m^i\mu_m^j\left[\psi\left(\frac{\bm \mu_m\cdot \bm \Phi}{2\pi}\right)+\psi\left(1-\frac{\bm \mu_m\cdot \bm \Phi}{2\pi}\right) \right]\\
&+& \mbox{other terms independent of}~\bm\mu\,.
\end{eqnarray}
The combination of the digamma functions $\psi(x)+\psi(1-x)$ blows up when $x$ is an integer including zero. Thus, the smallness of the coupling constant is guaranteed if the condition (\ref{strong coupling condition}) is respected. Now, from (\ref{simple fermion action}) and(\ref{unregulaized effective potential}) we see that the fermion mass is given by $m(p,\bm \mu_m)=\left|\frac{2\pi p}{L}+\frac{\bm\mu_m \cdot \bm \Phi_{\scriptsize \mbox{min}}}{L}\right|$, where $p \in \mathbb Z$ denotes the Kaluza-Klein mode. For $p=0$ and  $\frac{\bm\mu_m \cdot \bm \Phi_{\scriptsize \mbox{min}}}{2\pi}\neq q, q\in \mathbb Z$, we see that the mass of the zero mode is $\frac{\bm\mu_m \cdot \bm \Phi_{\scriptsize \mbox{min}}}{L}$.  Hence for $\frac{\bm\mu_m \cdot \bm \Phi_{\scriptsize \mbox{min}}}{2\pi}=q, q\in \mathbb Z$, we can shift $p=0$ by an integer to find $m(0,\bm \mu_m)=m(q,0)$, and thus, we conclude that the condition (\ref{strong coupling condition}) is equivalent to saying that non of the zero-mode fermions that are charged under the abelian subgroups $u(1)^{N_c-1}$ are massless.

\section{The admissible class of theories}
\label{The admissible class of theories}
As we pointed out above, finding a global minimum of a theory inside the affine Weyl chamber is not enough to conclude that the theory is weakly coupled in the IR. In fact, one also has to check that there are no light or massless charged modes under $u(1)^{N_c-1}$; otherwise the $3$-D effective theory is strongly coupled in the IR. This adds an extra constrain on the class of theories on $\mathbb R^3\times \mathbb S^1$ that are under analytical control. We call the theory that satisfies all the criteria:
\begin{enumerate}[(a)]
\item asymptotically free,
\item anomaly free,
\item has a global minimum (or a set of degenerate global minima, see Section \ref{Perturbative vacua and the role of discrete symmetries}) inside (and not on the boundary of) the affine Weyl chamber, and
\item has no light or massless charged fermions under $u(1)^{N_c-1}$ 
 \end{enumerate}
{\em admissible} in the sense that such theory is mathematically well-defined and amenable to semi-classical treatment at small circle radius, i.e., at $N_cL\Lambda_{\scriptsize\mbox{QCD}}\ll1$.

For every representation in Tables \ref{su2 table} to \ref{su8 table} we checked whether condition (\ref{strong coupling condition}) is  satisfied with six-digit accuracy, i.e., we declare theories with mass eigenvalues $\leq 10^{-6}$ as having massless modes. The $10^{-6}$ cutoff we choose is consistent with the analytical expressions we have for $su(2)$ and $su(3)$. The admissible theories are shown in blue bold face. In general, representations with at least one fermion that satisfy the above criteria are indicated by blue bold face (the second column in the tables). However, we warn the reader that he/she should also look at the fourth column to see how many fermions  are allowed in the representations, which is also indicated by blue bold face. Finally, it is interesting to note that all theories with two-index symmetric or two-index antisymmetric representations and satisfy ${\cal Z}\cong 0$ (very much respect the center symmetry) have massless modes in the infrared.

The admissible theories are summarized as follows:

\begin{enumerate}

\subsection*{General pattern}

\item Theories with pure $1\leq n_G \leq 5$ flavors of adjoint fermions. This class of theories has been extensively studied in the literature.  

\item Theories with even $N_c$ and fermions in the fundamental, $(100..00)$, and $2\leq n_G\leq 5$ and for all allowed range of $n_F$. However since theories with an odd number of $n_F$ suffer from anomalies, only theories with even $n_F$ are well defined.  

\item Theories with even $N_c>2$  and even number of fermions in the two-index symmetric representation, $S\equiv(200..00)$, and $2\leq n_G\leq 3$. None of these theories suffer from anomalies. 

\item Theories with odd $N_c$ and fermions in the two-index symmetric representation $S\equiv(200..00)$ for $2\leq n_G\leq 4$. However, theories with odd number of $n_{(200..00)}$ have anomalies and are excluded. 

\item Theories with odd $N_c>3$ and fermions in the fundamental representation, $F\equiv(100..00)$, for $2\leq n_G\leq 4$ and large number of $n_F$. Again, notice that theories with odd number of $n_F$ have anomalies. 

\subsection*{Exceptional theories}

\item $su(2)$ with a single fermion flavor in the representation ${\cal R}=(4)$. This theory was considered previously in \cite{Poppitz:2009tw}.

\item $su(2)$ with a single fermion flavor in the representation ${\cal R}=(3)$ and $n_G=1,2$. 

\item $su(4)$ with fermions in $(020)$ and $1\leq n_G\leq 3$.

\item $su(6)$ with fermions in $(00100)$ and $3\leq n_G\leq 5$. 

\end{enumerate}

Now, a few remarks are in order.
\begin{enumerate}[(a)]

\item Our classification is limited by our numerical capabilities to go beyond $su(8)$. In particular, some more admissible theories may be included beyond $su(8)$. We leave this for future investigation. 

\item Generally, one can add a small mass to the massless fermionic excitations or turn on a global $U(1)$ vector holonomy \cite{Cherman:2016hcd} (for theories with fermions in vector representations), which renders the theory weakly coupled in the IR. A small mass will not greatly affect the global minimum of the potential. In fact, adding a mass will only increase the chance of the theory to have a nontrivial minimum since a massive fermion in a representation ${\cal R}\neq \mbox{adj}$ will have less power, to fight against the adjoint fermions that prefer a nontrivial minimum, compared to a massless one. 

\item  Finally, we note that some of the admissible theories have degenerate global minima. This will have far-reaching consequences, as we will see in the next section and Section \ref{Monopole-instantons and fermions zero modes}.

\end{enumerate}

\subsection{Perturbative vacua and the role of discrete symmetries}
\label{Perturbative vacua and the role of discrete symmetries}

It has been understood for a long time that theories with adjoint fermions have a unique vacuum that preserves center symmetry, parity, and charge conjugations \cite{Unsal:2007jx,Davies:1999uw}.  In this section we perform a systematic analysis to shed light on the nature of  the perturbative vacua of the admissible theories with fermions in mixed representations.  One of the important tasks is to examine the uniqueness of the vacua we found in the previous section by means of a minimization procedure that aims to find all the degenerate global minima of the potential. 

An invaluable tool in our study is the Polyakov loop wrapping the $\mathbb S_1$ circle: $\Omega_{\cal R}=e^{i\bm H_{{\cal R}}\cdot \bm \Phi}$. In particular, the fundamental Polyakov loop $\Omega_F$ transforms under the center group $\mathbb Z_{N_c}$ of $SU(N_c)$ as $\Omega_F \rightarrow e^{i \frac{2\pi k}{N_c}}\Omega_F$, $k=1,2,...,N_c$. Using the Frobenius formula, one can determine whether the Lagrangian of fermions in a representation ${\cal R}$ is invariant under $\mathbb Z_{N_c}$ or a proper subgroup of it. For example, using the expressions in Appendix \ref{Using the Frobenius formula} it is trivial to show that the Lagrangian of adjoint matter  is invariant under $\mathbb Z_{N_c}$, while the Lagrangians of antisymmetric and two-index symmetric fermions are invariant under ${\mathbb Z_2}$ (for $N_c$ even), etc.

Under charge conjugation ${\cal C}$ and parity,  ${\cal P}:\bm r \rightarrow -\bm r$, the Cartan components of the gauge field transform as (we use the metric $\eta_{MN}=(1,-1,-1,-1)$)
\begin{eqnarray}
\nonumber
{\cal P} \bm A_{M}(t,\bm r) {\cal P}^\dagger \rightarrow \bm A^{M}(t,-\bm r)\,,\\
{\cal C} \bm A_{M}(t,\bm r) {\cal C}^\dagger \rightarrow -\bm A_{M}(t,\bm r)\,,
\end{eqnarray}
and therefore, the Polyakov loop transforms as
\begin{eqnarray}
\Omega_{\cal R} \rightarrow \Omega_{\cal R}^\dagger 
\end{eqnarray}
under both ${\cal P}$ and ${\cal C}$. Thus, we have 
\begin{eqnarray}
\mbox{Im}\Omega_{\cal R} \stackrel{{\cal C}~\mbox{or}~{\cal P}}{\longrightarrow} -\mbox{Im}\Omega_{\cal R}\,. 
\end{eqnarray}
If we draw the eigenvalues of $\Omega$ on the unit circle, then both   ${\cal P}$ and ${\cal C}$ send every eigenvalue to its complex conjugate.  Therefore, a theory with a unique vacuum must have complex conjugate pairs of eigenvalues. However, in a theory with spontaneously broken ${\cal P}$ or ${\cal C}$ symmetries the set of eigenvalues of one vacuum is the complex conjugate of the eigenvalues of the other vacuum.  Then  ${\cal P}$ and ${\cal C}$ operations send every eigenvalue of one vacuum to an eigenvalue of the other. Geometrically, this can be thought of as a reflection about the real axis.

Under time reversal, ${\cal T}: t\rightarrow -t$, we have
\begin{eqnarray}
{\cal T} \bm A_{M}(t,\bm r) {\cal T}^{-1} \rightarrow \bm A^{M}(-t,\bm r)\,,
\end{eqnarray}
and hence,
\begin{eqnarray}
\Omega_{\cal R} \rightarrow \Omega_{\cal R}\,. 
\end{eqnarray}
However, since ${\cal T}$ is antiunitary we find
\begin{eqnarray}
\mbox{Im}\Omega_{\cal R}&=&\frac{1}{2i}\left[\Omega_{\cal R}-\Omega_{\cal R}^\dagger  \right] \stackrel{{\cal T}}{\longrightarrow} -\mbox{Im}\Omega_{\cal R}\,.
\end{eqnarray}
Therefore, one can use $\mbox{Im}\left(\mbox{tr}_{\cal R}\Omega\right)$ as a gauge invariant order parameter for the breaking of ${\cal C}$, ${\cal P}$, ${\cal T}$, and ${\cal C}{\cal P}{\cal T}$ symmetries \cite{Unsal:2006pj}.

\subsubsection{Theories with a unique vacuum}

\begin{figure}[t] 
   \centering
	\includegraphics[width=2in]{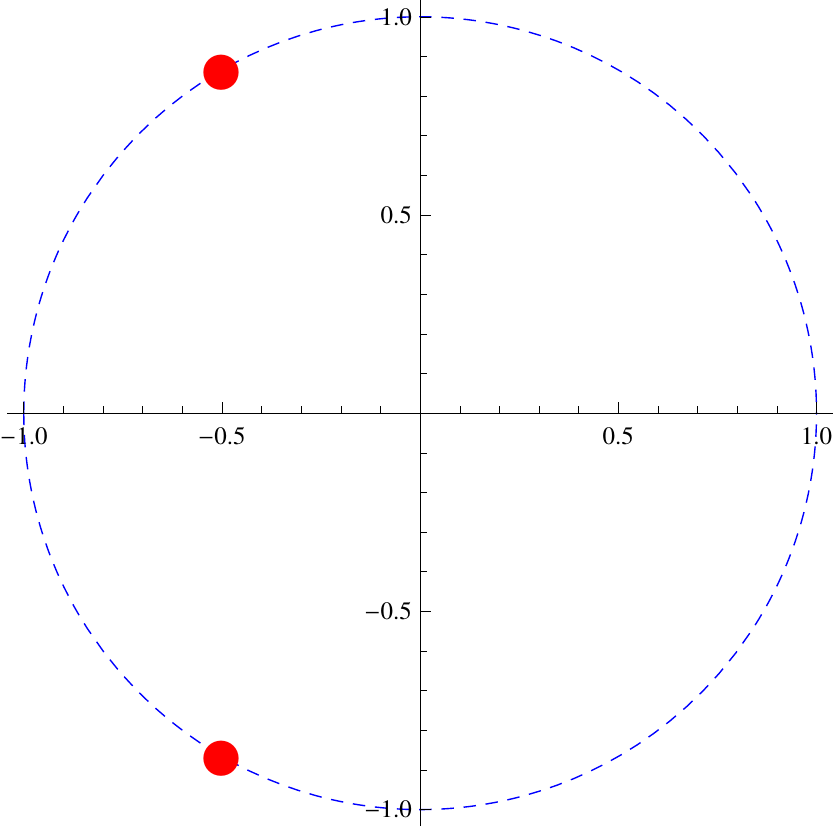}\quad\quad
   \includegraphics[width=2in]{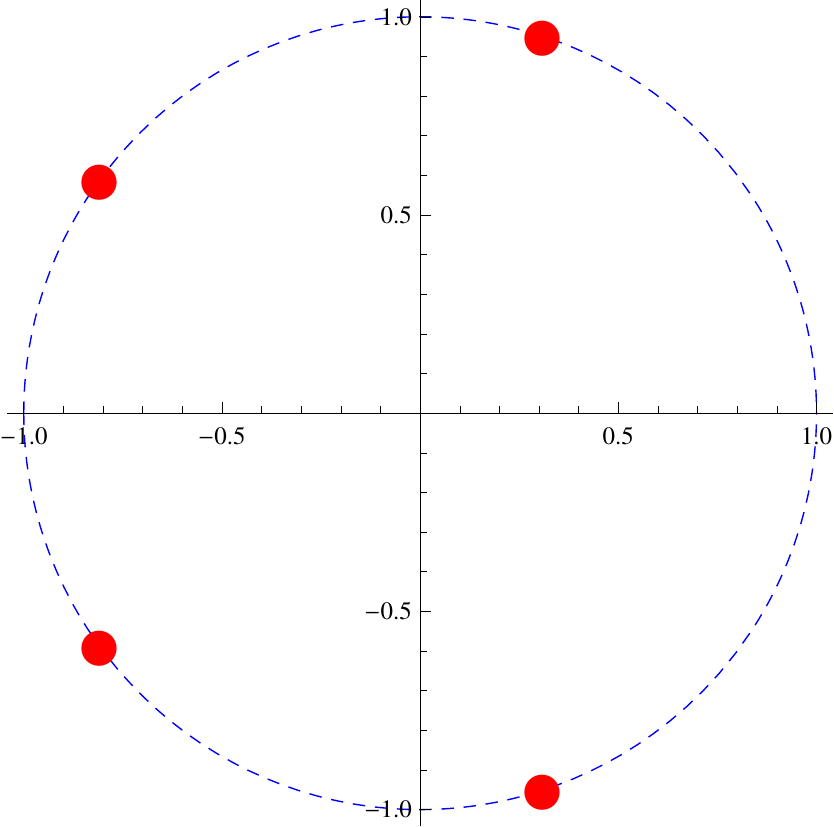} \quad\quad 
   \caption{The eigenvalues of the fundamental Polyakov loop. Left panel: $su(2)$ with two fundamental and two adjoint fermions. Right panel: $su(4)$ with two fundamental and two adjoint fermions. In both cases we find $\mbox{tr}_{\scriptsize F}\Omega=-1$. Since the trace is real, the unique vacuum of these theories respects ${\cal C}$, ${\cal P}$, and ${\cal T}$ symmetries.}
   \label{the unit circle with fundamentals in su2 and su4}
\end{figure}

 A typical distribution of the eigenvalues of the fundamental Polyakov loop is shown in Figure \ref{the unit circle with fundamentals in su2 and su4} for $su(2)$ and $su(4)$ with two adjoint and two fundamental fermions. Both of these theories have a unique vacuum. Also, since $\mbox{Im}\left(\mbox{tr}_F\Omega\right)=0$, both of them preserve ${\cal C}$, ${\cal P}$, and ${\cal T}$ symmetries. In fact, we find that all $su(N_c)$ theories with even $N_c$ and fundamental fermions (of course with an appropriate number of adjoint fermions as in Tables \ref{su2 table} to \ref{su8 table}) have a unique vacuum that preserves all the discrete symmetries.

\subsubsection{Theories with multiple vacua}

\begin{figure}[t] 
   \centering
	\includegraphics[width=2in]{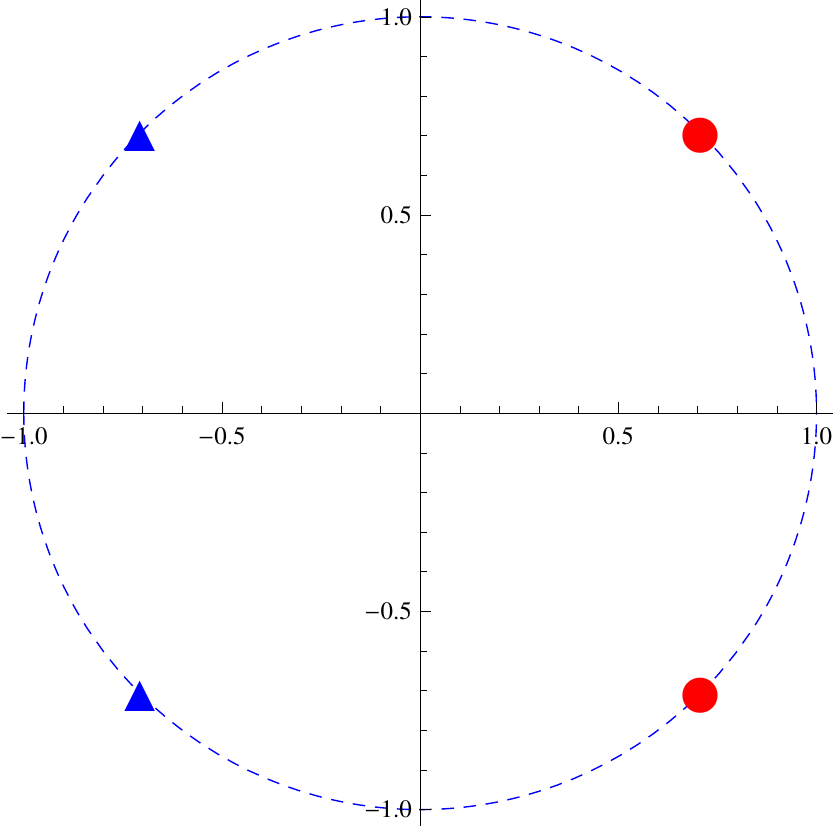} 
   \caption{The eigenvalue distribution of $\Omega_F$ for $su(2)$ with a single fermion in the $(4)$ representation. Different vacua are labeled by different markers and distinct colors.  The red (circle) vacuum has $\mbox{tr}_{F}\Omega=\sqrt 2$, while the blue (triangle) vacuum has $\mbox{tr}_{F}\Omega=-\sqrt 2$. Since $\Omega_F$ is real, this theory respects ${\cal C}$, ${\cal P}$, and ${\cal T}$ symmetries.}
   \label{the unit circle with su2 with 4 irrep}
\end{figure}

An interesting observation is that a subclass of the admissible theories have degenerate vacua. In order to search for the degenerate minima, we feed the minimization algorithm with different initial values of $\bm \Phi$ chosen randomly. We declare a set of minima degenerate when the difference of $V_{\scriptsize\mbox{eff}}$ computed at these minima is less than $10^{-10}$.  

 The simplest of these theories is $su(2)$ with a single fermion in the $(4)$ representation. The Lagrangian of this theory is invariant under $\mathbb Z_2$ center symmetry that negates the fundamental Polyakov loop: $\mbox{tr}_{F}\Omega\rightarrow  -\mbox{tr}_{F}\Omega$. This theory has two degenerate vacua as can be seen from the eigenvalue distribution of $\Omega_F$ shown in Figure \ref{the unit circle with su2 with 4 irrep}. The trace of the fundamental Polyakov loop is $\mbox{tr}_{F}\Omega=\pm \sqrt 2$ for the red (circle) and blue (triangle) vacua, respectively. Therefore, we see that every vacuum breaks the $\mathbb Z_2$ center and that the two vacua are exchanged under the application of the $\mathbb Z_2$ center transformation. However, since  $\mbox{tr}_{F}\Omega$ is real, both vacua respect ${\cal C}$, ${\cal P}$, and ${\cal T}$ symmetries. 

\begin{figure}[t] 
   \centering
   \includegraphics[width=2in]{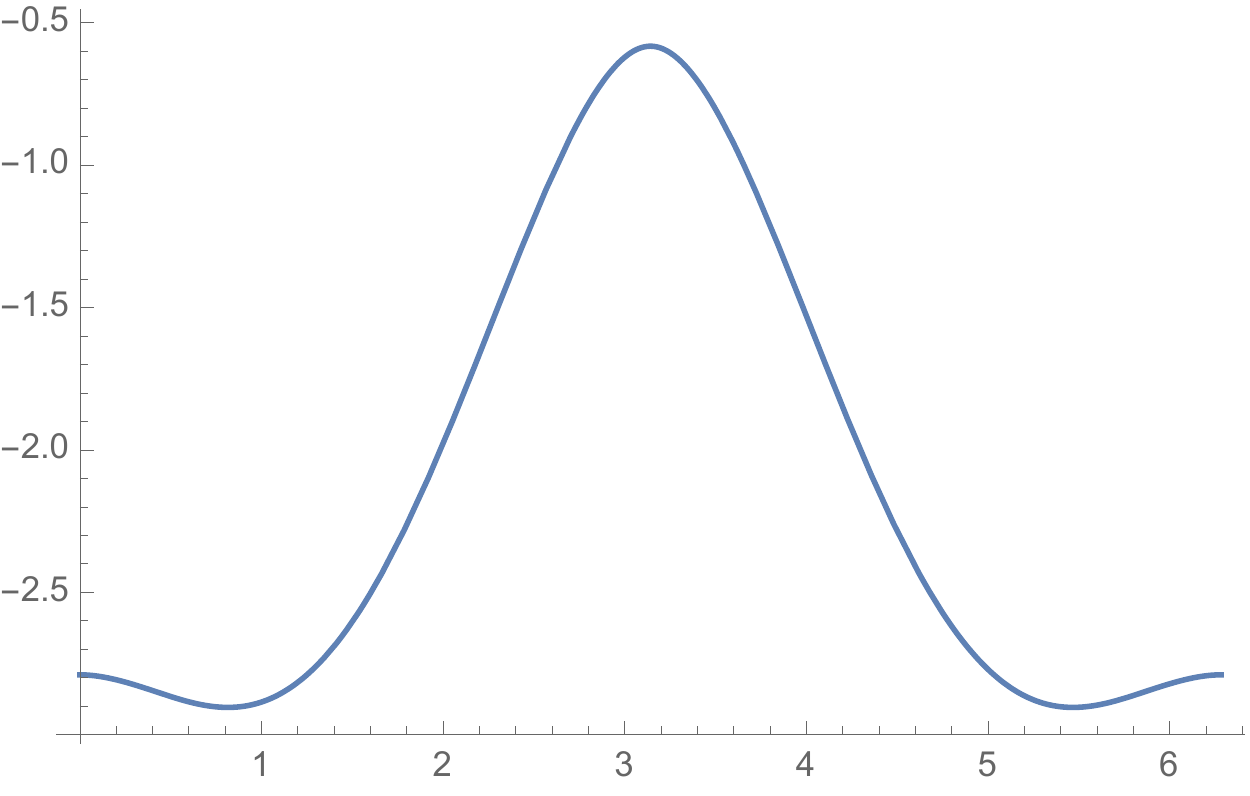} 
   \caption{A cross section of the effective potential of $su(4)$ with $n_{(200)}=2$ and $n_G=2$. We plot $V_{\scriptsize\mbox{eff}}(\Phi_1,\Phi_2,\Phi_3)$ as a function of $0\leq\Phi_1\leq 2\pi$ for constant values of $\Phi_2$ and $\Phi_3$.  The two minima are located at $\{5.468, 3.399,2.889\}$ and $\{0.816,3.399,2.889\}$. We use the simple roots $\bm \alpha_1=(1,0,0),~\bm \alpha_2=\left(-\frac{1}{2}, \frac{\sqrt 3}{2},0\right),~\bm \alpha_3\left(0,-\frac{1}{\sqrt 3},\sqrt{\frac{2}{3}}\right)$.}
   \label{degenerate effective potential}
\end{figure}

\begin{figure}[t] 
   \centering
 \includegraphics[width=2in]{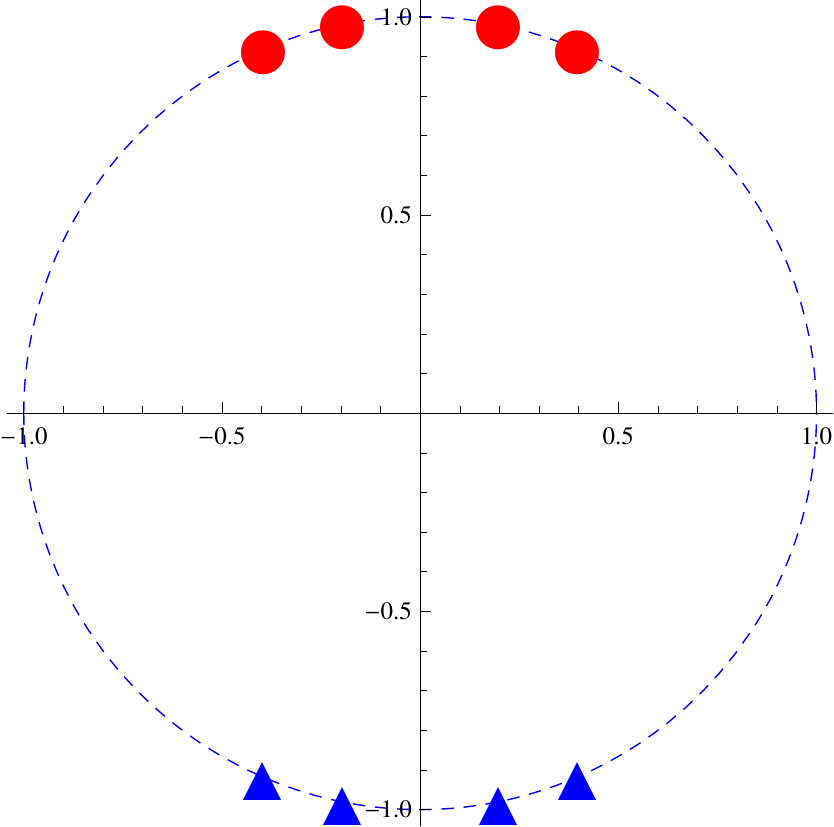} 
	\includegraphics[width=2in]{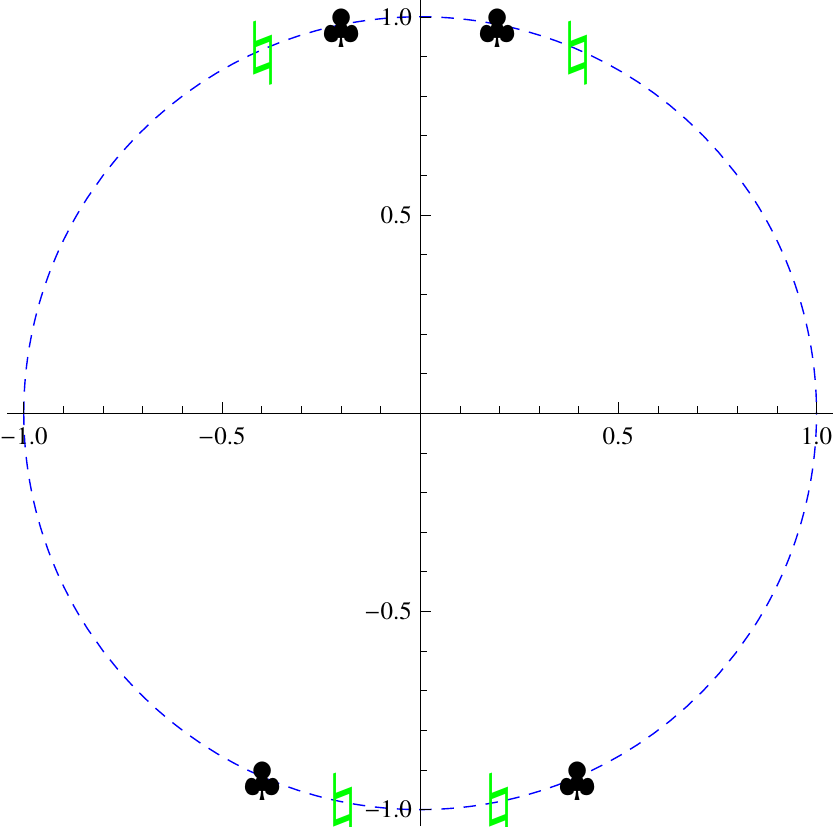} 
   \caption{The eigenvalue distribution of $\Omega_F$ for  $su(4)$ with two adjoint and two $(200)$ fermions.  Different vacua are labeled by different markers and distinct colors. The theory has $4$ degenerate minima and broken $\mathbb Z_2$ center. Every pair of minima is exchanged under $\mathbb Z_2$. In order to reduce clutter, we plot the eigenvalues of $\Omega_F$ for the related vacua (by $\mathbb Z_2$ transformation) on two separate panels. The right and left panels are not exchanged under any symmetry, and in general, the degeneracy between the two panels could be lifted after taking higher order loops into account.  In the left panel we have $\mbox{tr}_{F}\Omega=\pm i3.797$ for the red (circle) and blue (triangle) vacua, respectively.  In the right panel we have $\mbox{tr}_{F}\Omega=\pm i0.124$ for the Green (musical note) and black (club suit) vacua, respectively. All vacua break ${\cal C}$, ${\cal P}$, and ${\cal T}$ symmetries.}
   \label{the unit circle with su4 with ng 2 and n200 2 irrep}
\end{figure}

The second case is $su(4)$ with two adjoint and two $(200)$ (two-index symmetric) fermions. This theory enjoys a $\mathbb Z_2$ center symmetry. Here we have $4$ degenerate vacua that break the center symmetry. However, only each pair of them is exchanged under the $\mathbb Z_2$ center transformation. Thus, we have accidental degeneracy. In Figure \ref{degenerate effective potential} we plot a cross section of the effective potential showing $2$, out of the $4$, degenerate vacua, that are not related via a $\mathbb Z_2$ symmetry. We also plot the eigenvalue distribution of the fundamental Polyakov loop $\Omega_F$ in Figure \ref{the unit circle with su4 with ng 2 and n200 2 irrep}. In order to reduce clutter, we plot the eigenvalues of $\Omega_F$ on two separate panels. The right and left panels are not exchanged under any symmetry, and in general, the degeneracy between the two panels is expected to be lifted upon taking higher order loops into account. In addition, since $\mbox{tr}_{F}\Omega$ is imaginary, ${\cal P}$, ${\cal C}$, and ${\cal T}$ symmetries are spontaneously broken in the four different vacua. The accidental symmetry happens also in $su(8)$ with two adjoint and two $(200)$, and three adjoint and four $(200)$ fermions. In this case we find at least four degenerate vacua. However, our numerical method does not have enough resolution to check weather there are more degenerate vacua.

\begin{figure}[t] 
   \centering 
	\includegraphics[width=2in]{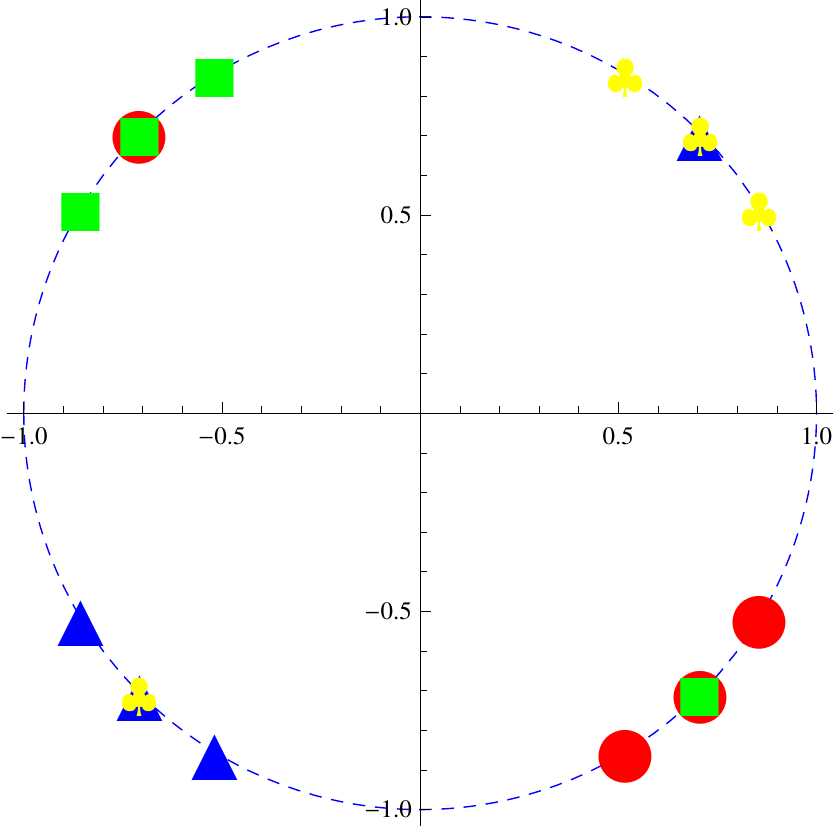} 
   \caption{The eigenvalue distribution of $\Omega_F$ for $su(4)$ with one adjoint and one $(020)$ fermions. The theory has four degenerate vacua labeled by different markers and distinct colors.  The red (circle), blue (triangle), green (square), and yellow (club suit)  vacua have $\mbox{tr}_{F}\Omega=1.373(1-i)\,,-1.373(1+i)\,,1.373(-1+i)\,,1.373(1+i)$, respectively. All the four vacua break ${\cal C}$, ${\cal P}$, and ${\cal T}$ symmetries.}
   \label{the unit circle with su4 with 022 irrep}
\end{figure}

Now, we move to the case of $su(4)$ with fermions in the $n_{G}\oplus n_{(020)}$ representation. This theory enjoys a $\mathbb Z_4$ center symmetry, which is completely broken in the four degenerate vacua. These vacua are shuffled under a $\mathbb Z_4$ transformation. In Figure \ref{the unit circle with su4 with 022 irrep} we plot the eigenvalue distribution of the fundamental Polyakov loop. Since  $\mbox{Im}\left(\mbox{tr}_{F}\Omega\right) \neq 0$, all the four vacua break  ${\cal C}$, ${\cal P}$, and ${\cal T}$ symmetries.  An application of any of the latter symmetries exchange the vacua by a complex conjugation of the eigenvalues.

\begin{figure}[t] 
   \centering
   \includegraphics[width=2in]{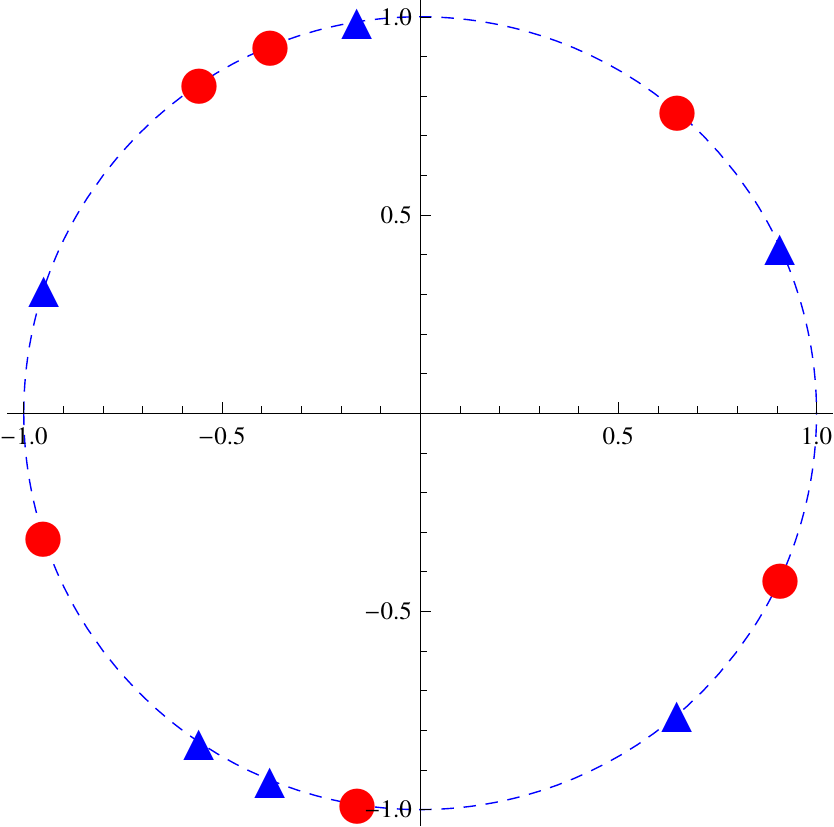} 
   \caption{The eigenvalues  of $\mbox{tr}_F\Omega$ for $su(3)$ with two $(200)$ and three adjoint fermions. This theory has two degenerate vacua, which are labeled by distinct colors and markers. We find  $\mbox{tr}_{F}\Omega=-0.487\pm 0.798I$ for the red (circle) and blue (triangle) vacua, respectively.  The two vacua break ${\cal C}$, ${\cal P}$, and ${\cal T}$, and therefore ${\cal C}{\cal P}{\cal T}$ symmetries.}
   \label{the unit circle with su3 with fundamentals}
\end{figure}

Our next theory is $su(3)$ with three adjoints and two $(200)$ fermions. This theory does not have a center. Yet, we find that the theory has two degenerate minima, see Figures \ref{the unit circle with su3 with fundamentals} and \ref{su3 Weyl chamber}. Upon computing the trace of the fundamental Polyakov loop we find $\mbox{tr}_{F}\Omega=-0.487\pm 0.798I$. Thus, the two vacua are exchanged under any of the discrete symmetries ${\cal C}$, ${\cal P}$, and ${\cal T}$. In fact, we find all $su(N_c)$ with odd $N_c$ and fermions in the $(200)$ representation share the same behavior. Theories with odd number of colors and large number of fundamental fermions belong to the same category.

In Table \ref{summary of admissible theories} we display all the admissible theories along with their center group $\mathbb Z_C$ and indicate whether $\mathbb Z_C$, ${\cal C}$, ${\cal P}$, ${\cal T}$ are broken. In all cases $\mathbb Z_C$ is either fully preserved or broken to unity. Theories with broken centers have as many vacua as elements of $\mathbb Z_C$, unless there is accidental degeneracy. Theories with no center and preserved ${\cal C}$, ${\cal P}$, ${\cal T}$ have a unique vacum, while those with no center and spontaneously broken ${\cal C}$, ${\cal P}$, ${\cal T}$  have doubly degenerate vacua.

\begin{table}
\caption{Summary of admissible theories}
\begin{center}
\tabcolsep=0.11cm
\footnotesize
\begin{longtable}[H]{|*{6}{c|}}
\hline 
Group & ${\cal R}$ & $n_G\oplus n_{\cal R}$ & $\mathbb Z_C$& ${\cal C}\,,{\cal P}\,,{\cal T}$ & Fermion zero modes of ${\cal R}$\\
\hline
$su(2)$ &$(1)$ &  $2\oplus \{2,4,6\}$  & --- & \checkmark& \{0,1\}\\
				&   & $3\oplus \{2,4,6,8,10\}$   & --- & \checkmark & \{0,1\}\\
				&   & $4 \oplus \{2,4,6\}$   & --- & \checkmark & \{0,1\}\\
				&  & $5 \oplus 2$  & --- & \checkmark& \{0,1\}\\
        & $(2)$ &  $[1,5]\oplus 0$ & $\mathbb Z_2$ & \checkmark& $\{2,2\}$\\
				&$(3)$ & $2\oplus 1$ & --- & \checkmark& \{6,4\}\\
				&      & $3\oplus 1$   & --- & \checkmark & \{6,4\}\\
        & $(4)$& $0\oplus 1$ & $\slashed{\mathbb Z_2}$ & \checkmark & $\{6,14\}$\\
		\hline		
	$su(3)$	& $(11)$ & $[1,5]\oplus 0$  & $\mathbb Z_3$ & \checkmark & $\{2,2,2\}$ \\
	        & $(20)$ & $3\oplus 2$  & --- & \ding{54} & $\{2,3,0\}$ \\
		\hline
		$su(4)$ & $(100)$  & $2 \oplus\{ n_{(100)} \in\mbox{even}\,, 2\leq n_{(100)}\leq 14\}$  & --- & \checkmark & $\{0,0,1,0\}$\\
		         &      & $3 \oplus\{n_{(100)} \in\mbox{even}\,, 2\leq n_{(100)}\leq 20\}$  & --- & \checkmark & $\{0,0,1,0\}$\\
						&      & $4 \oplus\{ n_{(100)} \in\mbox{even}\,, 2\leq n_{(100)}\leq 12\}$  & --- & \checkmark & $\{0,0,1,0\}$\\
						&      & $5 \oplus\{ 2, 4\}$  & --- & \checkmark & $\{0,0,1,0\}$\\
						&  $(101)$    & $[1,5] \oplus 0$  & $\mathbb Z_4$& \checkmark & $\{2,2,2,2\}$\\
						&  $(020)$    & $1 \oplus \{1,2\}$  & $\slashed{\mathbb Z_4}$ & \ding{54} & $\{6,6,2,2\}$\\
				    &      & $2 \oplus 1$  & $\slashed{\mathbb Z_4}$ & \ding{54} & $\{6,6,2,2\}$\\
				   &      & $3 \oplus 1$  & $\slashed{\mathbb Z_4}$ & \ding{54} & $\{6,6,2,2\}$\\
				&  $(200)$    & $2 \oplus 2$  & $\slashed{\mathbb Z_2}$ & \ding{54} & $\{2,2,2,0\}$ or $\{0,6,0,0\}$\\
		\hline		
			$su(5)$ & $(1000)$  & $2 \oplus\{ 6,8\}$  & --- & \ding{54} & $\{0,0,1,0,0\}$\\
			 &   & $3 \oplus\{12, 14,16,18\}$  & --- & \ding{54} & $\{0,0,1,0,0\}$\\
            &$(1001)$  & $[1,5]\oplus 0$  &$\mathbb Z_5$ & \checkmark & $\{2,2,2,2,2\}$\\
				&     $(2000)$ & $2 \oplus 2$  & --- & \ding{54} & $\{3,0,4,0,0\}$\\
				&      & $3 \oplus 2$  & --- & \ding{54} & $\{2,1,0,3,1\}$\\
				&     & $4 \oplus 2$  & ---& \ding{54} & $\{2,2,0,3,0\}$\\
		\hline		
				$su(6)$ & $(10000)$  & $2 \oplus\{ n_{(10000)} \in\mbox{even}\,, 2\leq n_{(10000)}\leq 14\}$ & --- & \checkmark & $\{0,0,0,1,0,0\}$\\
				&   & $3 \oplus\{ n_{(10000)} \in\mbox{even}\,, 2\leq n_{(10000)}\leq 28\}$ & --- & \checkmark & $\{0,0,0,1,0,0\}$\\
				&   & $4 \oplus\{ n_{(10000)} \in\mbox{even}\,, 2\leq n_{(10000)}\leq 18\}$ & --- & \checkmark & $\{0,0,0,1,0,0\}$\\
				&   & $5 \oplus\{ 2,4,6\}$ & --- & \checkmark & $\{0,0,0,1,0,0\}$\\
				& $(00100)$  & $3 \oplus 1$ & $\mathbb Z_3$& \checkmark & $\{0,2,0,2,0,2\}$\\
				&   & $4 \oplus \{1,2\}$ & $\mathbb Z_3$& \checkmark & $\{0,2,0,2,0,2\}$\\
				&   & $5 \oplus 1$ & $\mathbb Z_3$& \checkmark & $\{0,2,0,2,0,2\}$\\
				& $(10001)$  & $[1,5]\oplus 0$ & $\mathbb Z_6$& \checkmark & $\{2,2,2,2,2,2\}$\\
				& $(20000)$  & $2 \oplus 2$ & $\slashed{\mathbb Z_2}$& \ding{54} & $\{2, 0, 2, 2, 0, 2\}$\\
\hline
$su(7)$ & $(100000)$  & $2 \oplus\{ 6,8\}$ & --- & \ding{54} & $\{0,0,0,0,1,0,0\}$\\
&  & $3 \oplus\{ 12,14,16,18\}$ & --- & \ding{54} & $\{0,0,0,0,1,0,0\}$\\
&  & $4 \oplus 20$ & --- & \ding{54} & $\{0,0,0,0,1,0,0\}$\\
& $(100001)$  & $[1,5]\oplus 0$& $\mathbb Z_7$ & \checkmark & $\{2,2,2,2,2,2,2\}$\\
& $(200000)$  & $2\oplus 2$& --- & \ding{54} & $\{2, 1, 1, 3, 0,1,1\}$\\
&  & $3\oplus 2$& --- & \ding{54} & $\{2, 0, 1, 3, 0, 1, 2\}$\\
&  & $4\oplus 2$& --- & \ding{54} & $\{ 2, 0, 2, 0, 3, 0, 2\}$\\
\hline
$su(8)$ & $(1000000)$  & $2 \oplus\{ n_{(1000000)} \in\mbox{even}\,, 2\leq n_{(100000)}\leq 14\}$ & --- & \checkmark & $\{0, 0, 0, 0, 1, 0, 0, 0\}$\\
 &  & $3 \oplus\{ n_{(1000000)} \in\mbox{even}\,, 2\leq n_{(100000)}\leq 28\}$ & --- & \checkmark & $\{0, 0, 0, 0, 1, 0, 0, 0\}$\\
 &  & $4 \oplus\{ n_{(1000000)} \in\mbox{even}\,, 2\leq n_{(100000)}\leq 24\}$ & --- & \checkmark & $\{0, 0, 0, 0, 1, 0, 0, 0\}$\\
&  & $5 \oplus\{ n_{(1000000)} \in\mbox{even}\,, 2\leq n_{(100000)}\leq 8\}$ & --- & \checkmark & $\{0, 0, 0, 0, 1, 0, 0, 0\}$\\
& $(1000001)$  & $[1,5]\oplus 0$& $\mathbb Z_8$ & \checkmark & $\{2,2,2,2,2,2,2,2\}$\\
& $(200000)$  & $2\oplus 2$& $\slashed{\mathbb Z_2}$ & \ding{54} &  at least four degenerate vacua \\
&    & $3\oplus 4$& $\slashed{\mathbb Z_2}$ & \ding{54} & at least four degenerate vacua\\
\hline
\end{longtable}
\label{summary of admissible theories}
\end{center}
\end{table}
\clearpage

\section{Monopole-instantons and fermion zero modes on $\mathbb R^3 \times \mathbb S^1$}
\label{Monopole-instantons and fermions zero modes}

In this section we calculate the fermion zero modes attached to the fundamental saddles of Yang-Mills on $\mathbb R^3 \times \mathbb S^1$. This computation is essential to understand the structure of the topological molecules that proliferate in the vacuum and cause the theory to confine. The study of these molecules for a general representation $n_G\oplus n_{\cal R}$ will be pursued in a future work.  

The basic non-perturbative saddles in Yang-Mills theory on $\mathbb R^3 \times \mathbb S^1$ are monopole-instantons. The monopole action is $1/N_c$ the action of the BPST instanton. Hence, one can think of a single BPST instanton as being composed of $N_c$ monopole instantons. Each monopole carries a magnetic charge $\bm \alpha^*_a$, $a=0,1,2,...,N_c-1$, where $\alpha^*_a$ is the co-root defined as $\bm \alpha_a^*=\frac{2\bm \alpha_a}{\bm \alpha_a^2}$, $a=1,2,...,N_c-1$, and $\alpha^*_0=-\sum_{a=1}^{N_c-1}\bm \alpha_a^*$.\footnote{In our normalization $\bm \alpha^2=2$, and hence $\bm\alpha^*_a=\bm\alpha_a$ for all $a=0,1,2,...,N_c-1$.} Each monopole-instanton carries a number of fermion zero modes which can be computed using Singer-Poppitz-\"Unsal index  on $\mathbb R^3\times \mathbb S^1$ \cite{Nye:2000eg,Poppitz:2008hr}. The index computation was carried out for fermions in the adjoint and fundamental representations in \cite{Poppitz:2008hr}, and then was generalized for fermions in any representation in \cite{Anber:2014lba}.  The number of fermionic zero modes residing on a monopole-instanton with charge $\bm \alpha^a$, $a=1,2,...,N_c-1$, is given by
\begin{eqnarray}
{\cal I}_{f(\bm\alpha_a^*)}({\cal R})=n_{{\cal R}}\mbox{tr}_{\cal R}\left[ \lfloor{\frac{\bm \Phi\cdot \bm H}{2\pi}}\rfloor\bm \alpha^*_a\cdot \bm H \right]\,,
\label{fermion zero modes index}
\end{eqnarray} 
where $\lfloor x\rfloor$ is the floor function, which is the largest integer less than or equal to $x$.
The number of the zero modes attached to the affine monopole (the monopole corresponding to the root $\bm \alpha_0$) can be envisaged from the fact that a BPST instanton is made up of $N_c$ monopoles. Thus, we have
\begin{eqnarray}
{\cal I}_{f(\bm\alpha_0^*)}({\cal R})=n_{\cal R}\left[T({\cal R})-\sum_{a=1}^{N_c-1}{\cal I}_{f(\bm\alpha_a^*)}({\cal R})\right]\,.
\end{eqnarray}
The computation of the index can be carried out either by explicit sum over the weights of the representation ${\cal R}$ or using the Frobenius formula. The weights can be constructed using Verma bases as we discussed in Section \ref{Computation of the traces}. The use of the Frobenius formula is more involved and we discuss it in Appendix \ref{Computing the index using Frobenius formula}.

In the last column of Table \ref{summary of admissible theories} we list the number of the fermion zero modes attached to the monopole-instantons that correspond to the sequence of the roots $\{\bm \alpha_0, \bm \alpha_1,..., \bm \alpha_{N_c-1}\}$ for all admissible theories. 

Now we come to an important observation regarding the representation $(200...0)$ of $su(N_c)$ with even $N_c$. To be more specific, let us recall the theory $su(4)$ with $n_G=2$ and $n_{(200)}=2$ we discussed in Section \ref{Perturbative vacua and the role of discrete symmetries} (a similar behavior occurs in $su(8)$ with $n_G=2$ and $n_{(200)}=2$).   The effective potential of this representation  admits $4$ degenerate global minima. Each pair of them is related via a $\mathbb Z_2$ center transformation, as shown in the left and right panels of Figure \ref{the unit circle with su4 with ng 2 and n200 2 irrep}. However, every panel is not related to the other by any symmetry.  In fact, what is even more interesting is that the  ${\cal R}=(200)$ fermion zero modes of the two panels are different. The vacua of the left panel has $\{2,2,2,0\}$ while the vacua of the right panel has $\{0,6,0,0\}$  fermion zero modes attached to the monopoles. All four vacua, however, admit $\{2,2,2,2\}$ adjoint fermion zero modes. Thus, one expects that the two minima will have different topological molecules. The proliferation of these molecules could lift the degeneracy even before taking higher order loops into account.  This study will be left for a future work.

\section{$su(N_c)$ theories with fermions in $G\oplus F$: a detailed study}
\label{fermions in adj plus fund}

In this section we consider in great detail the case of mixed representation $G\oplus F$ on $\mathbb R^3\times \mathbb S^1$ and comment on connection between these theories and their cousins on $\mathbb R^4$. This works as an example for the rich phase structure of theories with fermions in mixed representations that will be pursued in great details in a future work.

\subsection*{Global symmetries}

 For convenience, let us take $N_F=n_F/2$ to denote the number of the Dirac fundamental fermions. These theories have a classical global symmetry $U(1)_G\times U(1)_A\times U(1)_B \times SU(n_G)\times SU(N_F)_L\times SU(N_F)_R$. The abelian group $U(1)_G$ is the global phase factor of the adjoint fermions, while $U(1)_B$ and $U(1)_A$ are respectively the baryon number and axial symmetries of the fundamentals. Quantum mechanically, only a diagonal subgroup of $U(1)_G\times U(1)_A$ survives as can be envisaged by studying the BPST instanton with the dressing zero modes (or 't Hooft vertex), which schematically takes the form\footnote{Recall that a BPST instanton consists of $N_c$ monopoles-instantons. Therefore, the total number of the BPST instanton zero modes can be found by summing up the zero modes of the individual monopoles.}:
\begin{eqnarray}
{\cal IT}=e^{-S_{I}}\left(\lambda_G\lambda_G\right)^{N_cn_g}\left(\lambda_L^1\lambda_R^1\right)...\left(\lambda_L^{N_F}\lambda_R^{N_F}\right)\,,
\end{eqnarray}   
where $S_I=\frac{8\pi^2}{g^2}$ is the BPST instanton action and $\lambda_G$ and $\lambda_{L,R}$ are the zero modes of the adjoint and fundamental fermions, respectively.
It is trivial to check that the 't Hooft vertex is invariant under the transformation $\lambda_G\rightarrow e^{i\alpha} \lambda_G\,,\lambda_L\rightarrow e^{i\beta} \lambda_L\,,\lambda_R\rightarrow e^{i\beta} \lambda_R$, where $\beta=-\frac{n_G N_c}{N_F}\alpha$, which is $U(1)_{A+G}$, the diagonal subgroup of $U(1)_G\times U(1)_A$. 

\subsection*{Mass gap and decompactification}

\begin{figure}[t] 
   \centering 
	\includegraphics[width=7in]{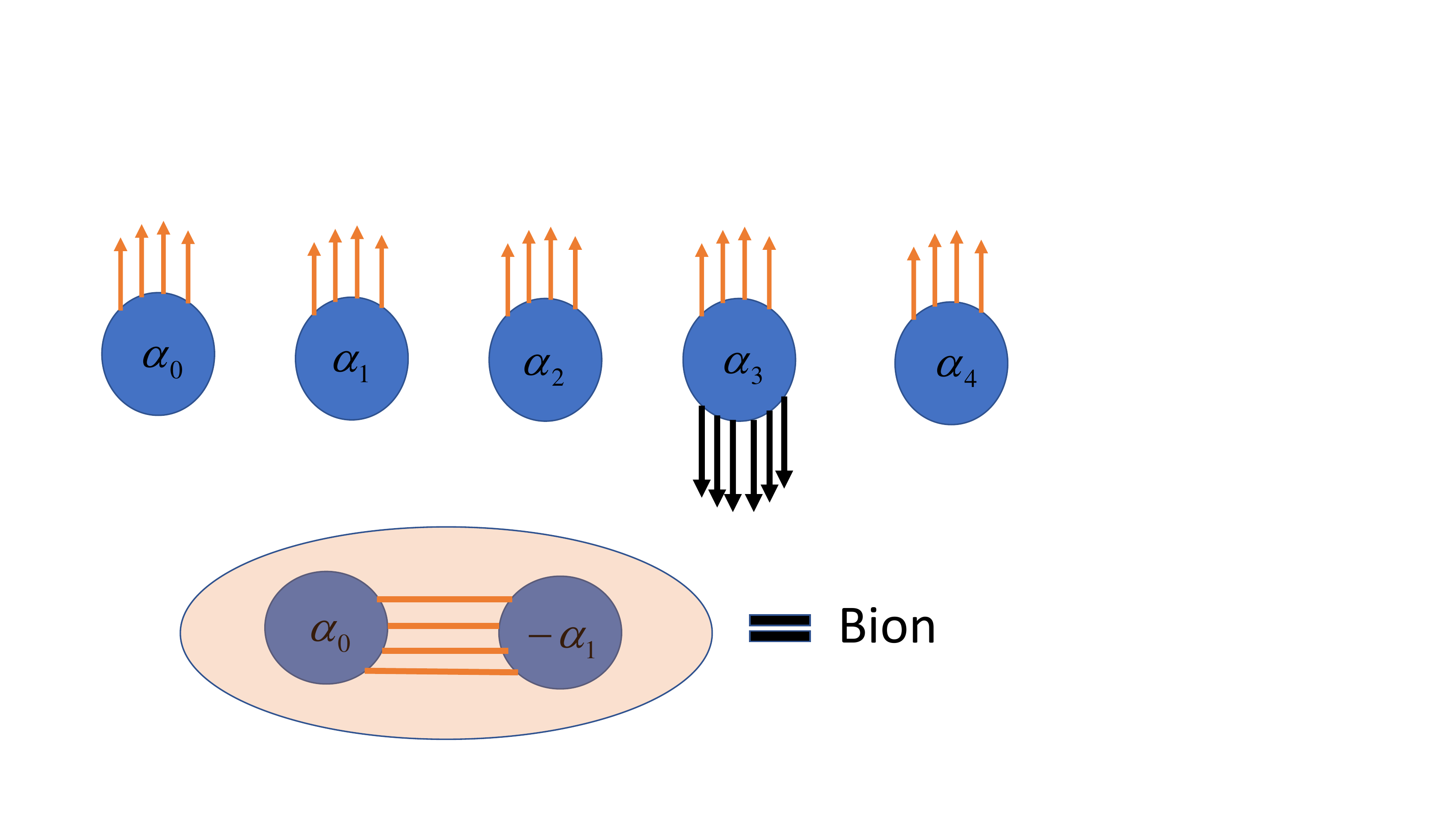}
   \caption{The monopoles and their zero modes in $su(5)$ with $n_G=2,N_F=3$. A monopole and anti-monopole can soak up their adjoint fermion zero modes to result in a magnetically charged bion. The  $\bm \alpha_3$ monopole has additional fundamental zero modes, and hence, it  does not participate in the formation of any magnetically charged objects.}
   \label{monopoles and attached zero modes}
\end{figure}

Both perturbative and nonperturbative spectra of the theory must be invariant under the non-anomalous global symmetries. This is particularly true for the monopole operators (see \cite{Poppitz:2009uq,Poppitz:2009tw,Anber:2014lba} for a lucid description of these operators):
\begin{eqnarray}
\nonumber
{\cal M}_1&=&e^{-S_{\bm \alpha_1}}e^{i\bm \sigma\cdot \bm \alpha_1}\left(\lambda_G\lambda_G\right)^{n_g}\,,{\cal M}_2=e^{-S_{\bm \alpha_2}}e^{i\bm \sigma\cdot \bm \alpha_2}\left(\lambda_G\lambda_G\right)^{n_g}\,,...\\
\nonumber
{\cal M}_a&=&e^{-S_{\bm \alpha_a}}e^{i\bm \sigma\cdot \bm \alpha_a}\left(\lambda_G\lambda_G\right)^{n_g}\left(\lambda_L^1\lambda_R^1\right)...\left(\lambda_L^{Nf}\lambda_R^{Nf}\right)\,,...,{\cal M}_0=e^{-S_{\bm \alpha_0}}e^{i\bm \sigma\cdot \bm \alpha_0}\left(\lambda_G\lambda_G\right)^{n_g}\,,\\
\end{eqnarray}
where $S_{\bm \alpha_a}=\frac{4\pi}{g^2}\left(2\pi\delta_{a,0}+\bm \alpha_a^*\cdot \bm \Phi_0\right)$ is the action of the $\bm \alpha_a$'s monopole at the vev $\bm \Phi_0$. Notice that the fundamental zero modes reside only on one of the monopoles according to the index theorem.
The invariance of the monopole operators demands that the dual photons transform as $\bm \sigma \rightarrow \bm \sigma -2n_g\alpha \bm \rho$ under $U(1)_{A+G}$, where $\bm \rho=\sum_{a=1}^{N_c-1}\omega_a$ is the Weyl vector. These monopoles cannot give rise to a mass gap since they are dressed with fermionic zero modes. However, larger molecules can be formed in the infrared as a result of soaking up the zero modes. In the theory under hand these molecules, we call them magnetic bions, are made of a monopole ${\cal M}_{a}$ and an anti-monopole $\overline{{\cal M}}_{a+1}$ such that none of them carry fundamental zero modes\footnote{There are also molecules that can be made up from monopoles with fundamental zero modes. Such molecules, however, do not carry a net magnetic charge, and hence, do not give rise to a mass gap.} (see \cite{Unsal:2007jx,Anber:2011de} for more details): 
\begin{eqnarray}
{\cal B}_a={\cal M}_{a}\overline{{\cal M}}_{a+1}=e^{-S_{\bm \alpha_a}-S_{\bm \alpha_{a+1}}}e^{i\bm \sigma \cdot \left(\bm \alpha_a-\bm \alpha_{a+1}\right)}\,,
\end{eqnarray} 
where $a=1,2,..,N_c-2$, see Figure \ref{monopoles and attached zero modes}. It is trivial to check that the bion operators are invariant under the non-anomalous global symmetries of the theory. These bions carry magnetic charges $\bm \alpha_a-\bm \alpha_{a+1}$ and their proliferation in the vacuum causes $N-2$ out of $N-1$ dual photons to acquire mass. Therefore, in any $su(N_c)$ theory with fundamental fermions there will be one massless dual photon in the spectrum. Mathematically, the effect of magnetic bions can be taken into account by inserting bion vertices in the partition function. The validity of the semi-classical description hinges on the assumption that the bion gas is very dilute, i.e., $S_{\bm \alpha_a}+S_{\bm \alpha_{a+1}}\gg1$, which is a very good assumption for a small compactification radius.  Finally, the bosonic part of the long-distance effective Lagrangian reads\footnote{The W-bosons and fluctuations of the holonomy field $\bm \Phi$ have masses of order $1/(N_cL)$ and $g/(\sqrt{N_c}L)$, respectively. These masses are hierarchically much larger than the photon mass $\sim e^{-2S_{\scriptsize\mbox{monopole}}}/L$, and hence, we neglect them in the IR description of the theory.}
\begin{eqnarray}
{\cal L}=\frac{1}{2}\left(\partial_\mu\bm \sigma\right)^2+\frac{1}{L^2}\sum_{a=1}^{N-2} {\cal C}_a e^{-\left(S_{\bm \alpha_a}+S_{\bm \alpha_{a+1}}\right)}\cos\left[\bm \sigma \cdot \left(\bm \alpha_a-\bm \alpha_{a+1}\right)\right]\,,
\end{eqnarray}
where ${\cal C}_a$ are ${\cal O}(1)$ dimensionless coefficients that are not important to our discussion. 
 
The mass gaps can be expressed in terms of the strong scale $\Lambda_{\scriptsize \mbox{QCD}}$ and $L$ using the $\beta$-function as\footnote{To one-loop order we have 
\begin{eqnarray}
\frac{4\pi}{g^2(L)}=\frac{\beta_0}{4\pi}\log\left(\frac{1}{L^2\Lambda^2}\right)\,.
\end{eqnarray}
}
\begin{eqnarray}
{\cal MG}_a=\Lambda_{\scriptsize \mbox{QCD}}\left(L\Lambda_{\scriptsize \mbox{QCD}}\right)^{\frac{\beta_0}{4\pi}\bm \Phi_0\cdot \left(\bm\alpha_a^*+\bm \alpha_{a+1}^*\right)-1}\,, \quad a=1,2,...,N_c-2\,,
\label{mass gaps}
\end{eqnarray} 
and $\beta_0$ is given by (\ref{beta function}). In a theory with a broken center symmetry, like in the case of $n_G\oplus N_F$, the mass gaps are not symmetric and rich structures in the theory is expected. Expression (\ref{mass gaps}) enables us to track the mass gaps as a function of the compactification radius all the way to $L\Lambda_{\scriptsize \mbox{QCD}}\gtrapprox 1$ and then\footnote{Strictly speaking, the borderline between the weak and strong coupling regimes is controlled by the parameter $N_cL\Lambda_{\scriptsize \mbox{QCD}}$, i.e., the W-boson mass, rather than $L\Lambda_{\scriptsize \mbox{QCD}}$. However, for a small number of colors this distinction is not of great importance, which makes our discussion simpler.} to $L \rightarrow \infty$. Four scenarios are possible as we decompactify the circle. (1) The mass gap increases as we increase $L$, and hence, the theory flows to the nonabelian confining  regime at $L\sim 1/\Lambda_{\scriptsize \mbox{QCD}}$. In this case spontaneous breaking of the continuous chiral symmetry happens on the way. (2) The mass gap is a monotonically decreasing function of $L$ and vanishes as $L \rightarrow \infty$, and therefore, the semi-classical description of the theory is valid all the way $L\rightarrow \infty$. (3) The mass gap decreases as we increase $L$ and then it saturates to a non-zero value as $L$ approaches $\Lambda_{\scriptsize \mbox{QCD}}$. This can happen if chiral symmetry breaking happens on the way. (4) The mass gap increases to some value as we increase $L$ until the theory hits a Banks-Zaks fixed point \cite{Banks:1981nn} before approaching the strong scale. After this point  the coupling constant ceases to run and the mass gap will decrease again as we decompactify the circle. The semi-classical description is also valid in this scenario all the way to $L\rightarrow \infty$. Theories with preserved center symmetry will fall into one of these four categories. However, mass gaps in theories with a broken center symmetry can enjoy a mix of these scenarios. 

Generally, theories with a small number of fundamental fermions belong to class (1), while theories with a large number of fundamentals belong to class (4). To be more specific, let us consider two examples of $N_c=5$ admissible theories with $n_G=2\,,N_F=3$ and then with $n_G=3\,,N_F=8$. As we see from Table \ref{summary of admissible theories} both theories have spontaneously broken discrete ${\cal C}$, ${\cal P}$, ${\cal T}$ symmetries. However, as we will see, they have different behaviors in the decompactification limit. 

\begin{figure}[t] 
   \centering 
	\includegraphics[width=3in]{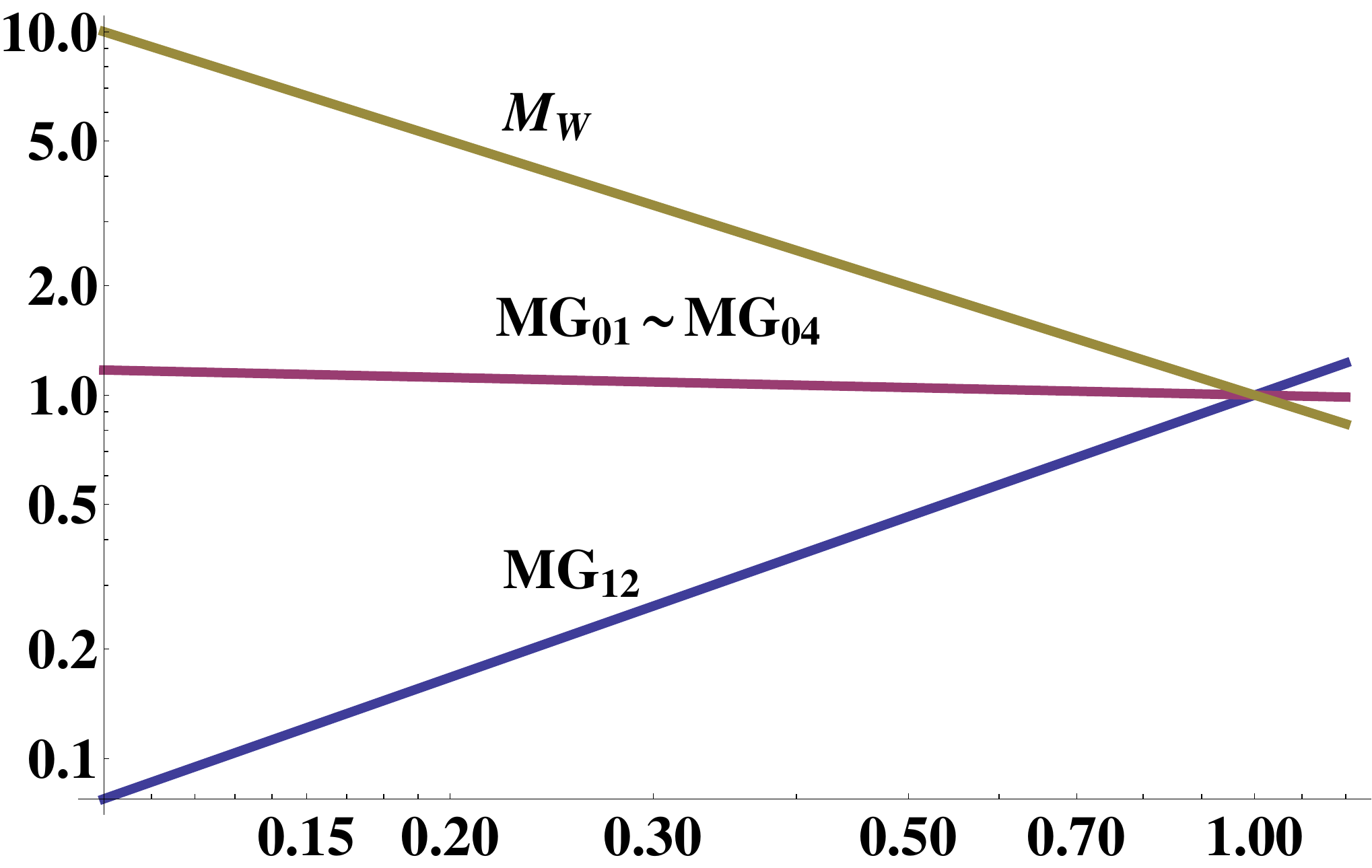}
	\includegraphics[width=3in]{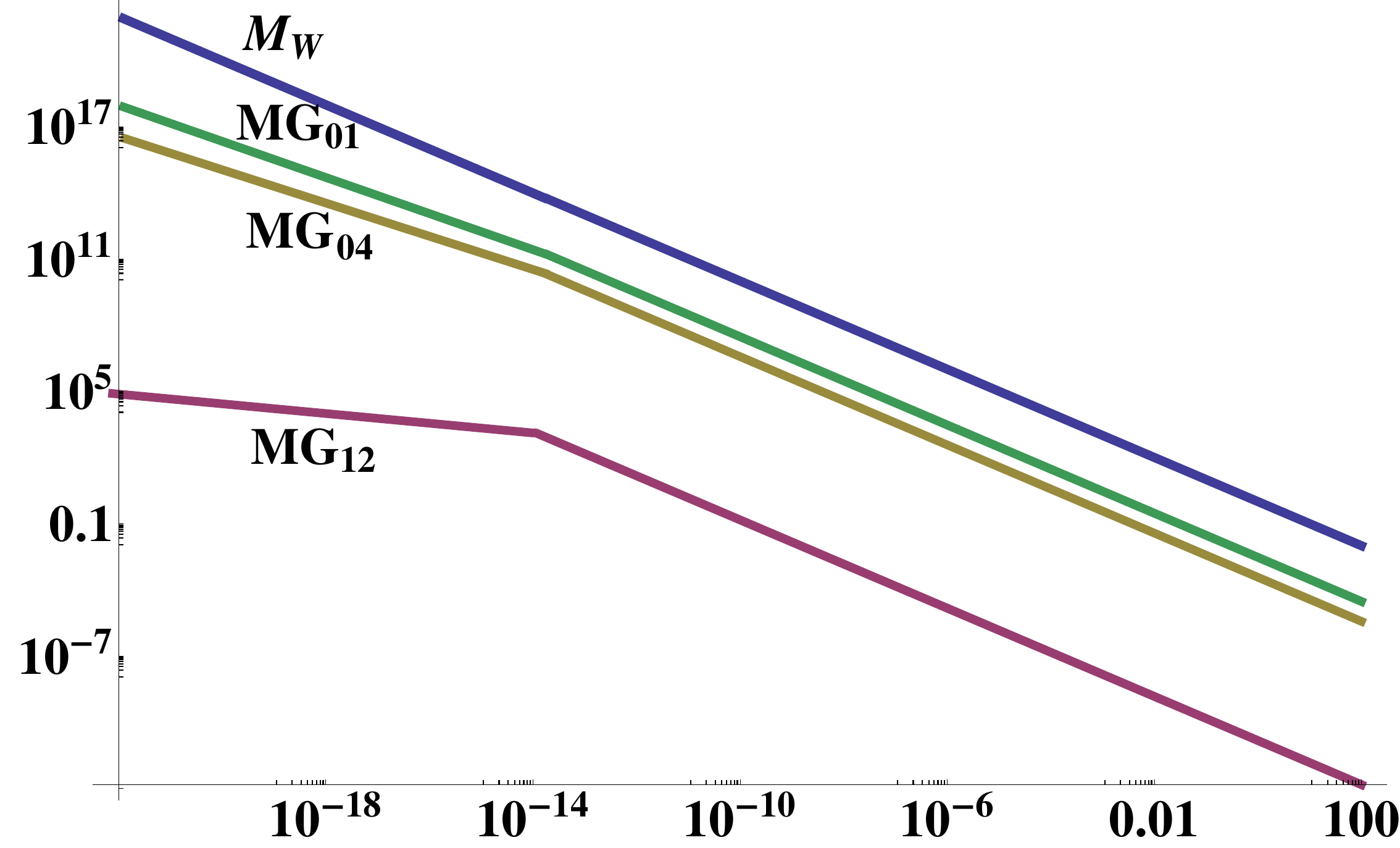} 
   \caption{The mass gaps ${\cal MG}$ (in units of $\Lambda_{\scriptsize\mbox{QCD}}$) as functions of $\Lambda_{\scriptsize\mbox{QCD}}L$ for two admissible theories with $N_c=5$. For comparison, we also plot the mass the W-boson. On the left: $n_G=2, N_F=3$. Weak coupling is lost at $L\lambda_{\scriptsize\mbox{QCD}}\gtrapprox 1$.  On the right: $n_G=3, N_F=8$. See the text for details.}
   \label{mass gaps as functions of L}
\end{figure}

The mass gaps of the $n_G=2\,,N_F=3$ theory are ${\cal MG}_{12}=\Lambda_{\scriptsize \mbox{QCD}}\left(L\Lambda_{\scriptsize \mbox{QCD}}\right)^{1.111}\,,{\cal MG}_{04}=\Lambda_{\scriptsize \mbox{QCD}}\left(L\Lambda_{\scriptsize \mbox{QCD}}\right)^{-0.069}\,,{\cal MG}_{01}=\Lambda_{\scriptsize \mbox{QCD}}\left(L\Lambda_{\scriptsize \mbox{QCD}}\right)^{-0.068}$. The subscript in ${\cal M}_{ab}$ denotes the monopoles that are used to make up the magnetic bion. For example, ${\cal MG}_{12}$  is made of the monopoles $\bm\alpha_1$ and $\bm\alpha_2$, etc.  Notice that in this example the monopole $\alpha_3$ carries fundamental zero modes, and therefore, does not participate in making any magnetic bions; see Figure \ref{monopoles and attached zero modes}. The behavior of the mass gaps as functions of the dimensionless parameter $L\Lambda_{\scriptsize \mbox{QCD}}$ is depicted in the left panel of Figure \ref{mass gaps as functions of L}. The mass gaps are asymmetric, as expected in a theory with a broken center symmetry. While ${\cal MG}_{12}$ is a monotonically increasing function of $L$, both ${\cal MG}_{04}$ and ${\cal MG}_{01}$ are monotonically decreasing functions of $L$. At any scale $L\ll\Lambda^{-1}_{\scriptsize \mbox{QCD}}$ the theory is weakly coupled, as can be seen from comparing the mass gaps with the W-mass $\sim 1/(N_cL)$, and hence, the semi-classical description is adequate. However, as $L$ approaches  $\Lambda_{\scriptsize \mbox{QCD}}$ the W-boson mass becomes comparable to ${\cal MG}_{12}$, we loose the hierarchy of scales, and the theory is expected to enter its strongly coupled nonabelian confining regime. In addition, spontaneous breaking of chiral symmetry is expected as we hit the strong scale. One can also check from the two-loop $\beta$ function (\ref{beta function}) that the theory does not develop an IR fixed point before hitting the strong scale. Thus, this theory belongs to a mix of classes (1) and (2) described in the previous paragraph.  Whether the theory really enters a strongly coupled confining regime can only be checked via lattice simulations.     

Now, let us move to the second example of $n_G=3,N_F=8$. The mass gaps of this theory are  ${\cal MG}_{12}=\Lambda_{\scriptsize \mbox{QCD}}\left(L\Lambda_{\scriptsize \mbox{QCD}}\right)^{-0.22}\,,{\cal MG}_{04}=\Lambda_{\scriptsize \mbox{QCD}}\left(L\Lambda_{\scriptsize \mbox{QCD}}\right)^{-0.753}\,,{\cal MG}_{01}=\Lambda_{\scriptsize \mbox{QCD}}\left(L\Lambda_{\scriptsize \mbox{QCD}}\right)^{-0.818}$. At very small $L$ we see that all mass gaps decrease with $L$.  The theory, however, admits a Banks-Zaks fixed point in the infrared at $\frac{g_*^{2}}{4\pi}=0.131 \ll 1$, as can be easily checked\footnote{A Banks-Zacks fixed point can be inferred from the two-loop $\beta$-function in  (\ref{beta function}): we require that the $\beta$-function vanishes at weak coupling, i.e. $g^2/(4\pi)\ll 1$. The existence of this fixed point hinges on the assumption that higher order loops do not bring in large numerical factors.} from (\ref{beta function}). The fixed point corresponds to $L_*\approx 10^{-14}/\Lambda_{\scriptsize \mbox{QCD}}$. Beyond this critical compactification radius the coupling constant ceases to run and all mass gaps decrease  as ${\cal MG}\approx C/L$, for some constant $C$ that can be determined from continuity across $L_*$. This intricate behavior is illustrated in the right panel of Figure \ref{mass gaps as functions of L}. Notice that the photons masses are always less than the W-boson mass, and therefore, the semi-classical description of the theory is valid \footnote{The fact that the mass gaps of this theory is much larger than $\Lambda_{\scriptsize \mbox{QCD}}$ should not come as a surprise since we have two scales: $\Lambda_{\scriptsize \mbox{QCD}}$ and $L$. What really matters for the validity of the semi-classical description is that ${\cal MG}\ll M_W$.} all the way up to $L \rightarrow \infty$. One can also check that the actions of the bions are much bigger than unity at the critical radius,  which lends more confidence in the dilute gas semi-classical description at all radii. In the decompactification limit $L\rightarrow\infty$ all the masses vanish and the theory flows to a conformal field theory.  Thus, $su(5)$ with $n_G=3, N_F=8$ belongs to a class of theories that are amenable to semi-classical description for all $0<L<\infty$. The only other two theories that are known to belong to this class are $su(5)$ with $n_G=5$ and $su(2)$ with $n_{(40)}=1$ \cite{Poppitz:2009uq,Poppitz:2009tw}. Analyzing the rest of the admissible theories found in this work to check whether there are more theories that belong to this class will be carried out in a future work.

\subsection*{${\cal C}\,,{\cal P}\,,{\cal T}$ symmetries}

\begin{figure}[t] 
   \centering 
	\includegraphics[width=5in]{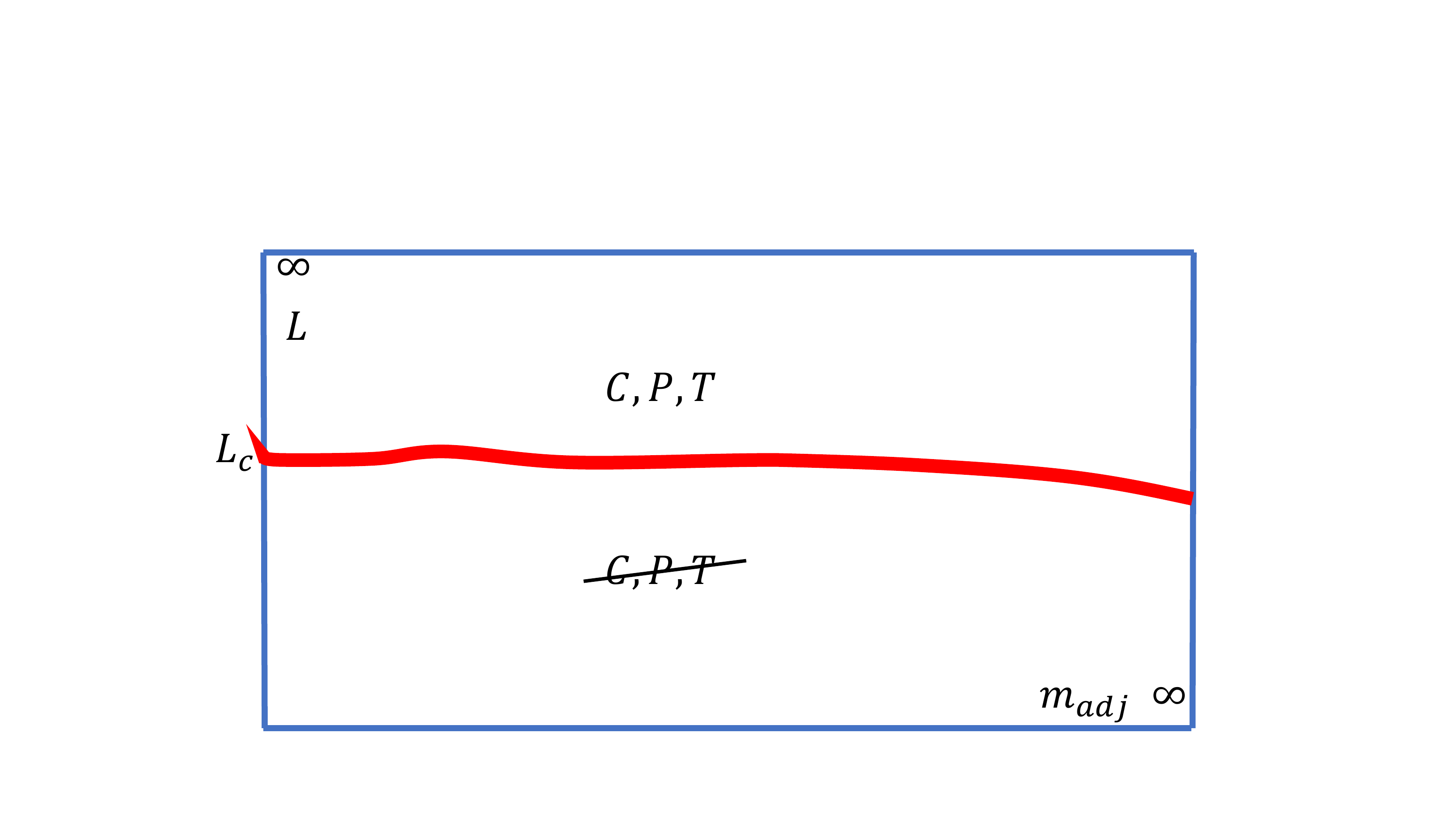}
	\includegraphics[width=5in]{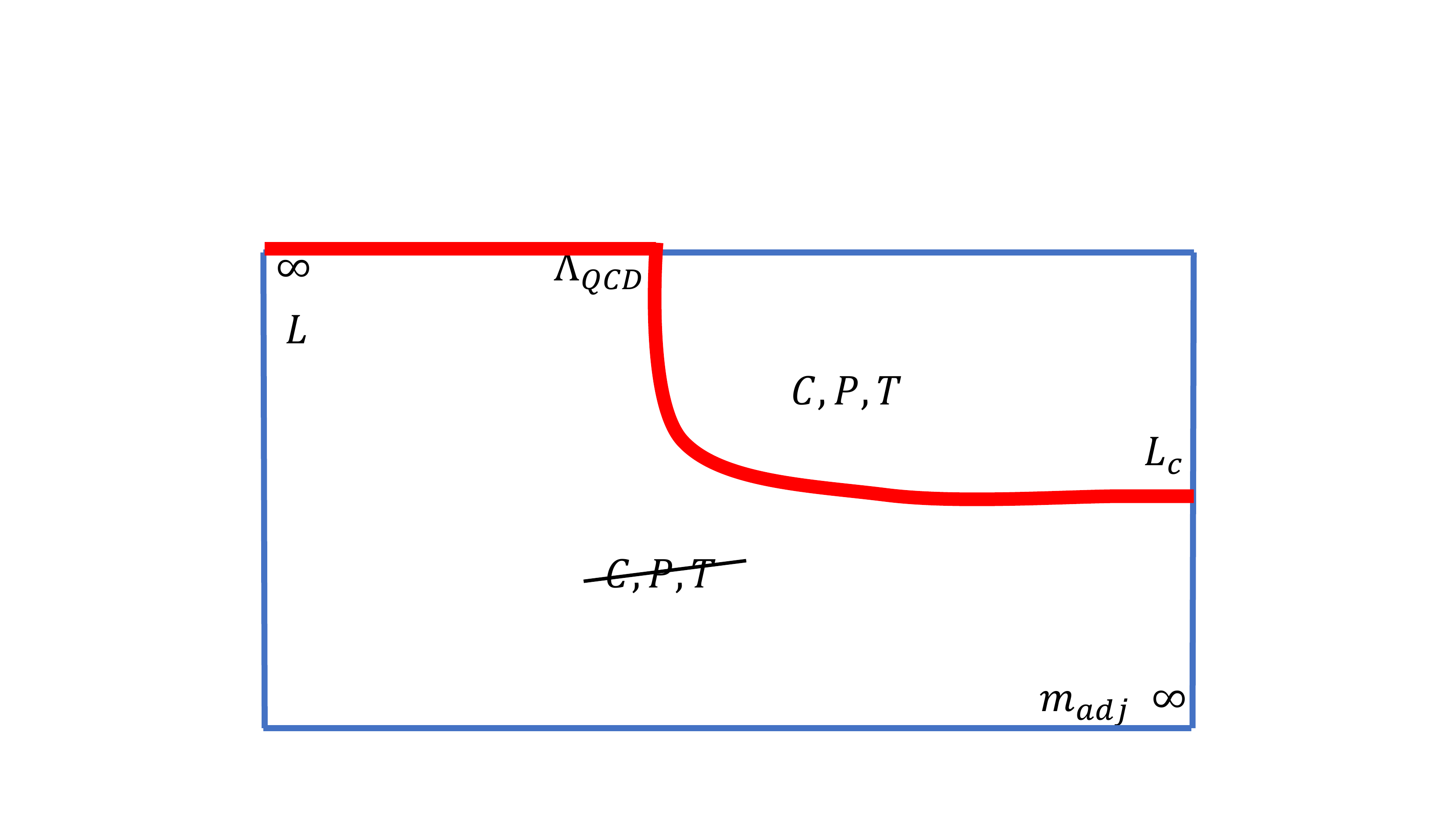} 
   \caption{$L-m_{adj}$ phase diagrams. On the top: $n_G=2, N_F=3$. Theories with a small number of fundamental fermions are expected to have a similar phase diagram. On the bottom: $n_G=3, N_F=8$. Theories with a large number of fundamental fermions are expected to have a similar phase diagram;  see the text for details. The red thick line separates between the ${\cal C}\,,{\cal P}\,,{\cal T}$ broken and restored phases. Notice that in the bottom diagram ${\cal C}\,,{\cal P}\,,{\cal T}$ are restored in the strict limit $L=\infty$. }
   \label{L madj phase diagram}
\end{figure}

Now, we comment on the fate of the spontaneously broken discrete  symmetries ${\cal C}\,,{\cal P}\,,{\cal T}$ in theories with $n_G\oplus N_F$ as we decompactify the circle. However, before doing that let us pause here to discuss the physical meaning and consequences of breaking these symmetries. 

First, one might wonder whether the spontaneously broken ${\cal P}$ symmetry we observe in our theories is in contradiction with Vafa-Witten theorem \cite{Vafa:1984xg}, which states that parity-conserving vector-like theories cannot have spontaneously broken parity. However, upon careful inspection of this theorem, one finds that Lorentz invariance is one of its assumptions. Compactifying a theory over a circle breaks its $4$-D Lorentz invariance, and hence, our findings are not in conflict with Vafa-Witten theorem \cite{Cohen:2001hf}. Second, an analysis of the spectrum of these theories does not reveal any unusual structures compared to theories with preserved  ${\cal C}\,,{\cal P}\,,{\cal T}$ symmetries. This is true for both the masses and charges (the perturbative spectrum) as well as monopoles (the nonperturbative spectrum). However, as we found in Section \ref{Perturbative vacua and the role of discrete symmetries}, $\mbox{Im}\left(\mbox{tr}_{\cal R}\Omega\right)$ is a gauge invariant order parameter that can signal the breaking of these symmetries. This order parameter, however, is nonlocal in nature since it wraps around the time circle. 

In fact,  $\mbox{Im}\left(\mbox{tr}_{\cal R}\Omega\right)$ is not the only gauge invariant order parameter that one can build to check the breaking of the discrete symmetries. In \cite{Lucini:2007as} a physical setup was proposed to check the breaking of charge conjugation via turning on a background $U(1)_B$ field ${\cal R}_M$ along the compact direction and then taking the limit ${\cal R}_M\rightarrow 0$. The additional term to the Lagrangian is
\begin{eqnarray}
\Delta {\cal L}=- \frac{\bar\Psi^I_F \gamma_M{\cal R}_M\Psi^I_F}{L}\,,
\end{eqnarray} 
where $\Psi_F^I$ are the fundamental Dirac fermions, $\gamma_M$ are the Dirac matrices, and the index $I$ is the flavor index which is summed over. The current $\langle J_M\rangle=\langle \bar\Psi^I_F \gamma_M\Psi_F^I\rangle$ is not invariant under ${\cal C}$ as $J_M \stackrel{{\cal C}}{\rightarrow}-J_M$, and therefore, it can serve as an order parameter for the spontaneous breaking of the charge conjugation symmetry. The finding  $\langle J_M\rangle \neq 0$ signals the breaking of ${\cal C}$ in the limit ${\cal R}_M\rightarrow 0$. An observer will see a flux of baryons flowing through the compact direction. $\langle J_M\rangle$ is calculated from the partition function ${\cal Z}[{\cal R}_M]$ as
\begin{eqnarray}
\langle J_M\rangle=-\left[\frac{\partial \log{\cal Z}[{\cal R}_M] }{\partial {\cal R}_M}\right]_{{\cal R}_M=0}\,.
\label{baryonic current}
\end{eqnarray}
The current $\langle J_M\rangle$ was used in \cite{Lucini:2007as} to study the charge conjugation symmetry breaking of QCD with fundamental fermions obeying periodic boundary conditions on the lattice. The presence of adjoint fermions in our setup provides a controlled way to study this current by analytical means in a semi-classical context. Recalling that ${\cal Z}[{\cal R}_M=0]$ is the effective potential $V_{\scriptsize\mbox{eff}}\left(\bm \Phi\right)$ in (\ref{final form of the potential in terms of the polyakov loop}), we immediately realize that ${\cal Z}[{\cal R}_M]$ can be obtained via the substitution $\bm \Phi\cdot \bm H \rightarrow \bm \Phi\cdot \bm H+{\cal R}_3$. We finally find:
\begin{eqnarray}
\langle J_3\rangle=\frac{4N_F}{\pi^2 L^3}\sum_{p=1}^\infty \frac{\mbox{Im}\left\{\mbox{tr}{\left[e^{ip\bf \Phi_0\cdot \bf H}\right]}\right\}}{p^3}\,.
\end{eqnarray}
The fact that $\mbox{Im}\left\{\mbox{tr}\left[e^{ip\bf \Phi_0\cdot \bf H}\right]\right\}\neq 0$ in $n_G\oplus N_F$ theories with odd number of colors indicates the presence of a baryonic current, and hence, the spontaneous breaking of ${\cal C}$. Interestingly, $\mbox{Im}\left\{\mbox{tr}\left[e^{ip\bf \Phi_0\cdot \bf H}\right]\right\}$ is the exact same order parameter that signals the breaking of all discrete symmetries, and therefore, the non-vanishing of the baryonic current is also an indication of breaking both ${\cal P}$ and ${\cal T}$ and in sequence ${\cal CPT}$ symmetry. 

Now, we discuss the fate of discrete symmetries in the $su(5)$ examples discussed above, namely, theories with $n_G=2\,,N_F=3$ and $n_G=3\,,N_F=8$, as we decompacftify the circle. As we explained in details, $su(5)$ with $n_G=2\,,N_F=3$ is expected to flow to a strongly coupled regime in the limit $L \gtrsim \Lambda_{\scriptsize\mbox{QCD}}$. Since in the strict limit $L \rightarrow \infty$ the parity symmetry is not spontaneously broken according to Vafa-Witten theorem, we expect ${\cal P}$ symmetry restoration to happen at some critical radius $L_c\gtrsim \Lambda_{\scriptsize\mbox{QCD}}$. Presumably, the restoration of the spontaneously broken $C$ and $T$ symmetries will happen at the same critical radius \footnote{Unlike the parity symmetry, there is no proof that a Lorentz-invariant vector-like theory cannot spontaneously break its ${\cal C}$ or ${\cal T}$ symmetries.}. If we give the adjoint fermions a mass, $m_{adj}$,  and send the mass to infinity, the adjoints decouple and we are left with a theory with only fundamentals. A simulation of $su(N_c)$ with odd $N_c$ and a small number of fundamental fermions on $\mathbb R^3 \times \mathbb S^1$ was performed in \cite{DeGrand:2006qb} and it was found that the theory experiences a spontaneous breaking of charge conjugation at $L_c\approx \Lambda_{\scriptsize\mbox{QCD}}$. Confronting the information we obtained from semi-classics with the available lattice simulations, we expect the $m_{adj}-L$ phase diagram for a small number of fundamentals to look like the top panel of Figure \ref{L madj phase diagram}. 

In the second example, $n_G=3\,,N_F=8$, the theory is under semi-classical control all the way to $L \rightarrow \infty$. All the discrete symmetries are broken at all radii except at the strict limit $L \rightarrow \infty$, where the theory flows to its conformal limit and ${\cal C}\,,{\cal P}\,,{\cal T}$ are restored. For a finite $m_{adj}$, this picture does not change as long as $m_{adj}< \Lambda_{\scriptsize\mbox{QCD}}$, since in this range of masses one can still identify a vev of $\bm \Phi$ inside the affine Weyl chamber. On the other hand, in the limit $m_{adj} \rightarrow \infty$ the theory experiences a spontaneous breaking/restoration of discrete symmetries at $L_c\approx \Lambda_{\scriptsize\mbox{QCD}}$, as can be inferred from lattice or weak-field computations, see \cite{Lucini:2007as}. The $m_{adj}-L$ phase diagram of this case is depicted in the bottom panel of Figure \ref{L madj phase diagram}.

\section{Summary and future directions}
\label{Discussion and Final Remarks}

In this work we have studied the general problem of classifying $su(N_c)$ gauge theories on $\mathbb R^3 \times \mathbb S^1$ endowed with $n_G\oplus n_{\cal R}$ fermions. Gauge theories on $\mathbb S^1$ are important class of theories since they provide a laboratory to study  interesting phenomena in a controllable way. In this regard, it is important to single out the weakly coupled theories that are amenable to semi-classical studies. We call these theories admissible in the sense that they are mathematically well defined (free from anomalies) asymptotically free theories with no massless modes in the IR. Our final results are displayed in Table \ref{summary of admissible theories}. We also computed the number of fermion zero modes in the background of monopole-instantons of these theories. Interestingly enough, some of the theories have degenerate vacua that may break the center, ${\cal P}$, ${\cal C}$, and ${\cal T}$ symmetries. One expects that the degenerate vacua will be separated by domain walls where these symmetries are restored.  

We have also studied in great details  theories with $n_G\oplus n_F$ mixed representations and found that such class of theories enjoys a plethora of new interesting phenomena. These theories can be categorized into two main groups as we decompactify the circle: (1) theories that flow to strongly coupled regime, and (2) theories that are amenable to semi-classical analysis all the way to $L\rightarrow \infty$, where $L$ is the circle circumference.  We also found that there exists a flux of baryonic current along the compact direction with theories that have broken Parity, which in the same time indicates the breaking of the charge and time-reversal symmetries. We finally studied the phase structure of this class of theories both in the small circle and decompactication limits.

\subsection{Future directions}

Theories with  $G\oplus F$ fermions serve as a prototype example of the rich phase structure of theories with mixed representations. Here, we describe possible future directions:

\begin{enumerate}

\item  The next step, which will be pursued in a future work, is to study the structure of the topological molecules that are responsible for the confinement in the infrared in the class of admissible theories.  In any $su(N_c)$ gauge theory, broken down to its maximum abelian subgroup, there are $N_c$ fundamental monopole-instantons dressed up with fermionic zero modes. Molecules made of these monopoles can form in the infrared given that (1) the molecules respect the fundamental symmetries of the theory and (2) an appropriate number of the constituent monopoles soak up their fermionic zero modes. If the resulting molecules carry a net  magnetic charge, e.g., bions, then they cause the theory to confine in the infrared. 

\item Understanding the nature of the molecules that are responsible for confinement in the admissible theories is crucial to track the mass gap as we decompactify the radius. This is especially important in order to compare and contrast theories on $\mathbb R^3 \times \mathbb S^1$ with those on $\mathbb R^4$.

\item It will be of immense importance to sort out all the admissible theories with Banks-Zaks fixed points. The existence of an IR fixed point before the theory enters its strongly coupled regime means that the semi-classical description of the theory is valid for all compactification radii.  

\item In \cite{Anber:2011gn}, a duality was established between two-dimensional XY-spin models with symmetry-breaking perturbations and QCD with adjoint fermions on a small circle and considered at temperatures near the deconfinement transition, i.e. QCD(adj) on $\mathbb R^2\times \mathbb S_\beta^1\times \mathbb S_L^1$, where $\mathbb S_\beta^1$ and $\mathbb S_L^1$ are respectively the thermal and spatial circles. The connection between the XY-spin model and QCD(adj) was made by mapping the partition functions of both theories to a  multi-component electric-magnetic Coulomb gas. This duality was also examined in \cite{Anber:2012ig,Anber:2013doa,Teeple:2015wca}. It will be interesting to further study this duality in theories with mixed representations.  

\item Confining strings in QCD with adjoint fermions on $\mathbb R^3\times S^1$ were studied in \cite{Anber:2015kea}. This study revealed that strings in this theory are made of two domain walls, which is attributed to the composite nature of the bions. The proliferation of more complex molecules in the vacuum will be accompanied by more complex string structure. Studying the nature of strings in the admissible theories found in this work will be pursued in a future work.

\item It has been known since the seminal work of \"Unsal and Yaffe \cite{Unsal:2006pj} (see also \cite{Hollowood:2006cq}) that Armoni-Shifman-Veneziano large-$N_c$ orientifold equivalence (which is an equivalence between QCD with adjoint fermions and QCD with two-index symmetric or antisymmetric representation \cite{Armoni:2003gp,Armoni:2003fb,Armoni:2004uu}) breaks down if the theory is put on $\mathbb R^3 \times \mathbb S^1$, for a sufficiently small circle. The breaking of the equivalence is a result of the spontaneous breaking of charge conjugation symmetry ${\cal C}$. A careful inspection of Tables \ref{su2 table} to \ref{su8 table} reveals that many theories in the mixed representations (adjoint)$\oplus$(two-index symmetric) and (adjoint)$\oplus$(two-index antisymmetric) have $\mbox{tr}\Omega=0$, and therefore, they do not break ${\cal C}$ spontaneously. Despite the fact that such theories do not have a semi-classical description (since theories with $\mbox{tr}\Omega=0$ have light or massless fermions charged under $u(1)^{N_c-1}$, see Section \ref{Numerical investigation}), one should still trust the effective potential calculations since it only requires weak coupling, which is always the case for a sufficiently small circle. In the large $N_c$ limit the dimensions of adjoint, two-index symmetric, and two-index antisymmetric representations scale as $N_c^2$, and the non-breaking of ${\cal C}$ is suggesting an equivalence between the adjoint representation on one hand and  adjoint$\oplus$two-index symmetric or adjoint$\oplus$two-index antisymmetric representations on the other hand on $\mathbb R^3 \times \mathbb S^1$. Such tantalizing equivalence should be taken with a great care in the light of \cite{Cherman:2016jtu} and will be pursued somewhere else.

\end{enumerate}

\acknowledgments
 
We would like to thank Aleksey Cherman, Erich Poppitz, and Mithat \"{U}nsal for valuable discussions. 
The work of M.A. is supported by Murdock Charitable Trust and by the NSF grant PHY-1720135. LVG  acknowledges support by the Swiss National Science Foundation. 

\appendix

\section{Lie Algebra and conventions}
\label{Lie Algebra and Conventions}

In this appendix we summarize important topics and set up the convention of the Lie algebra used throughout this work. See \cite{Slansky:1981yr,Georgi:1999wka} for reviews. 

\subsection*{Definitions}

We define the Lie Algebra $g$ of a group $G$ as a collection of elements $t^a, a=1,2,...d(G)$ that satisfy the following two conditions
\begin{eqnarray}
\nonumber
(i)\quad &&\left[t^a,t^b\right]=if_{abc} t^c\,,\\
(ii) \quad&& \left[\left[t^a,t^b\right], t^c \right]+\mbox{cyclic permutations}=0 \,\,\,( \mbox{the Jacobi identity})\,.
\label{Li algebra}
\end{eqnarray}
$f_{abc}$ are called the structure constants. If we assume that $t^{a}$ are Hermitian, then $f_{abs}$ are real. Instead of (\ref{Li algebra}), one can alternatively define the Lie algebra as a collection of elements that satisfy $\left[t^a,t^b\right]=if_{abc} t^c$ with the condition that $f_{abc}$ are totally anti-symmetric constants.  

\subsection*{Cartan-Weyl bases}

The classification of Lie algebra is obtained by finding a set of $r$ mutually commuting generators $H^i$ such that
\begin{eqnarray}
\left[ H^i, H^j\right]=0\,,\quad i,j=1,2,...,r\,,
\end{eqnarray}
where $r$ is the rank of the group. 
The rest of the Lie Algebra generators $t^a$ can be cast into raising and lowering operators $E_{\boldsymbol \beta}$ and $E_{-\boldsymbol \beta}\equiv E_{\boldsymbol \beta}^\dagger$ such that
\begin{eqnarray}
\nonumber
&&\left[H^i, E_{\boldsymbol \beta}\right]=\beta^i E_{\boldsymbol \beta}\,,\\
\nonumber
&&\left[ E_{\boldsymbol \beta}, E_{-\boldsymbol \beta} \right]=\beta_iH^i\,,\\
&&\left[ E_{\boldsymbol \beta}, E_{\boldsymbol \gamma} \right]={\cal N}_{\boldsymbol \beta, \boldsymbol\gamma}E_{\boldsymbol\beta, \boldsymbol\gamma}\,.
\label{the cartan algebra}
\end{eqnarray} 
$\boldsymbol\beta= (\beta^1,\beta^2,...,\beta^r)$ are $r$-dimensional vectors called the roots. There are $d(G)-r$ roots, half of them are positive and the other half is negative. Also,  there are $(d(G)-r)/2$ raising $E_{\boldsymbol \beta}$ and $(d(G)-r)/2$ lowering $E_{-\boldsymbol \beta}$ operators corresponding to the positive and negative roots, respectively. The constants ${\cal N}_{\boldsymbol \beta, \boldsymbol\gamma}$ can be determined using the above construction; however, we will not need them in the present work.

\subsection*{Simple roots, co-roots, and weights}

We define the weights $\bm \mu$ as the eigenvalues of the generators $H^i$ in any representation ${\cal R}$:
\begin{eqnarray}
H^i|\mu, {\cal R}\rangle=\mu_i |\mu,{\cal R}\rangle\,.
\end{eqnarray}
The number of these weights is the dimension of the representation ${\cal R}$. The roots $\{\bm \beta\}$ are the weights of the adjoint representation which has dimension $d(G)=N_c^2-1$ for $su(N_c)$.  

We say that a weight is positive if its first non-zero component is positive. Then, we define the simple roots as the positive roots that cannot be written as the sum of other positive roots and we denote them by $\{\bm\alpha\}$. The number of the simple roots is the rank of the group, which is $N_c-1$ for $su(N_c)$.  The affine root is given by
\begin{eqnarray}
\boldsymbol \alpha_{0}=-\sum_{a=1}^rk_a\boldsymbol\alpha_a\,,
\end{eqnarray}
where $k_a$ are the Kac labels. For $su(N_c)$ we have $k_a=1$ for all $a=1,2,...,N_c-1$.

We also define the co-roots $\boldsymbol \alpha^*$ as 
\begin{eqnarray}
\boldsymbol \alpha^*\equiv \frac{2}{\boldsymbol \alpha^2}\boldsymbol\alpha\,.
\end{eqnarray}  
For $su(N_c)$, which is a simple-laced algebra,  we normalize the simple roots as $\bm \alpha_a^2=2$ for all $a=1,2,..,N_c-1$, and find $\bm \alpha^*=\bm \alpha$. The Affine co-root $\boldsymbol\alpha_{r+1}^*\equiv \boldsymbol\alpha_{0}^*$ is given by:
\begin{eqnarray}
\boldsymbol \alpha_{r+1}^*=-\sum_{a=1}^rk_a^*\boldsymbol\alpha_a^*\,,
\end{eqnarray}
where $k_a^*$ are the dual Kac labels. For $su(N_c)$  we have $k_a^*=1$ for all $a=1,2,..,N_c-1$.

The fundamental weights (not the weights of the fundamental representation), $\bm \omega_a$, are given by
\begin{eqnarray}
\boldsymbol\omega_a\cdot\boldsymbol \alpha^*_b=\delta_{ab}\,,
\end{eqnarray}
where $a,b=1,2,...r$. The highest weight $\bm \mu_h$ of a  representation ${\cal R}$ is a linear superposition of the fundamental weights
\begin{eqnarray}
\bm \mu_h=\sum_{a=1}^rm_a \bm\omega\,,
\end{eqnarray} 
where $\{m_a\}\in \mathbb Z^+\cup {0}$, are called the Dynkin labels (or Dynkin indices). All other weights of ${\cal R}$ can be obtained from the highest weight by successive applications of annihilation operators $E_{-\bm \alpha}$.  Thus, a representation ${\cal R}$ is denoted by its Dynkin labels:
\begin{eqnarray}
{\cal R}=(m_1,m_2,...,m_r)\,, \mbox{or simply}~(m_1m_2...m_r)~\mbox{when no confusion can arise.}
\end{eqnarray}
In general,  $(m_r,m_{r-1},...,m_1)$ is the complex-conjugate representation of $(m_1,m_2,...,m_r)$. The representation is real if $(m_1,m_2,...,m_r)=(m_r,m_{r-1},...,m_1)$. For example, the adjoint representation $(1,0,0,...,0,1)$ is real. Some of the important representations are depicted in the following Table.

\vskip 1 cm
\begin{tabular}{|c|c|}
\hline
Representation & Dynkin indices\\
\hline
Fundamental (F) & $(100...00)$\\
\hline
Anti-fundamental ($\overline{F}$)  & $(00...001)$\\ 
\hline
Adjoint (adj)  & $(100...01)$\\
\hline
$n$-index symmetric & $(n00...00)$\\
\hline
Two-index anti-symmetric & $(010...00)$\\\hline
\end{tabular}
\vskip 1 cm

It is also useful to mention that the weights of the fundamental (defining) representation of $su(N_c)$ are given by
\begin{eqnarray}
\bm\nu_a=\bm\omega_1-\sum_{b=1}^{a-1}\bm \alpha_b\,,\quad a=1,2,...,N_c.
\end{eqnarray}
%

\subsection*{Tensors and Young tableau} 

The tensor associated with the representation $(m_1,m_2,...,m_r)$ has $m_{i}$ sets of indices, for each $i$ from $1$ to $r=N_c-1$,  that are anti-symmetric within each set.  The symmetry of this tensor can be obtained from Young tableau. For example, the Young tableau of the $(3,3,3)$ representation of $su(4)$ is 

\begin{eqnarray}
\ytableausetup{centertableaux}
\begin{ytableau}
{}& {} & {} & {} &{} &{} &{} & {} &{} \\
{}& {}& {} & {} &{} &{}  \\
{} & {} &{}
\end{ytableau}\,.
\end{eqnarray}

\subsection*{$\mathbb R^{N_c-1}$ root basis for $su(N_c)$}

A convenient choice of the simple and affine roots in $su(N_c)$ is given by
\begin{eqnarray}
\{\bm\alpha_a,=\bm e_a-\bm e_{a+1}\,,1\leq a \leq N_c-1\}\,,\quad
\bm \alpha_0=\bm e_{N_c}-\bm e_1\,,
\end{eqnarray}
where $\{\bm e_i\}$ is the set of unit bases in $\mathbb R^{N_c-1}$.  In this system the roots span a hyperplane in $\mathbb R^{N_c-1}$ given by
\begin{eqnarray}
\sum_{a=1}^{N_c-1}\Phi_a=0\,.
\end{eqnarray}
The fundamental weights are given by
\begin{eqnarray}
\bm \omega_b=\left(\sum_{a=1}^b\bm e_a\right)-\frac{b}{N_c}\sum_{a=1}^{N_c-1}\bm e_{a}\,. 
\end{eqnarray}
%

\section{The Casimir and trace operators, and the dimension of representation}
\label{The Casimir and trace operators, and the dimension of representation}

Computing the $\beta$ function requires the knowledge of the Casimir and trace operators. The quadratic Casimir operator of representation ${\cal R}$,  $C_2({\cal R})$, is defined as 
\begin{eqnarray}
t^a_{\cal R}t^a_{\cal R}=C_2({\cal R})\mathbb I\,.
\label{Casimir}
\end{eqnarray}
$C_2(G)$ is the quadratic Casimir of the adjoint representation. 

$T(\cal R)$ is the trace operator in the same representation which is defined by
\begin{eqnarray}
\mbox{tr}\left[t^a_{\cal R}t^b_{\cal R}\right]=T({\cal R})\delta^{ab}\,.
\label{trace}
\end{eqnarray}
From Eqs. (\ref{Casimir}) and (\ref{trace}) one can easily obtain the useful relation 
\begin{eqnarray}
T({\cal R})d(G)=C_2({\cal R})d(\cal R)\,,
\end{eqnarray}
where $d(\cal R)$ is the dimension of the $\cal R$ representation.

For a representation ${\cal R}$ with Dynkin indices $(a_1,a_2,...,a_{N_c-1},a_{N_c-2})$, the quadratic Casimir operator is given by \cite{White:1992aa}
\begin{eqnarray}
C_2({\cal R})=\frac{1}{N_c}\sum_{m=1}^{N_c-1}\left[N_c(N_c-m)ma_m+m(N_c-m)a_m^2+\sum_{n=0}^{m-1}2n(N_c-m)a_na_m \right]\,,
\end{eqnarray}
and the dimension of the representation is 
\begin{eqnarray}
d({\cal R})=\prod_{p=1}^{N_c-1}\left\{\frac{1}{p!}\prod_{q=p}^{N_c-1}\left[\sum_{r=q-p+1}^q(1+a_r) \right]\right\}\,.
\end{eqnarray}
In particular, we have $C_2(G)=2N_c$ and $d(G)=N_c^2-1$.

\section{Cubic Dynkin index}
\label{Cubic Dynkin index}

In this Appendix we list the values of the Cubic Dynkin index (or the anomaly of the representation) $A({\cal R})$ for a few important representations. For a complex representation ${\cal R}$ we have
\begin{eqnarray}
\mbox{tr}_{\cal R}\left[\left\{t_a,t_b \right\}t_c\right]=d_{abc}A({\cal R})\,,
\end{eqnarray}
where $d_{abc}$ is a third-rank tensor made out of the structure constants $f_{abc}$. Taking the complex conjugation can show that 
\begin{eqnarray}
\mbox{tr}_{\bar{\cal R}}\left[\left\{t_a,t_b \right\}t_c\right]=-d_{abc}A({\cal R})\,,
\end{eqnarray}
and therefore, real representations (the ones that satisfy $(m_1,m_2,...,m_r)=(m_r,m_{r-1},...,m_1)$) have vanishing Cubic Dynkin index. In Table \ref{cubic index of Dynkin table} we list $A({\cal R})$ for a few of the asymptotically free representations we encounter in this work, see e.g. \cite{Ramond:2010zz}.

\begin {table}
\begin{center}
\tabcolsep=0.11cm
\footnotesize
\begin{minipage}[t]{0.5\linewidth}
\caption{Cubic Dynkin index}
\begin{tabular}[t]{|*{3}{c|}}
\hline
Group & ${\cal R}$ & $A({\cal R})$\\
\hline
$su(3)$& $(10)$ & $1$ \\ 
& $(20)$ & $7$ \\ 
& $(11)$ & $0$ \\ 
& $(30)$ & $27$ \\ 
& $(21)$ & $14$ \\ 
 & &\\
& &\\
\hline
$su(4)$& $(100)$ & $1$\\
& $(010)$ & $0$\\
& $(200)$ & $8$\\
& $(101)$ & $0$\\
& $(020)$ & $0$\\
& $(110)$ & $7$\\
& $(300)$ & $29$\\
 & &\\
\hline
\end{tabular}
\begin{tabular}[t]{|*{3}{c|}}
\hline
Group & ${\cal R}$ & $A({\cal R})$\\
\hline
$su(5)$& $(1000)$ & $1$\\
& $(0100)$ & $1$\\
&$(0200)$ & $15$\\
&$(0100)$ & $1$\\
&$(1001)$ & $0$\\
&$(1100)$ & $16$\\
&$(1010)$ & $6$\\
\hline
$su(6)$& $(10000)$ & $1$\\
& $(01000)$ & $2$\\
& $(00100)$ & $0$\\
& $(20000)$ & $10$\\
& $(11000)$ & $27$\\
& $(10001)$ & $0$\\
& $(10010)$ & $4$\\
& $(10100)$ & $22$\\
\hline
\end{tabular}
\label{cubic index of Dynkin table}
\end{minipage}
\end{center}
\end{table}

\section{Constructing the weights using Verma bases}
\label{Constructing the wights using Verma bases}

One can use Verma bases to construct the weights of any representation of $su(N_c)$ in a systematic way. For representation ${\cal R}$ of $su(N_c)$, which we denote by ${\cal R}=(m_1,m_2,...,m_{N_c-1})$  the basis vectors are
\begin{eqnarray}
\nonumber
&&\left[\left(E_{-\bm \alpha_1} \right)^{a_N}\left(E_{-\bm \alpha_2} \right)^{a_{N-1}}...\left(E_{-\bm \alpha_{N_c-1}} \right)^{a_{N-N_c+2}} \right]\left[\left(E_{-\bm \alpha_1} \right)^{a_{N-N_c+1}}...\left(E_{-\bm \alpha_{N_c-2}} \right)^{a_{N-2N_c+4}} \right]\\
&&\times...\left[\left(E_{-\bm \alpha_1} \right)^{a_3}\left(E_{-\bm \alpha_2} \right)^{a_2}\right]\left(E_{-\bm \alpha_1} \right)^{a_1}|{\cal R}\rangle\,,
\label{verma labels}
\end{eqnarray}
where $N=N_c(N_c-1)/2$ and $\{E_{-\bm \alpha_a} \}$, $a=1,2,...,N_c-1$ is the set of the simple-root generators. The coefficients $\{a_i\}$ satisfy a set of inequalities that are given in Table \ref{verma basis table}, see \cite{:/content/aip/journal/jmp/27/3/10.1063/1.527222}.

 As an example, let us work out the weights of $su(3)$ algebra. For a given representation $(m_1,m_2)$, the bases are given according to (\ref{verma labels}) by
\begin{eqnarray}
\left(E_{-\bm \alpha_1} \right)^{a_3}\left(E_{-\bm \alpha_2} \right)^{a_2}\left(E_{-\bm \alpha_1} \right)^{a_1}|(1,1)\rangle,
\end{eqnarray}
such that $a_1$, $a_2$, and $a_3$ satisfy the inequalities
\begin{eqnarray}
0 \leq a_1\leq m_1\,,\quad 0 \leq a_2 \leq m_2+a_1\,,\quad 0\leq a_3\leq \mbox{min}\left[m_2,a_2\right]\,.
\end{eqnarray}
Then, for example, the basis of the adjoint representation, $G=(1,1)$,  are given by
\begin{eqnarray}
\nonumber
&&\left\{|(1,1)\rangle\,,E_{-\bm \alpha_1} |(1,1)\rangle\,,E_{-\bm \alpha_2} |(1,1)\rangle\,,E_{-\bm \alpha_2}E_{-\bm \alpha_1} |(1,1)\rangle \,,E_{-\bm \alpha_1}E_{-\bm \alpha_2} |(1,1)\rangle\right.\,,\\
&&\left.\left(E_{-\bm \alpha_2}\right)^2E_{-\bm \alpha_1} |(1,1)\rangle\,,E_{-\bm \alpha_1}E_{-\bm \alpha_2}E_{-\bm \alpha_1} |(1,1)\rangle\,,E_{-\bm \alpha_1} \left(E_{-\bm \alpha_2}\right)^2E_{-\bm \alpha_1} |(1,1)\rangle \right\}\,.
\label{1 1 representation}
\end{eqnarray}
The simple roots and fundamental weights of $su(3)$ are 
\begin{eqnarray}
\nonumber
\bm \alpha_1 = \left(\frac{1}{2}, \frac{\sqrt 3}{2}\right)\,,\quad \bm \alpha_2 = \left(\frac{1}{2}, -\frac{\sqrt 3}{2}\right)\,,\\
\bm \omega_1 = \left(\frac{1}{2}, \frac{1}{2\sqrt 3}\right)\,,\quad \bm \omega_2=\left(\frac{1}{2}, -\frac{1}{2\sqrt 3}\right)\,.
\end{eqnarray}
Now remembering that any representation $|(n_1,n_2)\rangle\equiv n_1 \bm \omega_1+n_2\bm \omega_2$, we can construct all the weights from (\ref{1 1 representation}) by subtracting $\bm\alpha_1$ and/or $\bm\alpha_2$ roots from $|(1,1)\rangle$. 

\begin {table}
\begin{center}
\tabcolsep=0.11cm
\footnotesize
\caption{Verma bases inequalities \cite{:/content/aip/journal/jmp/27/3/10.1063/1.527222}}
\begin{tabular}[t]{|*{2}{c|}}
\hline
$su(2)$ & $0\leq a_1 \leq m_1$ \\
         & $0\leq a_2\leq m_2+a_1$\\
$su(3)$  & $0\leq a_3\leq \mbox{min}\left[m_2,a_2\right]$\\
          &$0\leq a_4 \leq m_3+a_2$\\
					& $0\leq a_5\leq \mbox{min}\left[m_3+a_3,a_4\right]$\\
$su(4)$    & $0\leq a_6\leq \mbox{min}\left[m_3,a_5\right]$\\
. &  ...\\  
.&  ...\\ 
.&  ...\\	
$su(N_c)$ &  $0\leq a_{N_c-N}\leq m_N+a_{N_c+1-2N}$\\
          & $0\leq a_{N_c-N-3}\leq \mbox{min}\left[m_N+a_{N_c+2-2N},a_{N_c-N}\right]$	\\	
					& $0\leq a_{N_c-N+2}\leq \mbox{min}\left[m_N+a_{N_c-2N+3},a_{N_c-N+1} \right]$  \\	
					&...\\
					&...\\
					& $0\leq a_{N_c-1}\leq \mbox{min}\left[m_{N},a_{N_c-2}\right]$\\					
					\hline
\end{tabular}
\label{verma basis table}
\end{center}
\end{table}

\section{Frobenius formula and traces of the asymptotically free theories}
\label{Using the Frobenius formula}

In this Appendix we give examples that illustrate the usefulness of the Frobenius formula given by (\ref{Frobenius formula}). First, one needs to construct the vector  $\{j_1,j_2,...,j_n\}$, which is the permutations  of the symmetric group $S_n$. As we mentioned in the main text, the vector  $\{j_1,j_2,...,j_n\}$ can be obtained as the solution of the equation $1j_1+2j_2+...+nj_n=n$ for all integers $j_i\geq 0$.
For example, for $n=3$ we have $\{ j_1,j_2,j_3\}=\{3,0,0\},\{1,1,0\},\{0,0,1\}$ and for $n=4$ we have $\{j_1,j_2,j_3,j_4\}=\{4,0,0,0\},\{1,0,1,0\},\{0,2,0,0\},\{2,1,0,0\},\{0,0,0,4\}$, etc. Using this information, we obtain, for example,
\begin{eqnarray}
\nonumber
\mbox{tr}_{(2,\vec 0)}P&=&\frac{1}{2}\left[\left(\mbox{tr}_FP\right)^2+\left(\mbox{tr}_FP^2\right) \right]\,,\\
\mbox{tr}_{(3,\vec 0)}P&=&\frac{1}{3!}\left[ \left(\mbox{tr}_FP\right)^3+3\left(\mbox{tr}_FP\right)\left(\mbox{tr}_FP^2\right)+2\left(\mbox{tr}_FP^3\right)\right]\,,\mbox{etc,}
\label{traces of n rep}
\end{eqnarray}
where $\mbox{F}\equiv(1,\vec 0)$.
Other representations can be obtained from $(n,\vec 0)$ representations using the Young tableau. The idea is to express the direct products of two representations as the direct sum of other representations. Since the trace of the product is equal to the sum of the traces, then one can use this property to express the trace of a general representation  in terms of traces of $(n, \vec 0)$ representations. As an example, 
\begin{eqnarray}
(1,0,...,0)\otimes(1,0,...,0)=(0,1,0,...,0)\oplus(2,0,..,0)\,,
\end{eqnarray}
and hence 
\begin{eqnarray}
\mbox{tr}_{(0,1,0,...,0)}P=\left(\mbox{tr}_{F}P\right)^2-\left(\mbox{tr}_{(2,0,0,...,0)} P\right)\,.
\end{eqnarray}
Next, we can use Eq. (\ref{traces of n rep}) to express $\left(\mbox{tr}_{(2,0,0,...,0)}P\right)$ in terms of the trace of the fundamental representation to finally obtain:
\begin{eqnarray}
\mbox{tr}_{(0,1,0,...,0)}P\equiv\mbox{tr}_{AS}P=\frac{1}{2}\left[\left(\mbox{tr}_{F}P\right)^2-\left(\mbox{tr}_F P^2\right)\right]\,.
\end{eqnarray}
%

\subsection*{Traces of the asymptotically free representations}
\label{Append A}

Now, we list all the needed traces in the asymptotically free theories. As is explained above, the expressions of these traces in terms of the fundamental (defining) trace can be obtained from the Young tableau. 

First we list general expressions of traces that are valid for any number of colors $N_c\geq 3$:
\begin{eqnarray}
\mbox{tr}_{(0,1,0,...,0)}P&\equiv& \mbox{tr}_{\mbox{AS}}P =\frac{1}{2}\left[\left(\mbox{tr}_FP\right)^2-\left(\mbox{tr}_FP^2\right) \right]\,,\\
\mbox{tr}_{(1,0,0,...,1)}P&\equiv& \mbox{tr}_{\mbox{G}}P=\left|\mbox{tr}_FP\right|^2-1\,,\\
\mbox{tr}_{(2,0,0,...,1)}P&=&\left(\mbox{tr}_{(2,0,0,...,0)}P\right)\left(\mbox{tr}_FP\right)^*-\left(\mbox{tr}_FP\right)\,.
\label{important expressions using frobenuis formula}
\end{eqnarray}

Next, we list the needed traces in all other groups:
\begin{enumerate}
\item $su(4)$
\begin{eqnarray}
\mbox{tr}_{(1,1,0)}P&=&\left(\mbox{tr}_{AS}P\right)\left(\mbox{tr}_{F}P\right)-\left(\mbox{tr}_{F}P\right)^*\,\\
\mbox{tr}_{(0,2,0)}P&=&\left(\mbox{tr}_{AS}P\right)^2-\left(\mbox{tr}_{G}P\right)-1\,.
\end{eqnarray}
\item $su(5)$
\begin{eqnarray}
\mbox{tr}_{(0,1,0,1)}P&=&\left(\mbox{tr}_{G}P\right)\left(\mbox{tr}_{F}P\right)-\left(\mbox{tr}_{(2,0,0,1)}P\right)-\left(\mbox{tr}_{F}P\right)\,,\\
\mbox{tr}_{(0,2,0,0)}P&=&\left(\mbox{tr}_{AS}P\right)^2-\left(\mbox{tr}_{(0,1,0,1)}P\right)^*-\left(\mbox{tr}_{F}P\right)^*\,,\\
\mbox{tr}_{(0,1,1,0)}P&=&\left(\mbox{tr}_{AS}P\right)^2-\left(\mbox{tr}_{G}P\right)-1\,,\\
\mbox{tr}_{(0,1,1,0)}P&=&\left(\mbox{tr}_{(2,0,0,0)}P\right)\left(\mbox{tr}_{F}P\right)-\left(\mbox{tr}_{(2,0,0,0)}P\right)\,.
\end{eqnarray}
\item $su(6)$
\begin{eqnarray}
\mbox{tr}_{(1,0,0,1,0)}P&=&\left(\mbox{tr}_{AS}P\right)^*\left(\mbox{tr}_{F}P\right)-\left(\mbox{tr}_{F}P\right)^*\,,\\
\mbox{tr}_{(1,1,0,0,0)}P&=&\left(\mbox{tr}_{(2,0,0,0)}P\right)\left(\mbox{tr}_{F}P\right)-\left(\mbox{tr}_{(3,0,0,0)}P\right)\,,\\
\mbox{tr}_{(0,0,1,0,0)}P&=&\left(\mbox{tr}_{AS}P\right)\left(\mbox{tr}_{F}P\right)-\left(\mbox{tr}_{(1,1,0,0,0)}P\right)\,,\\
\mbox{tr}_{(1,0,1,0,0)}P&=&\left(\mbox{tr}_{(0,0,1,0,0)}P\right)\left(\mbox{tr}_{F}P\right)-\left(\mbox{tr}_{AS}P\right)^*\,,\\
\mbox{tr}_{(0,2,0,0,0)}P&=&\left(\mbox{tr}_{AS}P\right)^2-\left(\mbox{tr}_{(1,0,1,0,0)}P\right)-\left(\mbox{tr}_{AS}P\right)^*\,.
\end{eqnarray}
\item $su(7)$
\begin{eqnarray}
\mbox{tr}_{(1,0,0,0,1,0)}P&=&\left(\mbox{tr}_{AS}P\right)^*\left(\mbox{tr}_{F}P\right)-\left(\mbox{tr}_{F}P\right)^*\,.
\end{eqnarray}
The rest of the traces can be read from the $su(6)$ traces by adding $0$ to the last entry in any vector:  $(a_1,a_2,...,a_5)\rightarrow (a_1,a_2,...,a_5,0)$\,.
\item $su(8)$ and $su(9)$

The traces can be obtained from those of $su(6)$.

\end{enumerate}
%

\section{Computing the index using Frobenius formula}
\label{Computing the index using Frobenius formula}

In this Appendix we show how to compute the index (\ref{fermion zero modes index}) using Frobenius formula. First, we note that the quantity $\lfloor{\frac{\bm \Phi\cdot \bm H}{2\pi}}\rfloor\bm \alpha^*_a\cdot \bm H$ is not an element of $su(N_c)$, and hence, we cannot apply the Frobenius formula directly. In order to overcome this problem, we use the following definition of the floor function
\begin{eqnarray}
\lfloor x\rfloor=x-\frac{1}{2}+\frac{1}{\pi}\sum_{k=1}^\infty\frac{\sin(2\pi k x)}{k}\,,
\end{eqnarray}
provided that $x$ is not an integer. We also can write
\begin{eqnarray}
\bm \alpha^*\cdot \bm H=-i\left[\frac{\partial  e^{i \epsilon \bm \alpha^*\cdot \bm H}}{\partial{\epsilon}}\right]_{\epsilon=0}\,.
\end{eqnarray}
 Now the quantity $e^{i \epsilon \bm \alpha^*\cdot \bm H} \in su(N_c)$ and we can readily apply the Frobenius formula. Repeating this procedure, we finally obtain the index which can be written as
\begin{eqnarray}
\nonumber
{\cal I}_{f(\bm\alpha_a^*)}({\cal R})&=&n_{\cal R}\left\{-\frac{\partial}{\partial \epsilon_1\epsilon_2} \mbox{tr}_{\cal R}\left[e^{i\left(\epsilon_1 \bm \alpha^*_a\cdot \bm H+\epsilon_2\frac{\bm \Phi\cdot \bm H}{2\pi}\right)}\right]+\frac{i}{2}\frac{\partial}{\partial\epsilon_1}\mbox{tr}_{\cal R}\left[e^{i\epsilon_1\bm \alpha^*_a\cdot \bm H}\right]\right.\\
\nonumber
&&\left.-\mbox{Im}\left[\sum_{k=1}^{\infty}\frac{i}{\pi k}\frac{\partial}{\partial\epsilon_1}\mbox{tr}_{\cal R}\left[e^{i\left(\epsilon_1\bm \alpha_a^*\cdot \bm H+k\bm \Phi\cdot \bm H\right)}\right] \right]\right\}_{\epsilon_1=\epsilon_2=0}\,,\quad a=1,2,...,N_c-1\,.\\
\end{eqnarray}

\bibliography{References}

\bibliographystyle{JHEP}

\end{document}